\documentclass[12pt]{article}
\usepackage[margin=1in]{geometry}
\usepackage{times}
\usepackage{color}
\usepackage{graphicx}
\usepackage[colorlinks,citecolor=blue,urlcolor=blue,linkcolor=blue,bookmarks=false,hypertexnames=true]{hyperref} 

\usepackage{multirow}
\usepackage{amsmath}
\usepackage{amsthm}
\usepackage{amssymb}
\usepackage{natbib}
\usepackage{rotating}

\DeclareMathOperator*{\argmax}{arg\,max}
\DeclareMathOperator*{\argmin}{arg\,min}

\begin{document}

\title{On Weighted Trigonometric Regression for Suboptimal Designs in Circadian Biology Studies}
\author{%
	Michael T. Gorczyca \\
	MTG Research Consulting\\
	{\texttt{mtg62@cornell.edu}} 
	\and
	Justice D. Sefas \\
	University of British Columbia \\
	{\texttt{jsefas@cs.ubc.ca}} \\
}
\date{\today}

\maketitle

\begin{abstract}
    \noindent Studies in circadian biology often use trigonometric regression to model phenomena over time. Ideally, protocols in these studies would collect samples at evenly distributed and equally spaced time points over a 24 hour period. This sample collection protocol is known as an equispaced design, which is considered the optimal experimental design for trigonometric regression under multiple statistical criteria. However, implementing equispaced designs in studies involving individuals is logistically challenging, and failure to employ an equispaced design could cause a loss of statistical power when performing hypothesis tests with an estimated model. This paper is motivated by the potential loss of statistical power during hypothesis testing, and considers a weighted trigonometric regression as a remedy. Specifically, the weights for this regression are the normalized reciprocals of estimates derived from a kernel density estimator for sample collection time, which inflates the weight of samples collected at underrepresented time points. A search procedure is also introduced to identify the concentration hyperparameter for kernel density estimation that maximizes the Hessian of weighted squared loss, which relates to both maximizing the $D$-optimality criterion from experimental design literature and minimizing the generalized variance. Simulation studies consistently demonstrate that this weighted regression mitigates variability in inferences produced by an estimated model. Illustrations with three real circadian biology data sets further indicate that this weighted regression consistently yields larger test statistics than its unweighted counterpart for first-order trigonometric regression, or cosinor regression, which is prevalent in circadian biology studies.
\end{abstract}


\section{Introduction}\label{sec1}

The circadian clock gene network describes a set of timekeeping genes that operate in transcriptional and translational feedback loops with an approximate 24 hour cycle \citep{Gentry2021, Lowrey2004, Lowrey2011}. These timekeeping genes regulate the expression of many genes within an individual, which manifests in the expression levels of many genes displaying oscillatory behavior over a 24 hour period \citep{Cox2019, Koike2012, Mure2018, Ruben2018, Zhang2014}. The oscillatory behavior of these genes shape multiple behavioral and physiological facets of an individual's daily life, including their metabolism, their hormone balance, and their body temperature \citep{Dibner2015}. Notably, aberrations in these processes have been linked to an array of diseases, including cancer \citep{Schernhammer2003, Sigurdardottir2012}, heart disease \citep{Crnko2019}, and mental illness \citep{Walker2020}. The connection between the oscillatory behavior of many genes and an individual's health has motivated several circadian biology studies that study the behavior of these genes under different experimental conditions \citep{Ceglia2018, RijoFerreira2019}, often with the aim of identifying therapeutic treatments for diseases \citep{Chan2017, Chauhan2017, Halberg2013, Haus2009, Li2013}.

Many circadian biology studies are conducted with non-human species \citep{Almon2008, Lemos2006, Mure2018, Tovin2012, Yang2020, Zhang2014}, which gives an investigator control over the times at which samples are collected \citep{Hughes2017}. Ideally, these studies would employ an equispaced experimental design for sample collection, where samples are systematically gathered at evenly distributed and equally spaced intervals throughout a 24 hour period. This design satisfies multiple optimality criteria for the trigonometric regression model (\citealp[pages 94-97]{Federov1972}; \citealp[pages 241-243]{Pukelsheim2006}), which is frequently used to represent biological phenomena over time \citep{Cornelissen2014, Hughes2009, Tong1976}. 

Despite the advantages provided by an equispaced design for trigonometric regression, equispaced designs are difficult to employ in circadian biology studies conducted with human cohorts due to logistical constraints. One constraint can occur when a study cohort consists of individuals who are ill, where it is unethical to collect samples outside of their waking hours \citep{Kitsos1988}. A second constraint can occur when data are obtained from multi-cohort repositories \citep{gtex2020, Hoffman2019}, where the investigator does not have control of the time at which samples were collected. A third constraint can occur when the times of sample collection are mis-recorded. To clarify, the recorded times of sample collection are based on day-night cycle time. However, biological phenomena such as gene expression for individuals instead occurs at times that are based on individual-specific internal timing systems. In these scenarios, additional laboratory procedures are performed to determine the true times of sample collection for each individual \citep{Archer2014, Braun2019, Huang2024, Laing2019, MllerLevet2013, Wittenbrink2018}, which results in the true time points being distinct from the recorded time points specified by the investigator in their experimental design.

This paper is motivated by circadian biology studies where samples are not collected from an equispaced design, which could affect the statistical power of hypothesis tests computed with an estimated model. To mitigate the challenges arising from suboptimal sample collection, we propose a weighted trigonometric regression for estimation and inference. Specifically, the sample weights are the normalized reciprocals of estimates derived from a kernel density estimator for the time of sample collection, which inflates the contribution of underrepresented time points. We also propose an optimization procedure to identify the concentration hyperparameter for kernel density estimation based on the $D$-optimality criterion from experimental design literature \citep{Kitsos1988}.

The remainder of this paper is organized as follows. In Section \ref{sec:2}, a motivating example and the weighted regression are presented. In Section \ref{sec:3}, Monte Carlo simulation studies are performed to assess the utility of this weighted regression relative to an unweighted regression. In Section \ref{sec:4}, illustrations with three circadian biology data sets are provided. Finally, in Section \ref{sec:5}, we conclude this paper with some discussion on the weighted regression proposed and directions for future work.

\section{Methodology} \label{sec:2}

\subsection{Preliminaries} \label{sec:2.1}

Suppose an investigator performs a circadian biology experiment in which data are collected for multiple genes. Specifically, the investigator obtains data $\{(X_i,  Y^{(1)}_i,\ldots, Y^{(G)}_i), \ i = 1,\ldots, N\}$ that consist of $N$ independent realizations of the random variables $(X, Y^{(1)},\ldots, Y^{(G)})$. Here, $X_i$ denotes the time that the $i$-th sample was collected, and $Y_i^{(g)}$ denotes the expression levels of the $g$-th gene recorded at $X_i$. It is noted that this paper adopts a notation convention where estimators, statistics, and other quantities related to the $g$-th gene will be denoted with $(g)$ in their superscript.

In this paper, it is assumed that the trigonometric regression model of order $K$ is correctly specified for modelling each gene's expression levels given time, or
\begin{equation} \label{eq:1}
    Y^{(g)} = (\theta^{(g)})^Tf(X) + \epsilon = \theta^{(g)}_0 + \sum_{k=1}^K\left\{\theta^{(g)}_{2k- 1}\text{\text{sin}}\left(\frac{\pi k X}{12}\right) + \theta^{(g)}_{2k}\text{\text{\text{cos}}}\left(\frac{\pi k X}{12}\right)\right\} + \epsilon^{(g)}.
\end{equation}
Here, $\theta^{(g)}$ denotes a vector of unknown model parameters, and
\begin{align}
    f(X) &= \{f_0(X),\ldots,f_{2K}(X)\}^T \nonumber \\
    &= \left\{1, \text{sin}\left(\frac{\pi X}{12}\right), \text{cos}\left(\frac{\pi X}{12}\right), \ldots, \text{sin}\left(\frac{\pi K X}{12}\right),\text{cos}\left(\frac{\pi K X}{12}\right)\right\}^T \label{eq:f}
\end{align}
is a vector of regression functions. The random noise $\epsilon^{(g)}$ is generated by a probability distribution with mean zero and finite, constant variance that only relates to the response $Y^{(g)}$ \citep{Tong1976}. Each $\theta^{(g)}$ is estimated by minimizing squared loss, or
\begin{align*}
\hat{\theta}^{(g)} =& \argmin_{\theta \in \Theta} \ \ 
\frac{1}{N}\sum_{i=1}^N\mathcal{L}(Y_{i}^{(g)}, X_i; \theta) = \argmin_{\theta \in \Theta} \ \ \frac{1}{N}\sum_{i=1}^N\left\{Y_{i}^{(g)}-\theta^Tf(X_i)\right\}^2,
\end{align*}
which corresponds to the assumption that the random noise is generated from a standard normal distribution. Here, $\hat{\theta}^{(g)}$ denotes the parameter vector estimate that minimizes squared loss, and $\Theta\subseteq \mathbb{R}^{2K+1}$ denotes a compact subset of a Euclidean space, with each corresponding $\theta^{(g)}$ a point in the interior of $\Theta$. It is emphasized that many circadian biology studies specify $K=1$ \citep{Archer2014, delolmo2022, MllerLevet2013}, which is referred to as cosinor regression in circadian biology \citep{Cornelissen2014}. To the best of our knowledge, no circadian biology study has considered $K>3$ \citep{Hughes2009}. It is also noted that the assumption random noise is generated from a normal distribution is maintained for inference \citep{Carlucci2019, Zong2023}, where the empirical variance of the parameter vector is defined as
\begin{align}
\text{Var}_N(Y^{(g)}, X; \hat{\theta}^{(g)}) &= (\hat{\sigma}^{(g)})^2 A_N(Y^{(g)}, X; \hat{\theta}^{(g)})^{-1}, \label{eq:asy_var}
\end{align}
with 
\begin{align*}
 A_N(Y^{(g)}, X; \theta) = \frac{d^2}{d\theta d\theta^T}\left\{\frac{1}{N}\sum_{i=1}^N\mathcal{L}(Y_{i}^{(g)}, X_i; \theta)\right\} = \frac{1}{N}\sum_{i=1}^Nf(X_i)f(X_i)^T.
\end{align*}
The estimator of the variance for gene-specific random noise is defined as
\begin{align*}
(\hat{\sigma}^{(g)})^2 = \frac{1}{N-2K-1}\sum_{i=1}^N\left\{Y_i-(\hat{\theta}^{(g)})^Tf(X_i)\right\}^2.
\end{align*}

In studies where samples are collected from non-human species, it would be optimal under multiple statistical criteria to collect data once every $j$ hours such that $j$ is evenly distributed and equally spaced, or to collect data from an equispaced design \citep[pages 241-243]{Pukelsheim2006}. Specifically, an equispaced design is known to satisfy every statistical criterion within the class of $\phi_p$-criteria for trigonometric regression when $p$ is a constant within the interval $[-\infty, 1]$ \citep[page 241]{Pukelsheim2006}. To clarify, let $\lambda_j$ denote the $j$-th eigenvalue of $A_N(Y^{(g)}, X; \theta)$. Then the class of $\phi_p$-criteria is defined as
\begin{equation}
    \phi_p\left\{A_N(Y^{(g)}, X; \theta)\right\} = 
    \begin{cases}
        \max_{j \in \{1,\ldots,2K+1\}} \lambda_j & \text{for \ $p=\infty$,} \\
       \min_{j \in \{1,\ldots,2K+1\}} \lambda_j & \text{for \ $p=-\infty$,} \\
       \prod_{j=1}^{2K+1}\lambda_j & \text{for \ $p=0$,} \\
       \left(\frac{1}{2K+1}\sum_{j=1}^{2K+1}\lambda_j^p \right)^{1/p} &\text{for $p \in (-\infty, 0) \cup (0, \infty)$.}
    \end{cases} \label{eq:phi_def}
\end{equation}
with respect to the matrix $A_N(Y^{(g)}, X; \theta)$ \citep[page 141]{Pukelsheim2006}. Three popular optimality criteria belonging to this class are the $A$-optimality criterion ($p=-1$), the $D$-optimality criterion ($p=0$), and the $E$-optimality criterion ($p=-\infty$) \citep[pages 135-138]{Pukelsheim2006}. $A$-optimality indicates that the trace of $A_N(Y^{(g)}, X; \theta)^{-1}$ is minimized, or the average variance of the model parameter estimates is minimized \citep[page 137]{Pukelsheim2006}. $D$-optimality indicates that the determinant of $A_N(Y^{(g)}, X; \theta)$ is maximized, or equivalently, the generalized variance is minimized \citep{Kitsos1988}. Finally, $E$-optimality indicates that the smallest eigenvalue of $A_N(Y^{(g)}, X; \theta)$ is maximized, which is associated with minimizing the worst possible variance of an estimator for all possible linear combinations of $c^T\theta$, where $c \in \mathbb{R}^{2K+1}$ has a Euclidean norm of one \citep{Dette1993}. Note that in an equispaced design, 
\begin{align}
    A_N(Y^{(g)}, X; \theta)& = \text{diag}(1, 1/2,\ldots, 1/2). \label{eq:equi_mat}
\end{align}
Here, diag($m_1,\ldots, m_{2K+1}$) denotes a $(2K+1) \times (2K+1)$ diagonal matrix, with the $j$-th argument $m_j$ representing the $j$-th element along the diagonal \citep[page 359]{Pukelsheim2006}.

\subsection{Motivating example} \label{sec:2.2}

This paper is motivated by circadian biology studies where data are not collected from an equispaced design. To provide an example of how data not collected from an equispaced design can affect hypothesis testing, suppose an investigator gathers data in a circadian biology study, and correctly specifies the first-order trigonometric regression model ($K=1$) for modelling every gene's expression profile over time.  The parameter vector estimates for the first three genes ($g=1,2,3$) equal their estimands, which are 
\begin{align}
\hat{\theta}^{(1)}=\theta^{(1)}=(4,1,1), \quad \hat{\theta}^{(2)}=\theta^{(2)}=(4,0,\sqrt{2}), \quad \hat{\theta}^{(3)}=\theta^{(3)}=(4,\sqrt{2}, 0). \label{eq:est} 
\end{align}
The variance estimates for gene-specific random noise also equal their estimands, which are
\begin{align*}
(\hat{\sigma}^{(g)})^2 = (\sigma^{(g)})^2 = 1.
\end{align*}

After estimating a first-order trigonometric regression model for each gene, the investigator decides to perform a Wald test \citep{Boos1992, Boos2013} to identify genes with oscillatory behavior. Specifically, the investigator defines the null hypothesis $H_0: \gamma^{(g)} = (0, 0)$, where $\gamma^{(g)}$ denotes the subvector $(\theta_1^{(g)}, \theta_{2}^{(g)})$, and partitions the corresponding estimand of the variance matrix for the parameter vector such that
\begin{align*}
\text{Var}(Y^{(g)}, X; {\theta}^{(g)}) &= \left[
\begin{array}{c|c c}
\left\{\text{Var}(Y^{(g)}, X; {\theta}^{(g)})\right\}_{1,1} & 
\left\{\text{Var}(Y^{(g)}, X; {\theta}^{(g)})\right\}_{1,2} & 
\left\{\text{Var}(Y^{(g)}, X; {\theta}^{(g)})\right\}_{1,3} \\[0.5ex]
\hline & \\[-2.25ex]
\left\{\text{Var}(Y^{(g)}, X; {\theta}^{(g)})\right\}_{2,1} & 
\left\{\text{Var}(Y^{(g)}, X; {\theta}^{(g)})\right\}_{2,2} & 
\left\{\text{Var}(Y^{(g)}, X; {\theta}^{(g)})\right\}_{2,3} \\[0.5ex]
\left\{\text{Var}(Y^{(g)}, X; {\theta}^{(g)})\right\}_{3,1} & 
\left\{\text{Var}(Y^{(g)}, X; {\theta}^{(g)})\right\}_{3,2} & 
\left\{\text{Var}(Y^{(g)}, X; {\theta}^{(g)})\right\}_{3,3} \\
\end{array}
\right] \\
&= \begin{bmatrix}
\left\{\text{Var}(Y^{(g)}, X; {\theta}^{(g)})\right\}_{\theta_0^{(g)}, \theta_0^{(g)}} &  \left\{\text{Var}(Y^{(g)}, X; {\theta}^{(g)})\right\}_{\theta_0^{(g)}, \gamma^{(g)}} \\[0.5ex]
\left\{\text{Var}(Y^{(g)}, X; {\theta}^{(g)})\right\}_{\gamma^{(g)}, \theta_0^{(g)}} & \left\{\text{Var}(Y^{(g)}, X; {\theta}^{(g)})\right\}_{\gamma^{(g)}, \gamma^{(g)}}
\end{bmatrix}.
\end{align*}
It is also assumed that $\text{Var}_N(Y^{(g)}, X; \hat{\theta}^{(g)}) = \text{Var}(Y^{(g)}, X; {\theta}^{(g)})$. Under this matrix partitioning, the Wald test statistic $\tau^{(g)}_{\text{W}}$ is expressed as
\begin{align} \label{eq:2}
\tau^{(g)}_{\text{W}} = N(\hat{\gamma}^{(g)})^T\left[\left\{\text{Var}_N(Y^{(g)}, X; \hat{\theta}^{(g)})\right\}_{\gamma^{(g)}, \gamma^{(g)}}\right]^{-1}(\hat{\gamma}^{(g)})
\end{align}
in finite-sample settings, which follows a central chi-squared distribution with $2$ degrees of freedom when $K=1$. The null hypothesis would be rejected at a pre-determined $\alpha$-level if $\tau_{\text{W}}^{(g)}$ surpasses the $1-\alpha$ percentile of the central chi-squared distribution \citep{Kent1982, Boos1992}. 

Now, suppose each $X_i \sim \text{VM}(0, 1)$, where $\text{VM}(0, 1)$ denotes a von Mises distribution with mean $\mu=0$ and concentration $\xi=1$, and the generated $X_i$ is then multiplied by $12/\pi$. It is noted that time is often considered a circular random variable, and the von Mises distribution is recognized as the circular counterpart to the normal distribution \citep{Lee2010}. In Appendix \ref{app:A}, it is shown that 
\begin{eqnarray*}
\left[\left\{\text{Var}(Y^{(g)}, X; \hat{\theta}^{(g)})\right\}_{\gamma^{(g)}, \gamma^{(g)}}\right]^{-1}&=&\begin{bmatrix}
    \frac{1}{2}-\frac{I_2(1)}{2I_0(1)} & 0 \\ 
    0 & \frac{1}{2}+\frac{I_2(1)}{2I_0(1)} -\frac{I_1(1)^2}{I_0(1)^2}
    \end{bmatrix},
\end{eqnarray*}
where $I_{\psi}(\nu)$ denotes the modified Bessel function of the first kind \citep[page 376]{Abramowitz1965}, which numerically approximates to
\begin{eqnarray*}
\left[\left\{\text{Var}(Y^{(g)}, X; \hat{\theta}^{(g)})\right\}_{\gamma^{(g)}, \gamma^{(g)}}\right]^{-1}&\approx& \begin{bmatrix}
    0.446 & 0 \\ 
    0 & 0.354
    \end{bmatrix}.
\end{eqnarray*}
Subsequently computing $\tau_{\text{W}}^{(g)}$ for the first three genes with this approximated matrix yields
\begin{align*}
\tau^{(1)}_{\text{W}} = 0.8N, \quad \tau^{(2)}_{\text{W}} = 0.708N, \quad \tau^{(3)}_{\text{W}} = 0.892N.
\end{align*}
Notice that if $X_i$ was instead sampled in an equispaced design,
\begin{align*}
\text{Var}_N(Y^{(g)}, X; \hat{\theta}^{(g)})= \begin{bmatrix} 
1 & 0 & 0 \\
0 & 2 &  0 \\
0 & 0 & 2
\end{bmatrix},
\end{align*}
and each $\tau^{(g)}_{\text{W}}$ would equal $N$. In other words, an equispaced design would result in all three Wald test statistics equalling each other, and in this example would result in test statistics that are greater than those obtained from the von Mises distribution specified. 

To provide insight into this phenomenon, consider the amplitude-phase representation of a trigonometric regression model,
\begin{equation} \label{eq:3}
    Y^{(g)} = \theta^{(g)}_0 + \sum_{k=1}^K\mu^{(g)}_{k}\text{\text{cos}}\left(\frac{\pi k X}{12}+\phi^{(g)}_k\right) + \epsilon.
\end{equation}
Here, $\mu^{(g)}_k$ and $\phi^{(g)}_k$ are interpreted as the amplitude and phase-shift of the $k$-th order harmonic, respectively, where each pair of $k$-th order terms $(\theta^{(g)}_{2k-1}, \theta^{(g)}_{2k})$ can be expressed as $\theta^{(g)}_{2k-1} = -\mu^{(g)}_k\sin(\phi^{(g)}_k)$ and $\theta^{(g)}_{2k} = \mu^{(g)}_k\cos(\phi^{(g)}_k)$ \citep{Tong1976}. In this representation, $\hat{\mu}_1^{(g)} = \sqrt{2}$ for each estimated model, which indicates that each gene displays the same oscillatory behavior in how it deviates from its mean expression levels. The phase-shift estimand for the second gene is $\phi_1^{(2)}=0$, which indicates that the mode of the von Mises density aligns with the time point where the sinusoidal curve represented by the second model reaches its peak. As a consequence, the generated covariate data are most concentrated within an interval where the magnitude of the rate of change for the sinusoidal curve represented by the second model is minimized, which also minimizes the value of the Wald test statistic in (\ref{eq:2}). On the other hand, the phase-shift estimand for the third gene is $\phi_1^{(3)}=\pi/2$, which indicates that the mode of the von Mises density aligns with a time point situated between the peak and nadir of the sinusoidal curve represented by the third model. This alignment implies the generated covariate data are most concentrated within an interval where the magnitude of the rate of change for this sinusoidal curve represented by the third model is maximized, which also maximizes the value of the Wald test statistic in (\ref{eq:2}). It is noted that $\phi_1^{(1)}=\pi/4$ is in between $\phi_1^{(2)}$ and $\phi_1^{(3)}$, and yields a test statistic of $\tau_{\text{W}}^{(1)} = 0.800$, which is also in between $\tau_{\text{W}}^{(2)} = 0.708$ and $\tau_{\text{W}}^{(3)} = 0.892$. 

\subsection{Weighted regression overview} \label{sec:2.3}

The motivating example presented in Section \ref{sec:2.2} highlights how the marginal density of sample collection time, denoted as $\rho(X)$, can influence the numeric output from Wald test statistic calculation. Further, this influence vanishes when the sampling design is equispaced. To mitigate the variability in Wald test statistic calculation due to $\rho(X)$, this paper aims to identify a non-negative weight function $w(X)$ for weighted parameter vector estimation and weighted Wald test statistic calculation that yield the same quantities as those computed from an equispaced design. To clarify, one could instead estimate each $\theta^{(g)}$ by minimizing weighted squared loss, or
\begin{align*}
\hat{\theta}^{(g)} =& \argmin_{\theta \in \Theta} \ \ 
\frac{1}{\sum_{j=1}^Nw(X_j)}\sum_{i=1}^Nw(X_i)\left\{Y_{i}^{(g)}-\theta^Tf(X_i)\right\}^2.
\end{align*}
The corresponding empirical weighted variance of this parameter vector for Wald test statistic calculation is defined as
\begin{align}
\text{Var}_N(Y^{(g)}, X; \hat{\theta}^{(g)}) &= (\hat{\sigma}^{(g)}_{w})^2 W_N(Y^{(g)}, X; \hat{\theta}^{(g)})^{-1}, \label{eq:wasy_var}
\end{align}
with 
\begin{align*}
 W_N(Y^{(g)}, X; \theta) = \frac{1}{\sum_{j=1}^Nw(X_j)}\sum_{i=1}^Nw(X_i)f(X_i)f(X_i)^T
\end{align*}
and 
\begin{align*}
(\hat{\sigma}_w^{(g)})^2 = \frac{N}{(N-2K-1)\sum_{j=1}^Nw(X_i)}\sum_{i=1}^Nw(X_i)\left\{Y_i-(\hat{\theta}^{(g)})^Tf(X_i)\right\}^2.
\end{align*}

To identify $w(X)$, it is noted that when $\rho(X)$ follows a uniform density over a $24$ hour interval, the expected Hessian of squared loss 
\begin{align*}
\mathbb{E}\left\{A_N(Y^{(g)}, X; {\theta}^{(g)})\right\} = \text{diag}(1, 1/2,\ldots, 1/2).
\end{align*}
This observation implies that an initial candidate for a weight function would be $w(X) = 1/\{24\rho(X)\}$, as it would transform $\rho(X)$ to a uniform density during estimation and inference. A challenge with employing $w(X) = 1/\{24\rho(X)\}$ in finite-sample settings, however, is that an investigator typically does not know the true marginal density for the time of sample collection. In this paper, we instead propose the normalized weight function
\begin{align}
    w(X_i) = \frac{1/\hat{\rho}(X_i; \kappa)}{\sum_{j=1}^N\left\{1/ \hat{\rho}(X_j; \kappa)\right\}} \label{eq:weight_def}
\end{align}
for finite-sample settings, where $\hat{\rho}(X_i; \kappa)$ denotes a kernel density estimator for the time of sample collection, or
\begin{equation*}
\hat{\rho}(Z; \kappa) = \frac{1}{N} \sum_{i=1}^{N} K_{\kappa}\left(Z - \frac{\pi X_i}{12}\right).
\end{equation*}
To be precise, $K_{\kappa}(Z)$ denotes a periodic, circular kernel with a symmetric density function that has a univariate argument $Z$. The concentration hyperparameter $\kappa > 0$ controls the contribution of observations within a neighborhood of an estimation point for the kernel density. To clarify, for a circular kernel, smaller values of $\kappa$ indicate that a wider neighborhood of data points is used for each estimation point \citep{DiMarzio2021}. It is noted that as the sample size and $\kappa$ tend to infinity, the kernel density estimator converges to the true, circular density \citep{DiMarzio2009}. Further, (\ref{eq:weight_def}) can be interpreted as normalizing $w(X) = 1/\{24\hat{\rho}(X; \kappa)\}$ given observed data.

Kernel density estimation relies on an investigator specifying a concentration hyperparameter $\kappa$. In this paper, the hyperparameter specified for estimation, $\kappa_{\text{opt}}$, is defined as the argument that maximizes the optimization problem
\begin{align}
    \kappa_{\text{opt}} = \argmax_{\kappa \in \mathbb{R}_{>0}} \ \ \text{det}\left\{\sum_{i=1}^N\frac{f(X_i)f(X_i)^T/\hat{\rho}(X_i; \kappa)}{\sum_{j=1}^N\left\{1/ \hat{\rho}(X_j; \kappa)\right\}}\right\}, \label{eq:opt}
\end{align}
which can be interpreted as identifying the weight function that maximizes the $D$-optimality criterion in (\ref{eq:phi_def}) when $p = 0$. Given the assumption that the random noise is homoscedastic, this optimization problem can also be interpreted as identifying the concentration hyperparameter that minimizes the generalized variance, or the determinant of the variance matrix for the estimated regression parameters \citep{Kitsos1988}. 

It is noted (\ref{eq:equi_mat}) implies that when data are sampled from an equispaced design, the determinant $\text{det}\{A_N(X, Y; \hat{\theta})\} = 1/4^K$. A benefit of using the optimization problem in (\ref{eq:opt}) to identify $\kappa_{\text{opt}}$ is that the maximum value this optimization problem can attain is $1/4^K$, which is derived in Appendix \ref{app:B}. However, a limitation with the optimization problem in (\ref{eq:opt}) is that it naively uses every sample to identify $\kappa_{\text{opt}}$, which could result in a kernel density estimator that overfits to the data. To reconcile this issue, one can instead utilize a cross-validation scheme for identifying the concentration hyperparameter that maximizes (\ref{eq:opt}). Specifically, suppose the data are randomly divided into $M$ folds, with $\mathcal{I}_m$ denoting the index set of samples that belong to the $m$-th fold, and let $\hat{\rho}^{(-m)}(X; \kappa)$ denote a kernel density estimator obtained from data excluding the $m$-th fold during estimation. One could instead identify $\kappa_{\text{opt}}$ by solving the optimization problem
\begin{align}
\kappa_{\text{opt}} = \argmax_{\kappa \in \mathbb{R}_{>0}} \ \ \text{det}\left[\sum_{m=1}^M \sum_{j \in \mathcal{I}_m} \frac{f(X_j)f(X_j)^T/\hat{\rho}^{(-m)}(X_j; \kappa)}{\sum_{m=1}^M\sum_{j \in \mathcal{I}_m}\left\{1/ \hat{\rho}^{(-m)}(X_j; \kappa)\right\}}\right]. \label{eq:opt_CV}
\end{align}
This cross-validation scheme will be utilized in subsequent analyses for identifying $\kappa_{\text{opt}}$.

\section{Simulation Study}  \label{sec:3}

\subsection{Simulation study setup} \label{sec:3.1}

A simulation study is conducted to assess the weighted regression proposed. This simulation study utilizes sample collection time data from three publicly available circadian biology studies \citep{Archer2014, Braun2018, MllerLevet2013}, which have each been described in detail in their respective studies and previously summarized by \cite{Gorczyca2024}. Briefly, the ``Archer data set'' was compiled from a study where individuals underwent a sleep desynchrony protocol, where each individual was required to stay awake for 20 hours and sleep for eight hours each wake-sleep cycle over a duration of 96 hours. This data set consists of two sets of samples: one set obtained before the sleep desynchrony protocol (control samples), and another set obtained after the sleep desynchrony protocol (experimental samples). The ``Braun data set'' is derived from individuals with similar sleep schedules, health statuses, and ages, which consists of a single set of samples. Finally, the ``M\"{o}ller-Levet data set'' was generated during a study on the effects of insufficient sleep on gene expression. This data set consists of two sets of samples: one where individuals were allowed to sleep for up to ten hours each night (control samples), and another where individuals were allowed to sleep for up to six hours each night (experimental samples). The sample collection times for each data set were not evenly distributed over a 24 hour period. Additionally, in each study, the investigators initially collected samples based on day-night cycle time, and then performed laboratory procedures to ascertain each individual's internal time of sample collection. It is noted that the internal times of sample collection for certain individuals could not be determined, leading to the exclusion of these samples from the simulation study.

\begin{figure*}[!h]
\centering
\includegraphics[scale=0.175]{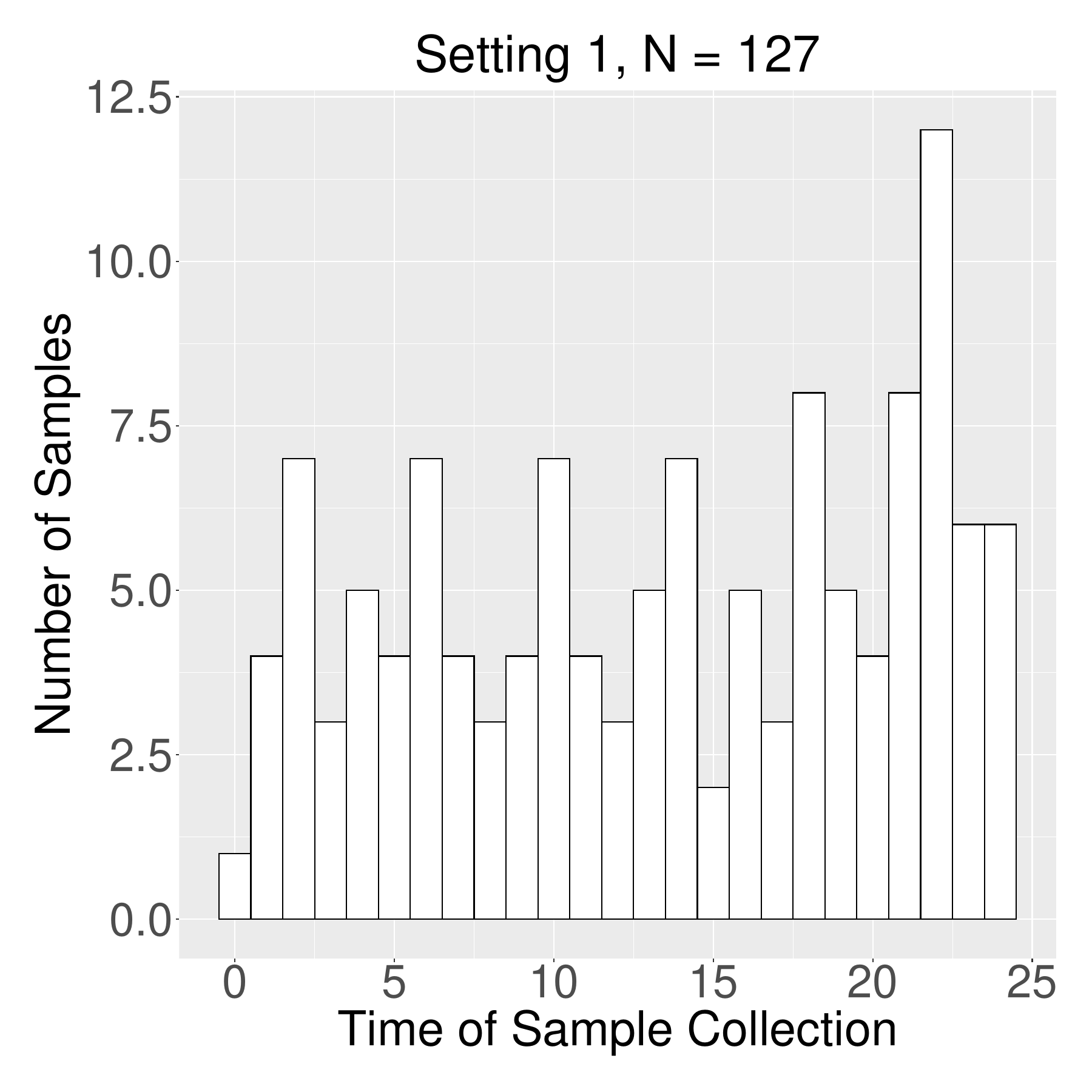}
\includegraphics[scale=0.175]{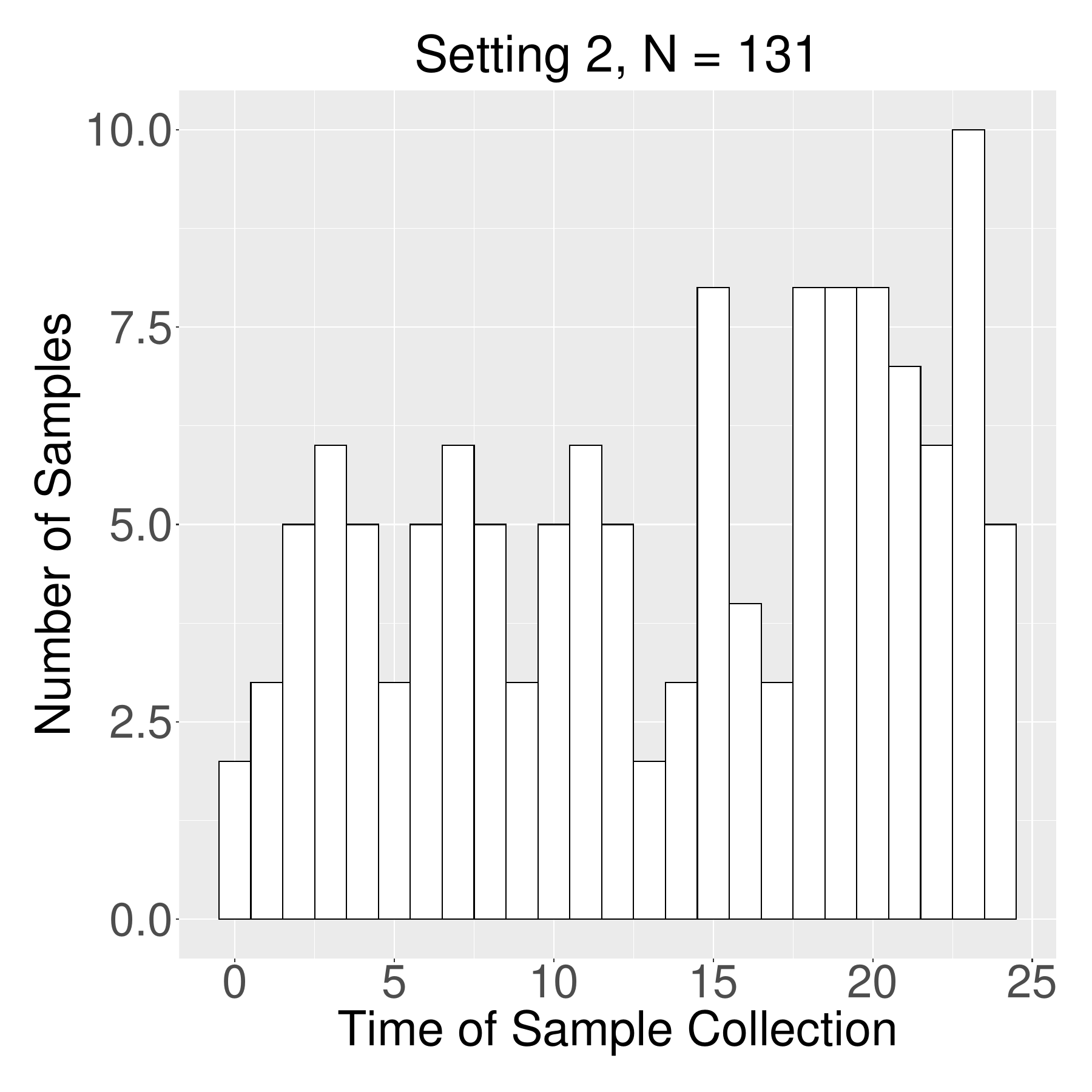}
\includegraphics[scale=0.175]{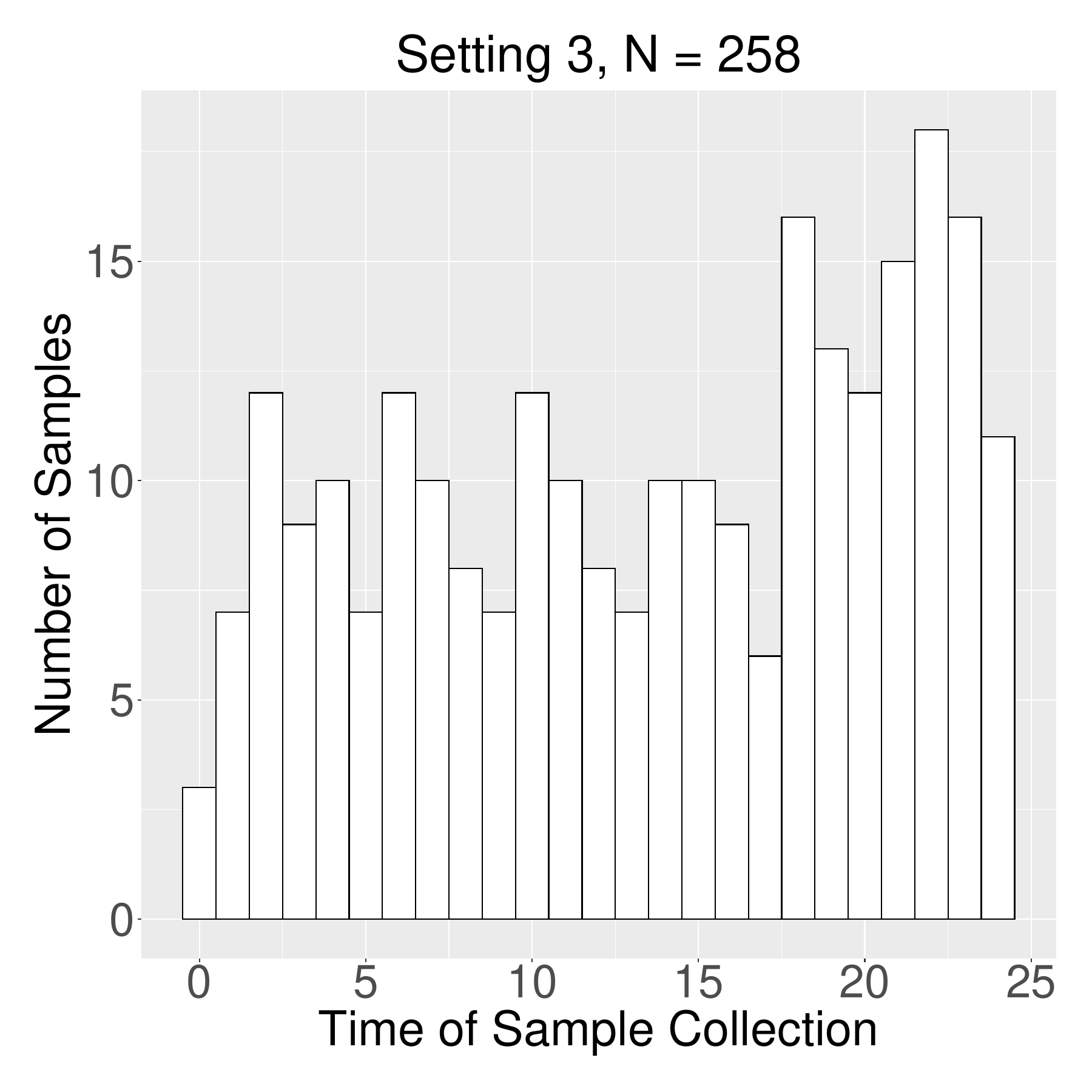} \\
\includegraphics[scale=0.175]{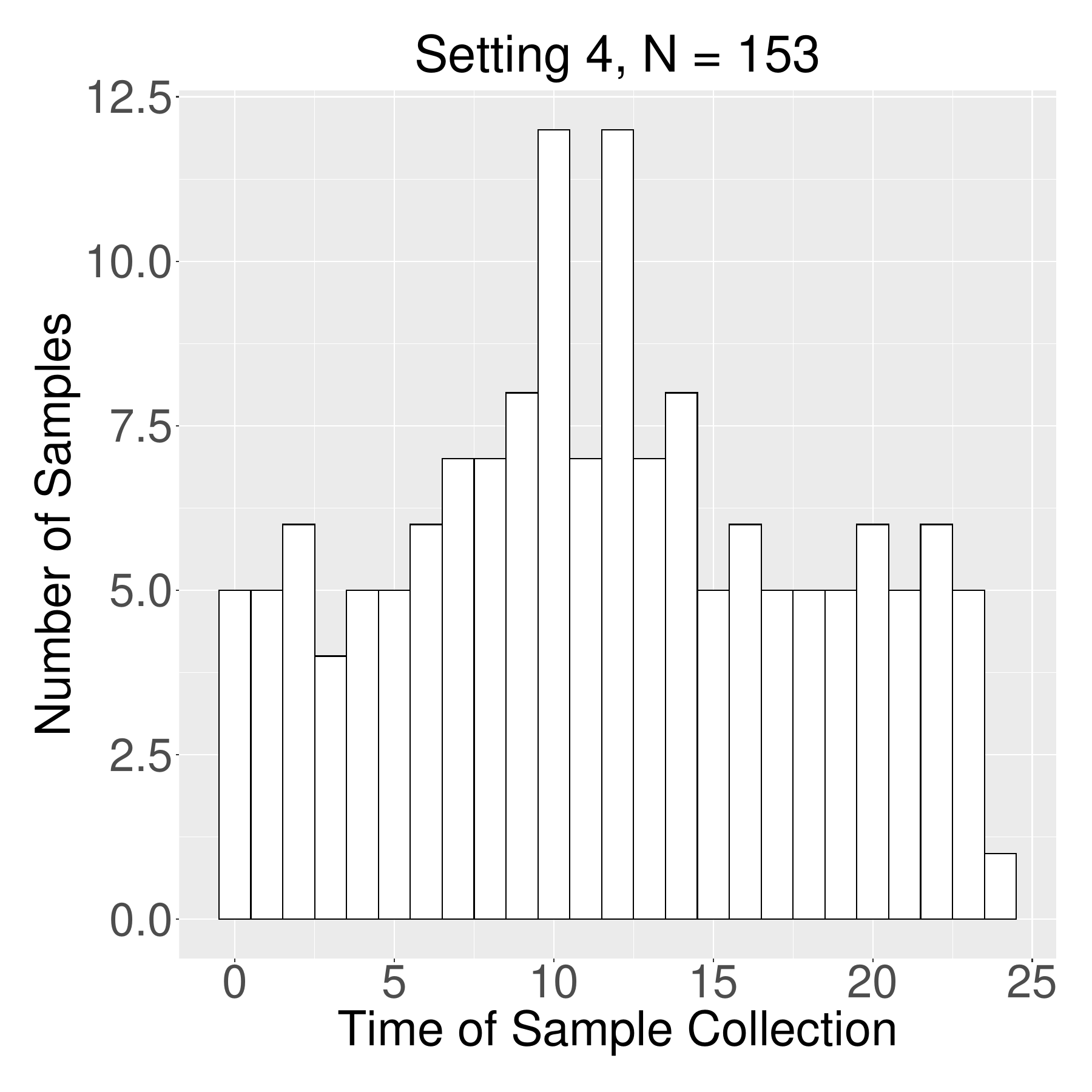}
\includegraphics[scale=0.175]{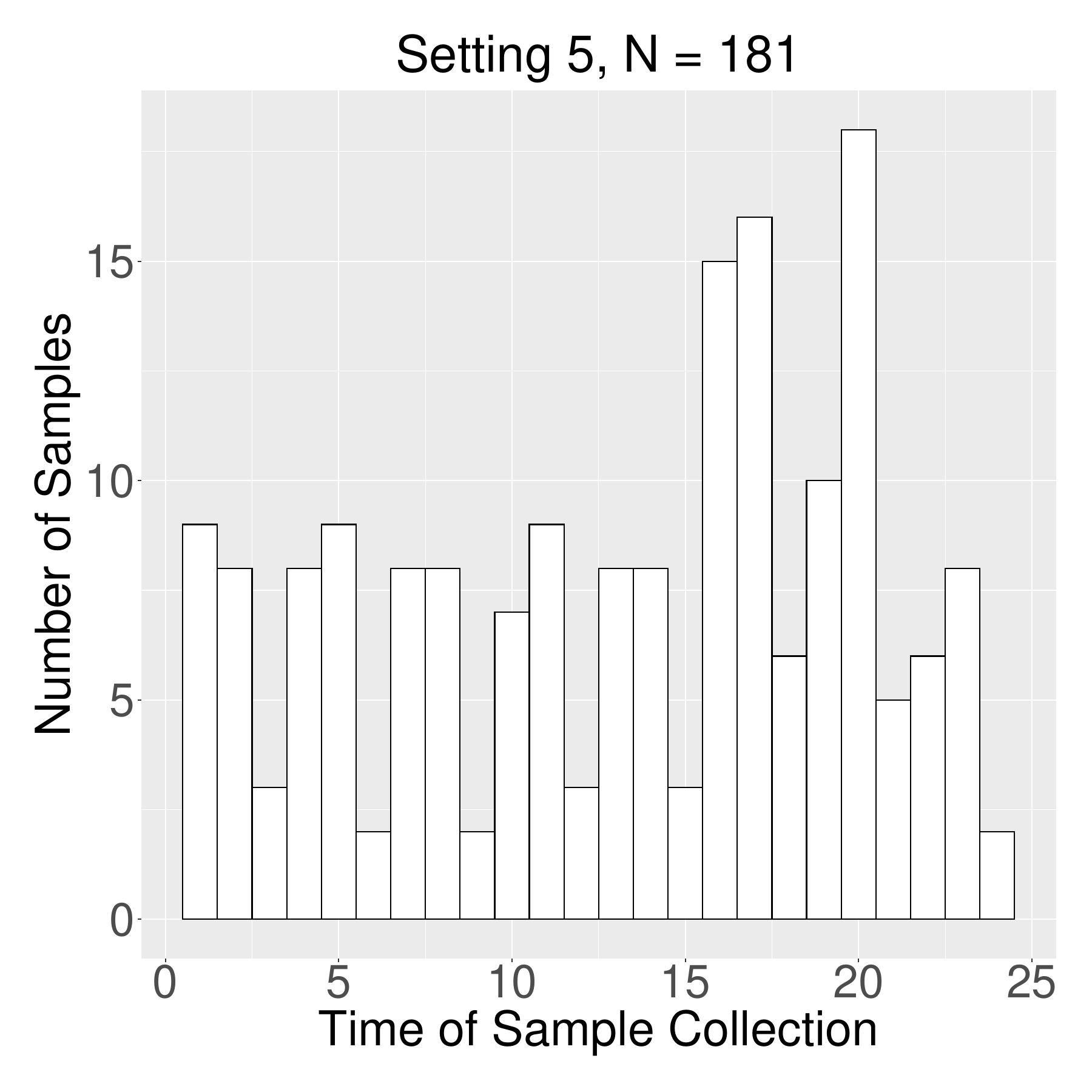} \\
\includegraphics[scale=0.175]{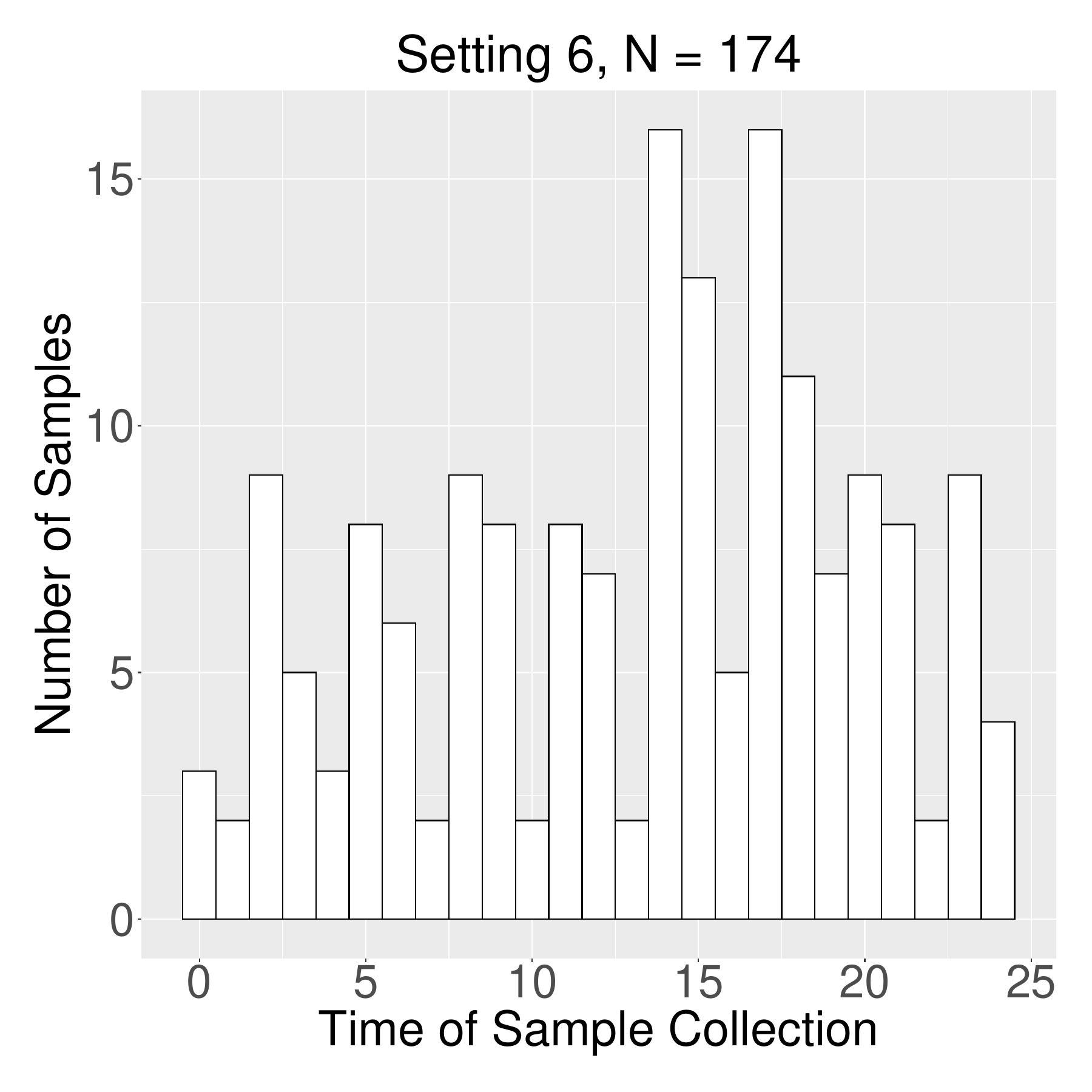} 
\includegraphics[scale=0.175]{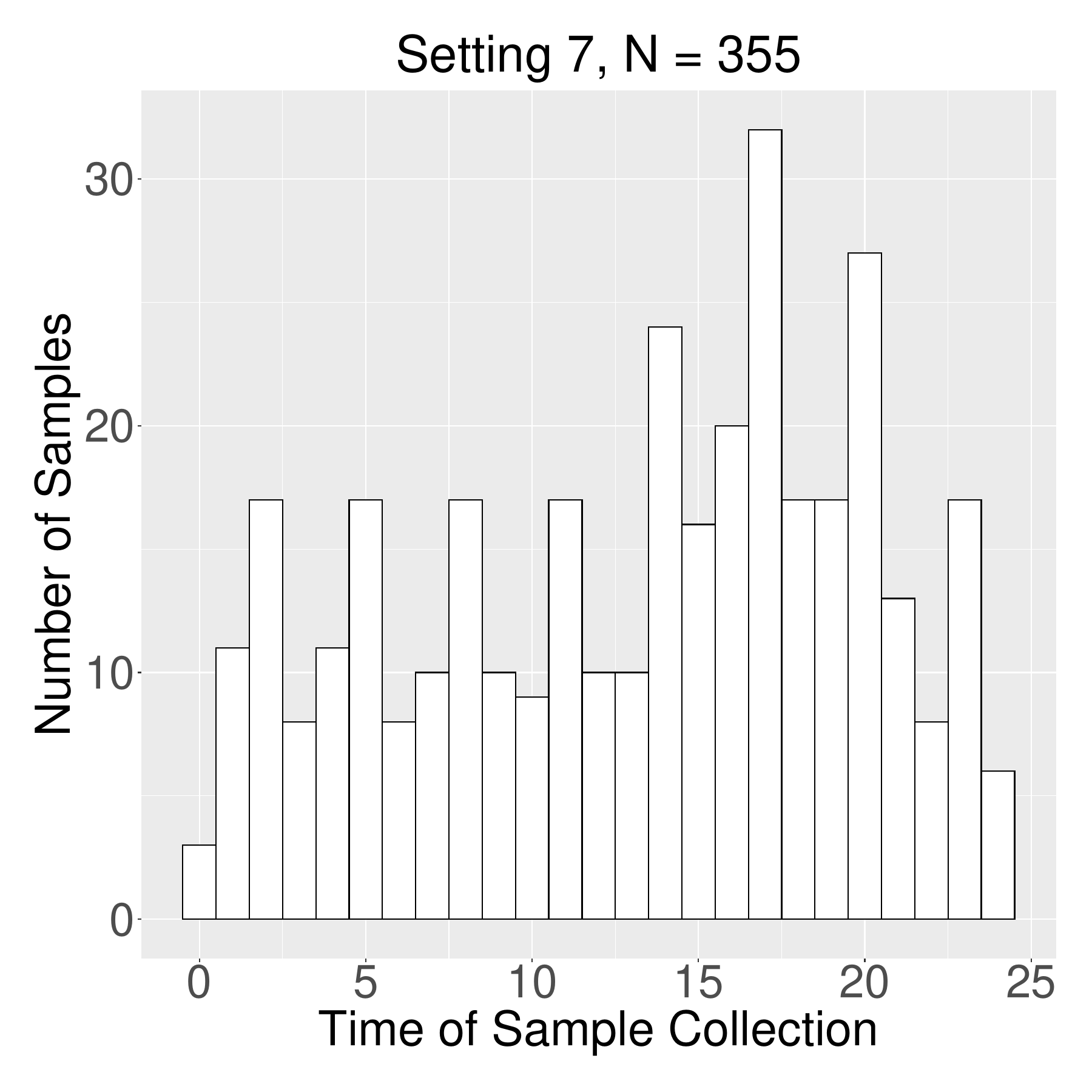}

\caption{Histograms of sample collection times for each of the seven simulation settings. The binwidth of each bar in a plot is one hour.}
    \label{fig:hists}
\end{figure*}

In this simulation study, seven simulation settings are constructed for generating gene expression $Y$ using the model in (\ref{eq:3}). Multiple runs of each simulation setting are performed, where two parameters, $\phi_{\text{base}}$ and $K$, are varied each run. Specifically, the parameter $\phi_{\text{base}}$ is a quantity that represents how every phase-shift estimand changes for a $K$-th order trigonometric regression model in (\ref{eq:3}). For this simulation study, $\phi_{\text{base}}$ takes on different values in the set $\{2\pi j/20: j=1,\ldots 20\}$ for each run. The parameter $K$ represents the order of the trigonometric regression model specified. For this simulation study, $K$ is varied to take on different values in the set $\{1,2, 3\}$ for each run, as previous circadian biology studies have not considered values of $K$ greater than three for trigonometric regression \citep{Hughes2009}.  Each run consists of 250,000 simulation trials, where two data sets are generated in each trial. The first data set is generated as follows: 
\begin{description}
    \item[Setting 1.] $\theta^{(g)}_0=6$; $\mu^{(g)}_k = 0.5$ for all $k=1,\ldots,K$; $\phi^{(g)}_k = \phi_{\text{base}}$ for all $k=1,\ldots,K$; $X_i$ equals the control sample collection times of the Archer data set; and $\epsilon_i \sim \text{N}(0, 1)$. 
    \item[Setting 2.] $\theta^{(g)}_0\sim \text{TN}(6, 1, 4, 8)$ for each sample; $\mu^{(g)}_k = 0.5$ for all $k=1,\ldots,K$; $\phi^{(g)}_k = \phi_{\text{base}}$ for all $k=1,\ldots,K$; $X_i$ equals the experimental sample collection times of the Archer data set; and $\epsilon_i \sim \text{N}(0, 1)$. 
    \item[Setting 3.] $\theta^{(g)}_0\sim \text{TN}(6, 1, 4, 8)$ for each sample; $\mu^{(g)}_k = 0.5$ for all $k=1,\ldots,K$; $\phi^{(g)}_k = \phi_{\text{base}}$ for all $k=1,\ldots,K$; $X_i$ equals the control  and experimental sample collection times of the Archer data set combined; and $\epsilon_i \sim \text{N}(0, 1)$. 
    \item[Setting 4.] $\theta^{(g)}_0=6$; $\mu^{(g)}_k \sim \text{TN}(0.5, 0.25, 0, 1)$ for all $k=1,\ldots,K$ and each sample; $\phi^{(g)}_k = \phi_{\text{base}}$ for all $k=1,\ldots,K$; $X_i$ equals the sample collection times of the Braun data set; and $\epsilon_i \sim \text{N}(0, 1)$. 
    \item[Setting 5.] $\theta^{(g)}_0=6$; $\mu^{(g)}_k \sim \text{TN}(0.5, 0.25, 0, 1)$ for all $k=1,\ldots,K$ and each sample; $\phi^{(g)}_k = 0$ for all $k=1,\ldots,K$; $X_i$ equals the control sample collection times of the M\"{o}ller-Levet data set; and $\epsilon_i \sim \text{N}(0, 1)$. 
    \item[Setting 6.] $\theta^{(g)}_0\sim \text{TN}(6, 1, 4, 8)$ for each sample; $\mu^{(g)}_k \sim \text{TN}(0.5, 0.25, 0, 1)$ for all $k=1,\ldots,K$ and each sample; $\phi^{(g)}_k = \phi_{\text{base}}$ for all $k=1,\ldots,K$; $X_i$ equals the experimental sample collection times of the M\"{o}ller-Levet data set; and $\epsilon_i \sim \text{N}(0, 1)$. 
    \item[Setting 7.] $\theta^{(g)}_0\sim \text{TN}(6, 1, 4, 8)$ for each sample; $\mu^{(g)}_k \sim \text{TN}(0.5, 0.25, 0, 1)$ for all $k=1,\ldots,K$ and each sample; $\phi^{(g)}_k = \phi_{\text{base}}$ for all $k=1,\ldots,K$; $X_i$ equals the control and experimental sample collection times of the M\"{o}ller-Levet data set combined; and $\epsilon_i \sim \text{N}(0, 1)$. 
 \end{description}
Here, $\text{N}(\mu, \sigma^2)$ denotes a normal distribution with mean $\mu$ and variance $\sigma^2$; and $\text{TN}(\mu, \sigma^2, a, b)$ a truncated normal distribution with mean $\mu$, variance $\sigma^2$, lower bound $a$, and upper bound $b$. It is emphasized that the first data set represents a scenario where samples are not collected from an equispaced design. The second data set is generated in the same manner as the first data set, except the sample collection times are obtained from an equispaced design, with $X_i = 24(i-1)/N$ for the $i$-th sample. Histograms of the times of sample collection for the first data set in each simulation setting are provided in Figure \ref{fig:hists}. 

It is noted that in Setting 1, the estimands $\theta_{k}^{(g)}$, $k=0,\ldots K$ are the same for every sample. Settings 2 through 7, on the other hand, represent scenarios where there are individual-specific variation in gene expression \citep{Cheung2003, Rockman2006, Stranger2007}. Specifically, Settings 2 and 3 represent scenarios where the average expression level of a gene is different for each individual, but the amplitude variation is the same. Settings 4 and 5 represent scenarios where the average expression level of a gene is the same for each individual, but the amplitude variation is different. Finally, Settings 6 and 7 represent scenarios where both the average expression level of a gene and the amplitude variation are different for each individual. 

For the two data sets generated in a trial, three types of regressions are considered:
\begin{description}
    \item[Unweighted Regression.] A parameter vector estimate $\hat{\theta}^{(g)}$ is obtained from an unweighted regression using the first data set (time collection data from a sample population). 
    \item[Unweighted Regression (Equispaced).] A parameter vector estimate $\hat{\theta}^{(g)}$ is obtained from an unweighted regression using the second data set (time collection data from an equispaced design). 
    \item[Weighted Regression.] A parameter vector estimate $\hat{\theta}^{(g)}$ is obtained from a weighted regression using the first data set.
 \end{description}
It is noted that the weighted regression in this simulation study uses a von Mises kernel for kernel density estimation \citep{DiMarzio2021}, and the concentration hyperparameter is determined to be the quantity that maximizes (\ref{eq:opt_CV}) with leave-one-out cross-validation, or where $M$ in (\ref{eq:opt_CV}) is defined to be the total sample size for a sample population. 

The following two quantities are computed after each simulation trial, and are reported after every run to assess how they vary for different values of $\phi_{\text{base}}$:
\begin{itemize}
    \item[1.] $\tau^{(g)}_{\text{W}}$, or the Wald test statistic computed with an estimated model and data used for model estimation.
    \item[2.] $\tau^{(g)}_F$ or the $F$-statistic computed with an estimated model and data used for model estimation.
\end{itemize} 
For computing $\tau^{(g)}_{\text{W}}$ in (\ref{eq:2}), $\gamma^{(g)}$ is defined as the subvector $(\theta_1^{(g)}, \ldots, \theta_{2K}^{(g)})$, and the null hypothesis $H_0: \gamma^{(g)} = (0, \ldots, 0)$. The test statistic $\tau_{W}^{(g)}$ would follow a central chi-squared distribution with $2K$ degrees of freedom, and the null hypothesis would be rejected at a pre-determined $\alpha$-level if $\tau_{\text{W}}^{(g)}$ surpasses the $1-\alpha$ percentile of the central chi-squared distribution \citep{Kent1982, Boos1992}. It is noted that when computing the Wald test statistic with parameters obtained from the weighted regression, $\text{Var}_N(Y^{(g)}, X; \hat{\theta}^{(g)})$ in (\ref{eq:asy_var}) would be replaced with the quantity in (\ref{eq:wasy_var}). Further, the test statistic $\tau^{(g)}_F$ is defined as
\begin{align*}
\tau^{(g)}_F = \frac{\frac{1}{2K}\sum_{i=1}^N w(X_i)\left[\left\{Y^{(g)}_i - \sum_{j=1}^N w(X_j)Y^{(g)}_j\right\}^2 - \left\{Y^{(g)}_i - X_i^T\hat{\beta}^{(g)} \right\}^2 \right]}{\frac{1}{N-2K-1}\sum_{i=1}^Nw(X_i)\left\{Y^{(g)}_i - \sum_{j=1}^N w(X_j)Y^{(g)}_j\right\}^2}
\end{align*}
given non-negative weights $w(X_i)$ such that $\sum_{i=1}^Nw(X_i)=1$ \citep{Moser1992}. Here, $w(X_i) = 1/N$ for all $i$ for both unweighted regressions, and $w(X_i)$ equals (\ref{eq:weight_def}), with $\kappa=\kappa_{\text{opt}}$ from (\ref{eq:opt_CV}), for the weighted regression. It is assumed that $\tau^{(g)}_F \sim F(2K-1, N-2K-1)$ under the null hypothesis, or a central $F$ distribution with degrees of freedom parameters $d_1=2K-1$ and $d_2 = N-2K-1$. It is noted that the $F$-statistic is also considered due to its availability in software \citep{Carlucci2019}.

\subsection{Simulation study results}

\begin{figure*}[!h]
\centering
\includegraphics[scale=0.175]{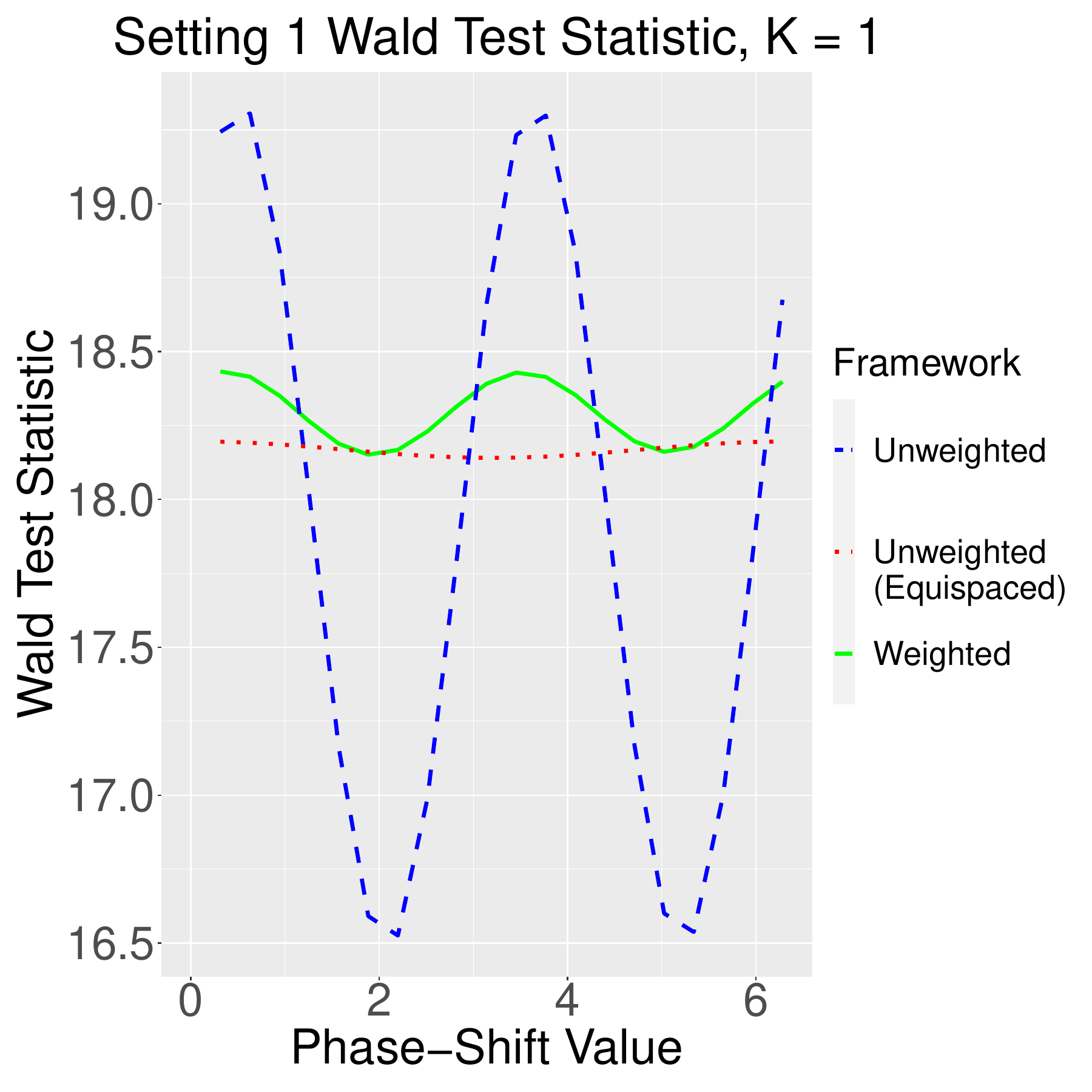}
\includegraphics[scale=0.175]{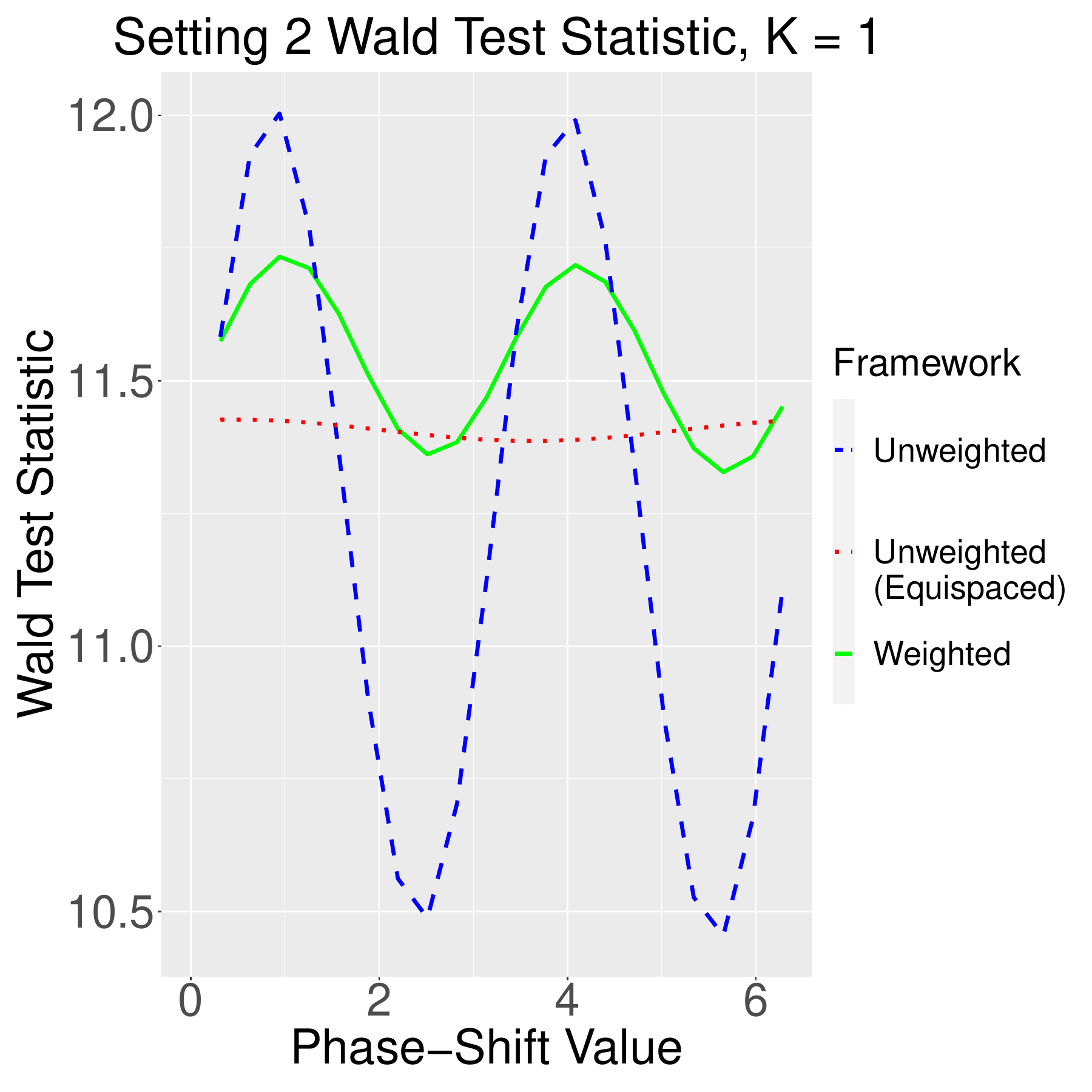} 
\includegraphics[scale=0.175]{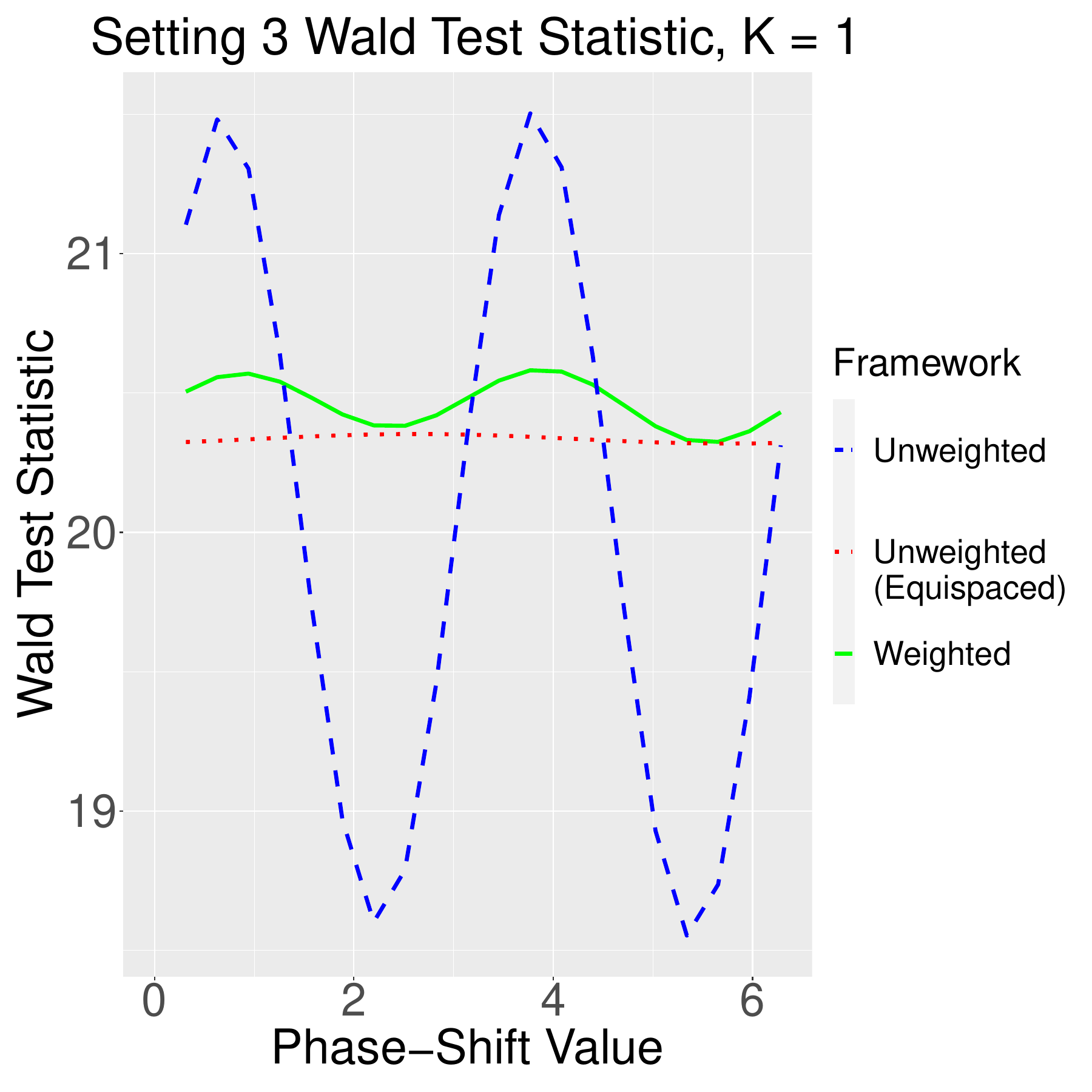} \\
\includegraphics[scale=0.175]{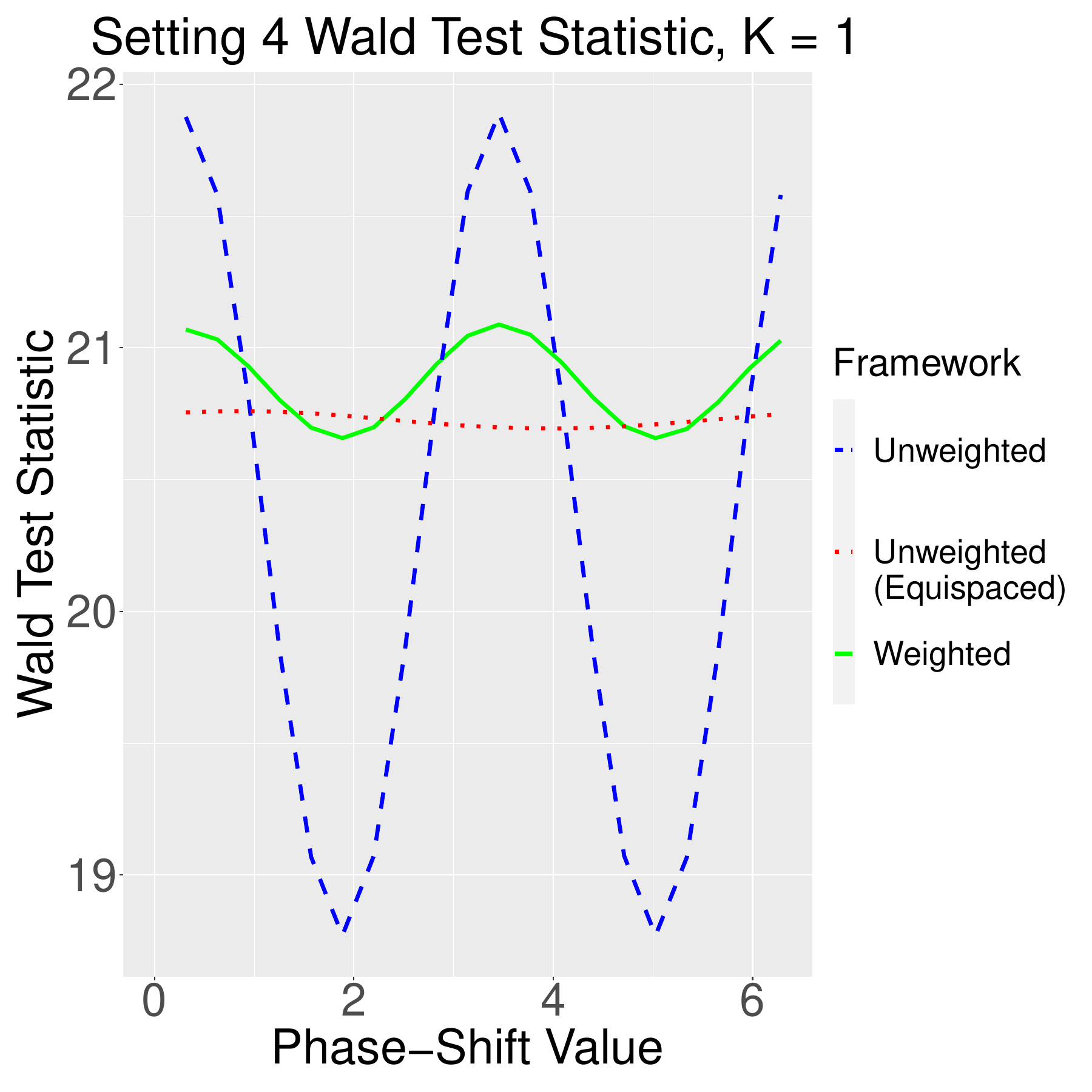} 
\includegraphics[scale=0.175]{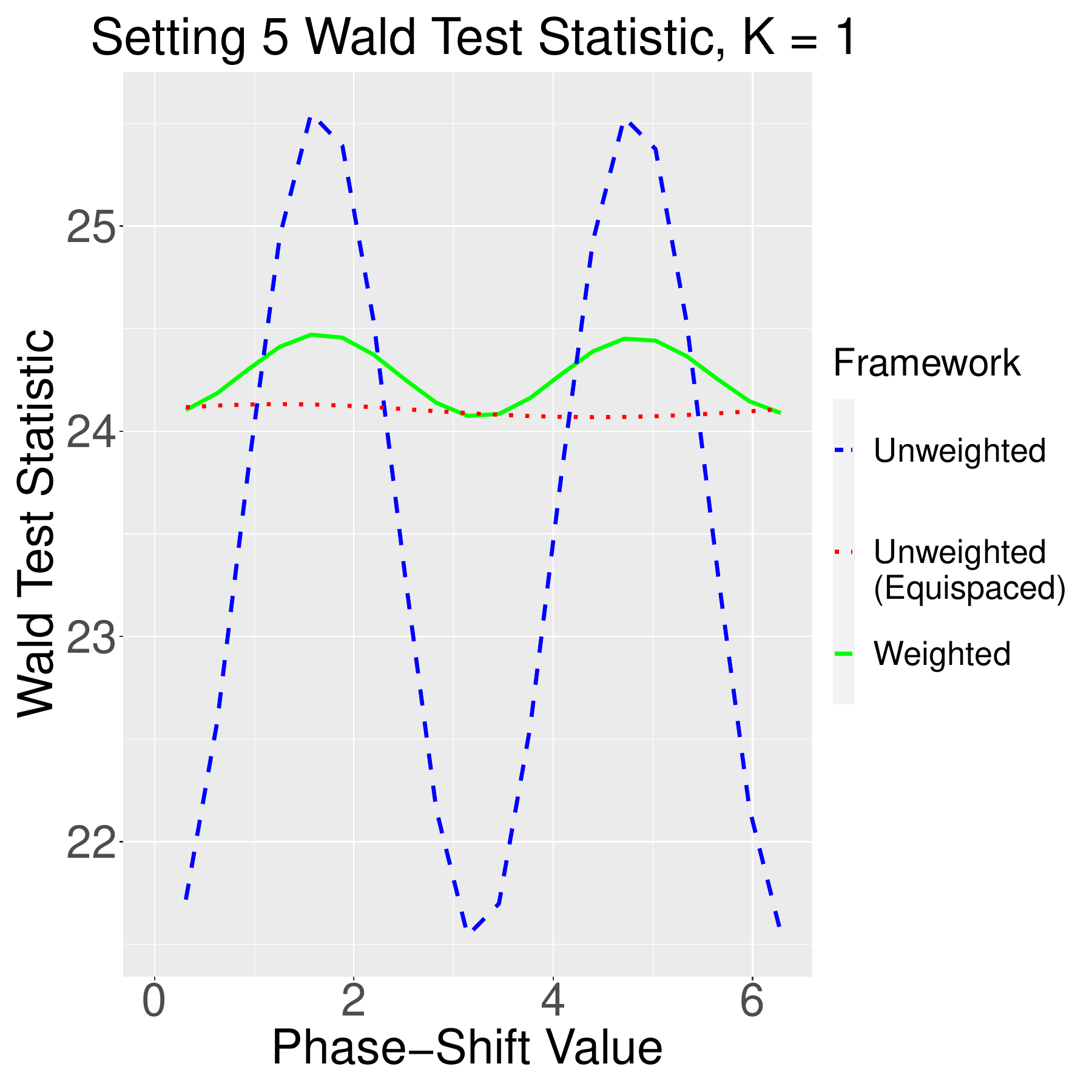} \\
\includegraphics[scale=0.175]{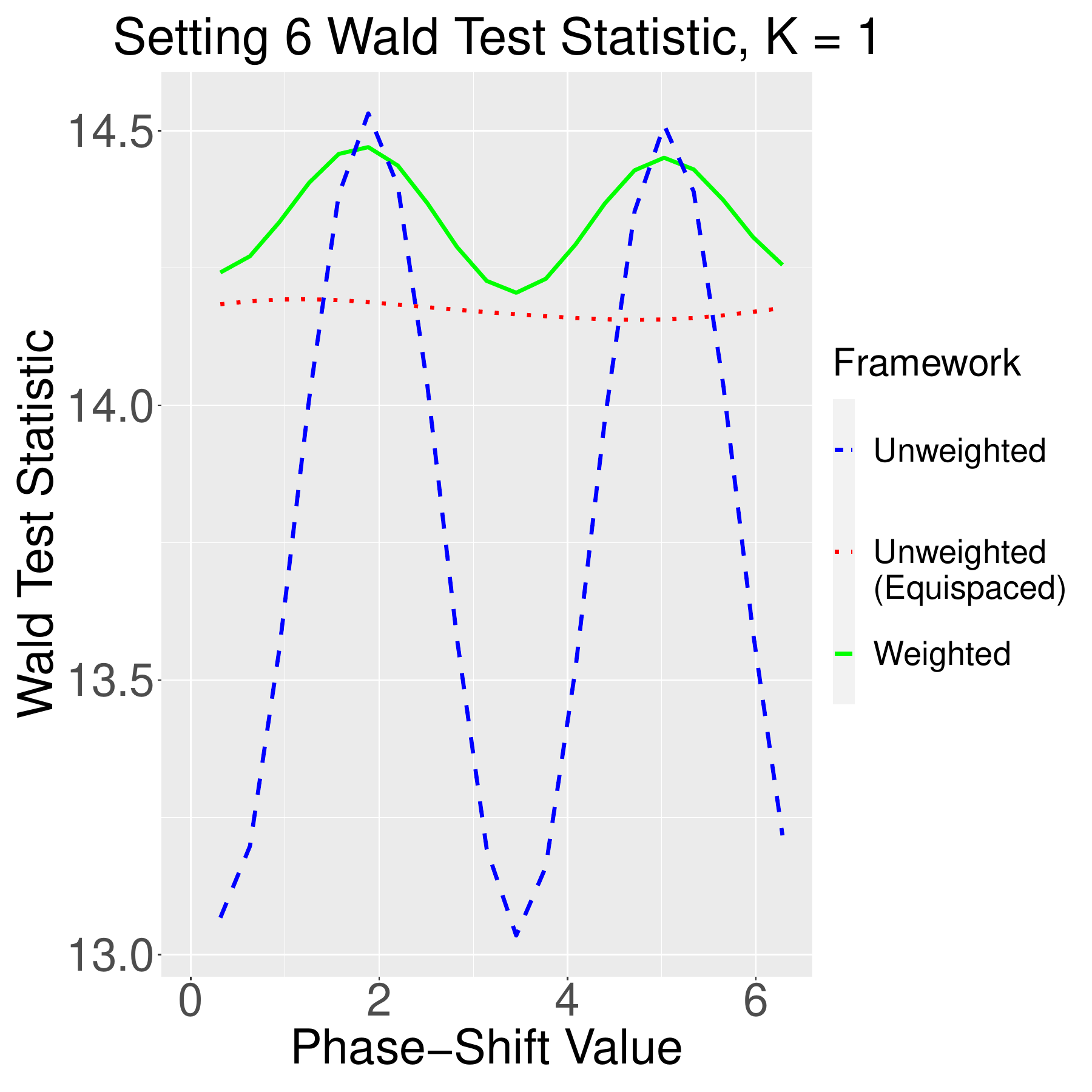}
\includegraphics[scale=0.175]{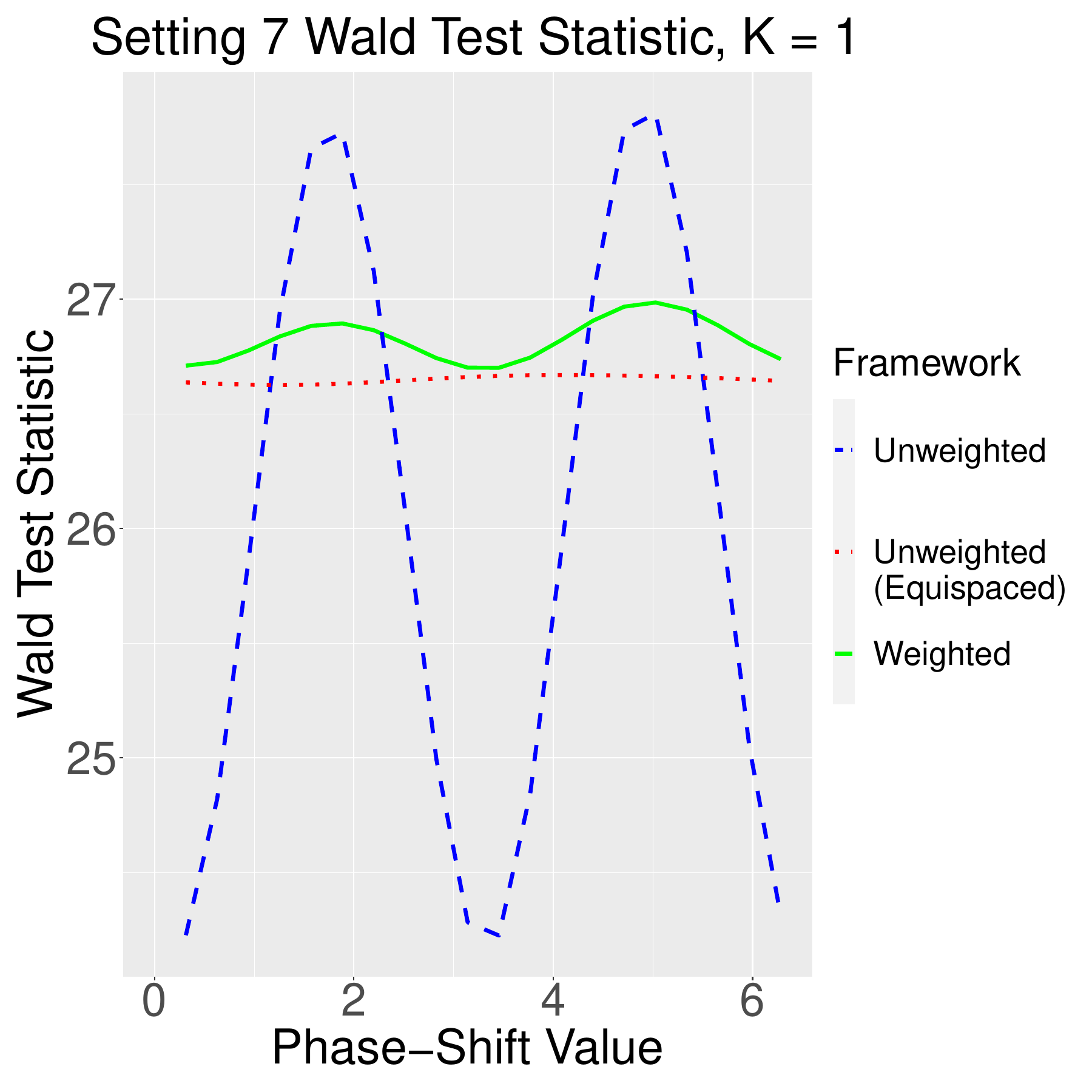}
\caption{Wald test statistics divided by sample size across 250,000 trials for first-order trigonometric regression ($K=1$). Empirically, the unweighted regression with sampled data has greater variability in numeric value for the computed statistic than the corresponding statistic computed from the weighted regression as the phase-shift estimand varies from $0$ to $2\pi$. }
    \label{fig:phases1}
\end{figure*}

\begin{figure*}[!h]
\centering
\includegraphics[scale=0.175]{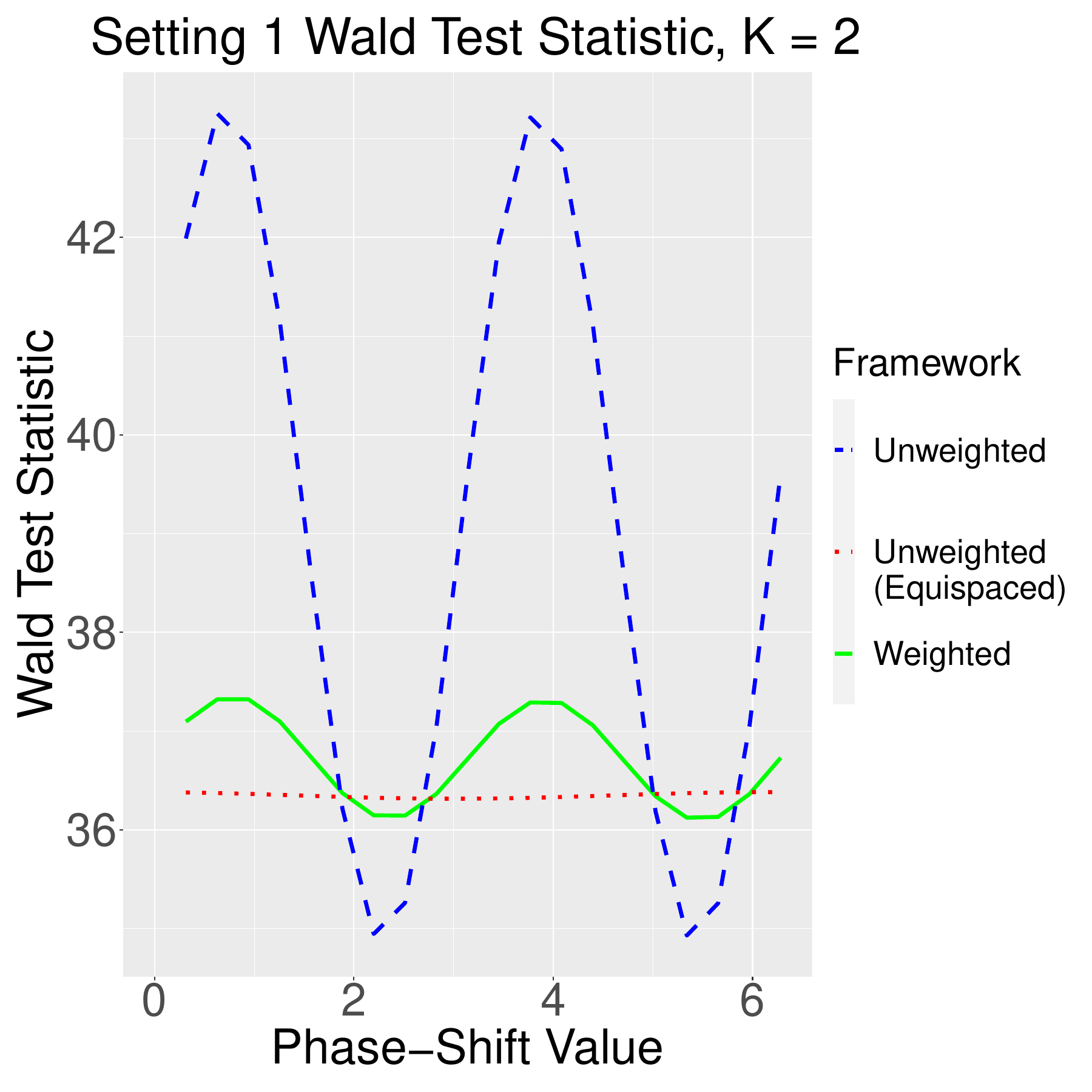}
\includegraphics[scale=0.175]{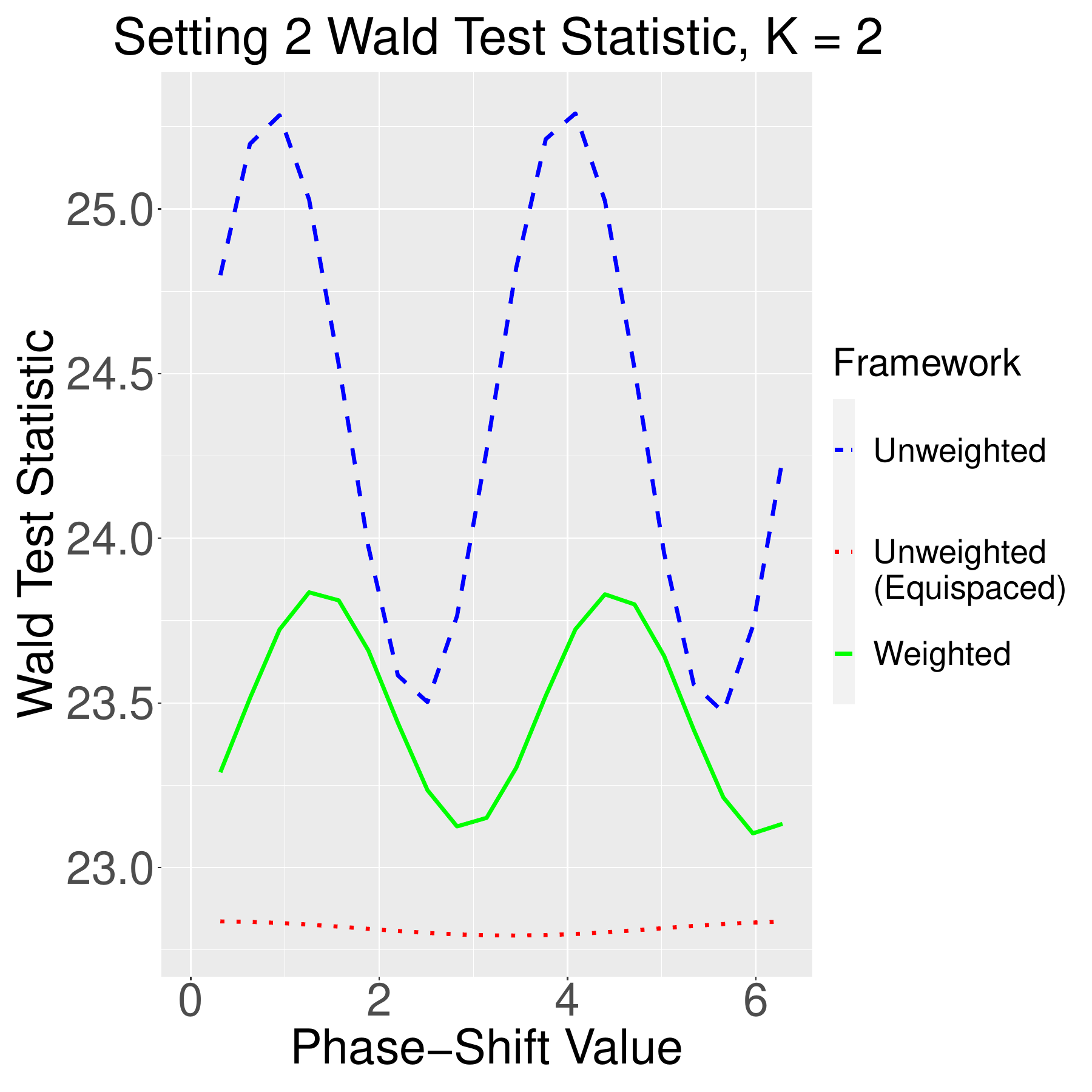} 
\includegraphics[scale=0.175]{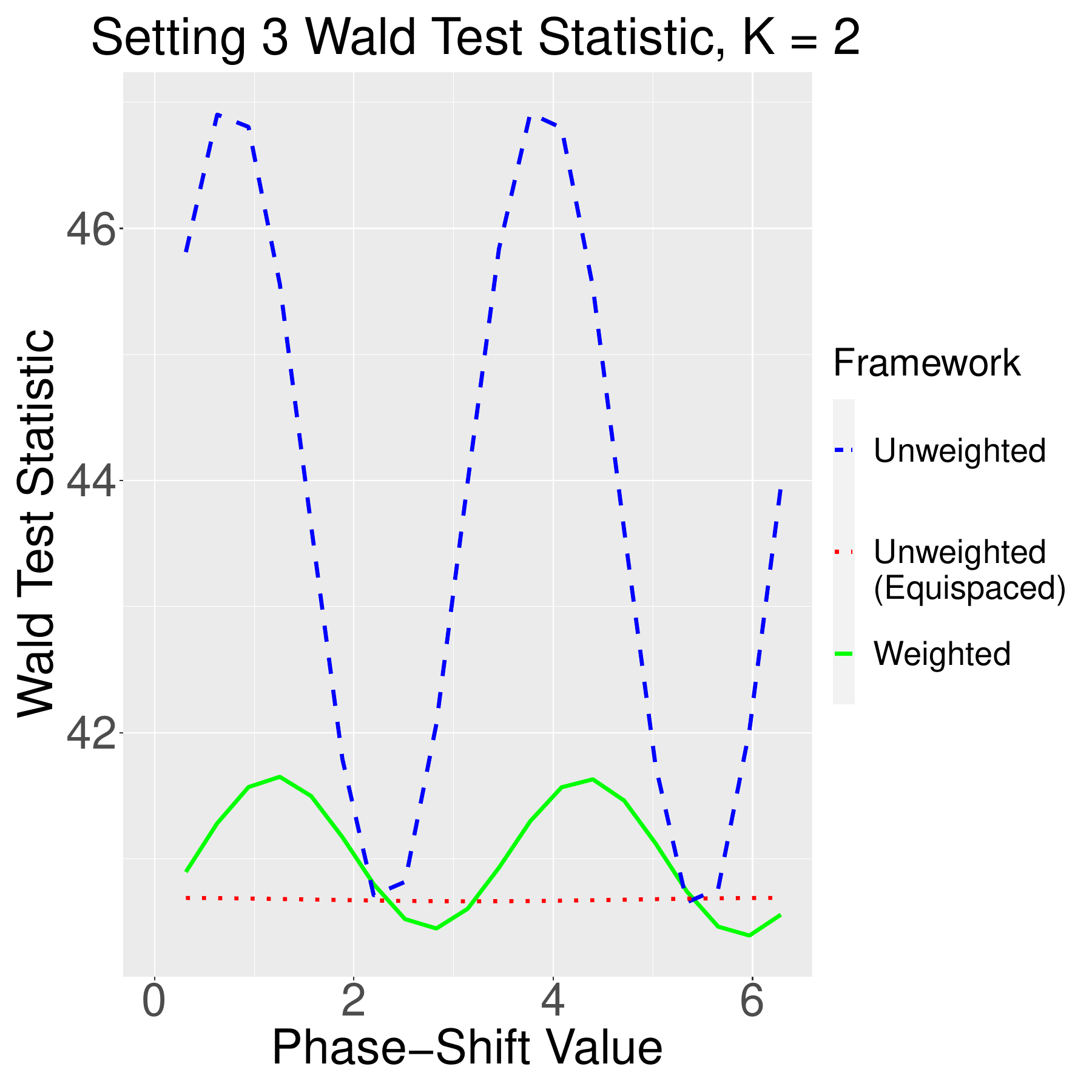} \\
\includegraphics[scale=0.175]{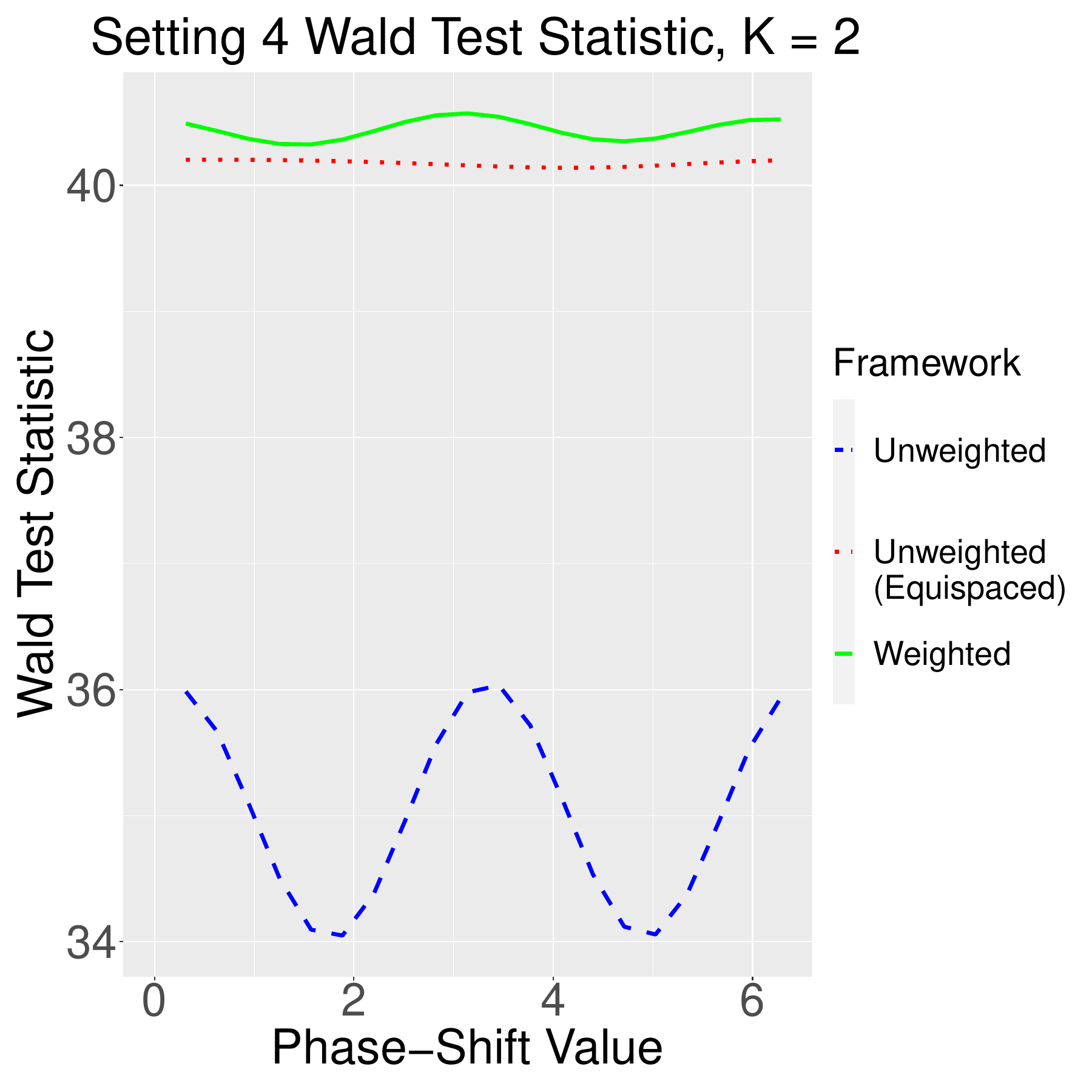} 
\includegraphics[scale=0.175]{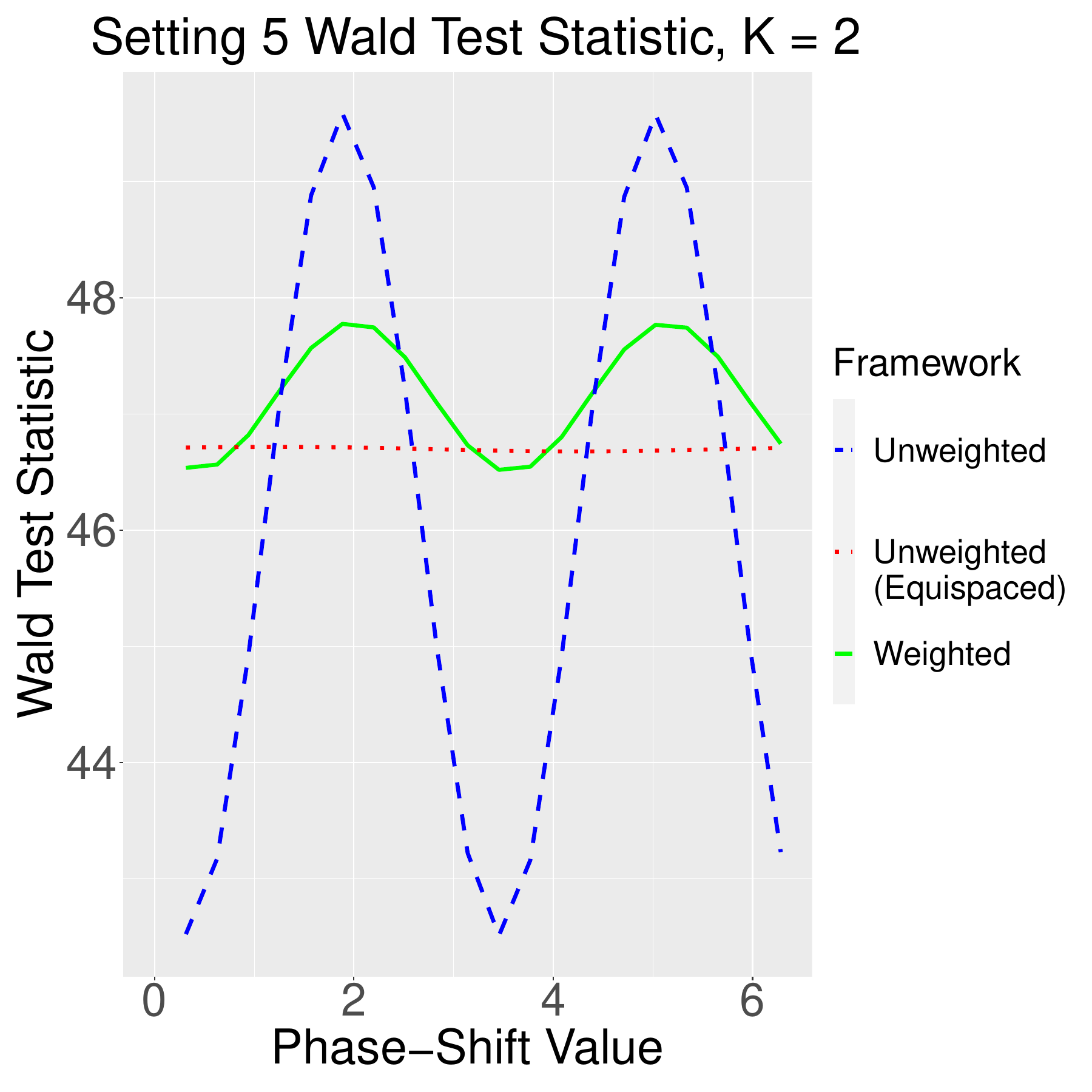} \\
\includegraphics[scale=0.175]{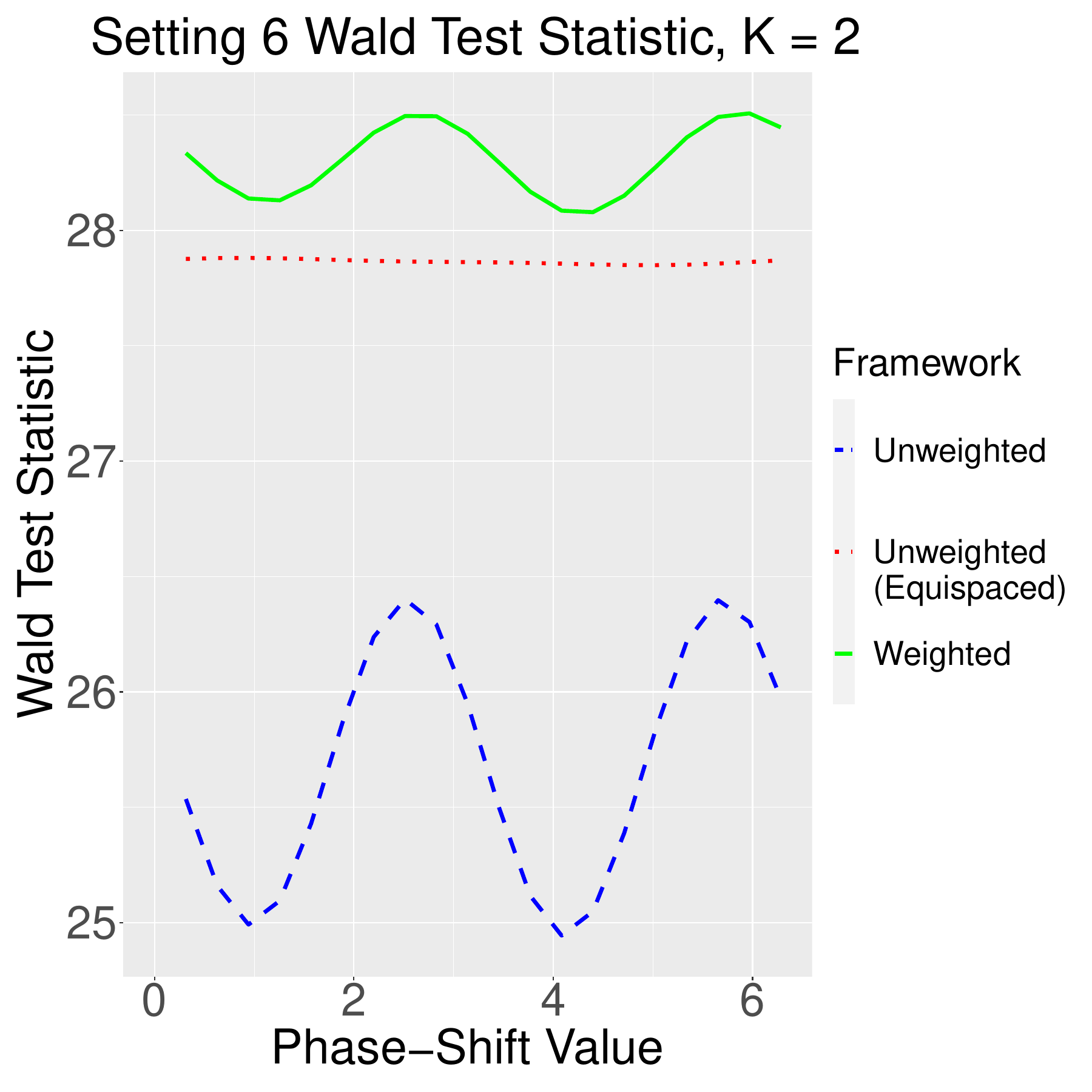}
\includegraphics[scale=0.175]{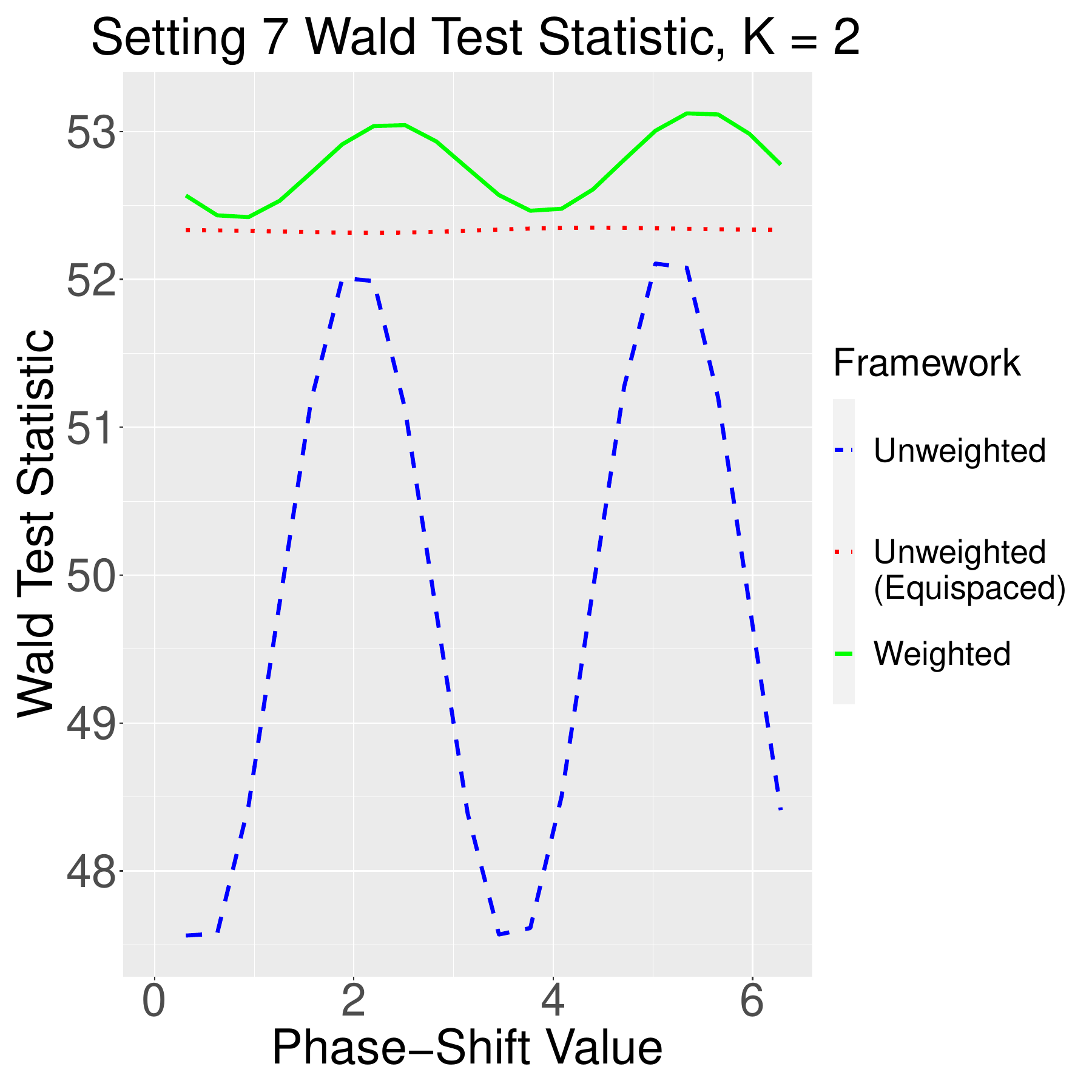}
\caption{Wald test statistics divided by sample size across 250,000 trials for second-order trigonometric regression ($K=2$). Empirically, the unweighted regression with sampled data has greater variability in numeric value for the computed statistic than the corresponding statistic computed from the weighted regression as the phase-shift estimand varies from $0$ to $2\pi$. }
    \label{fig:phases2}
\end{figure*}

\begin{figure*}[!h]
\centering
\includegraphics[scale=0.175]{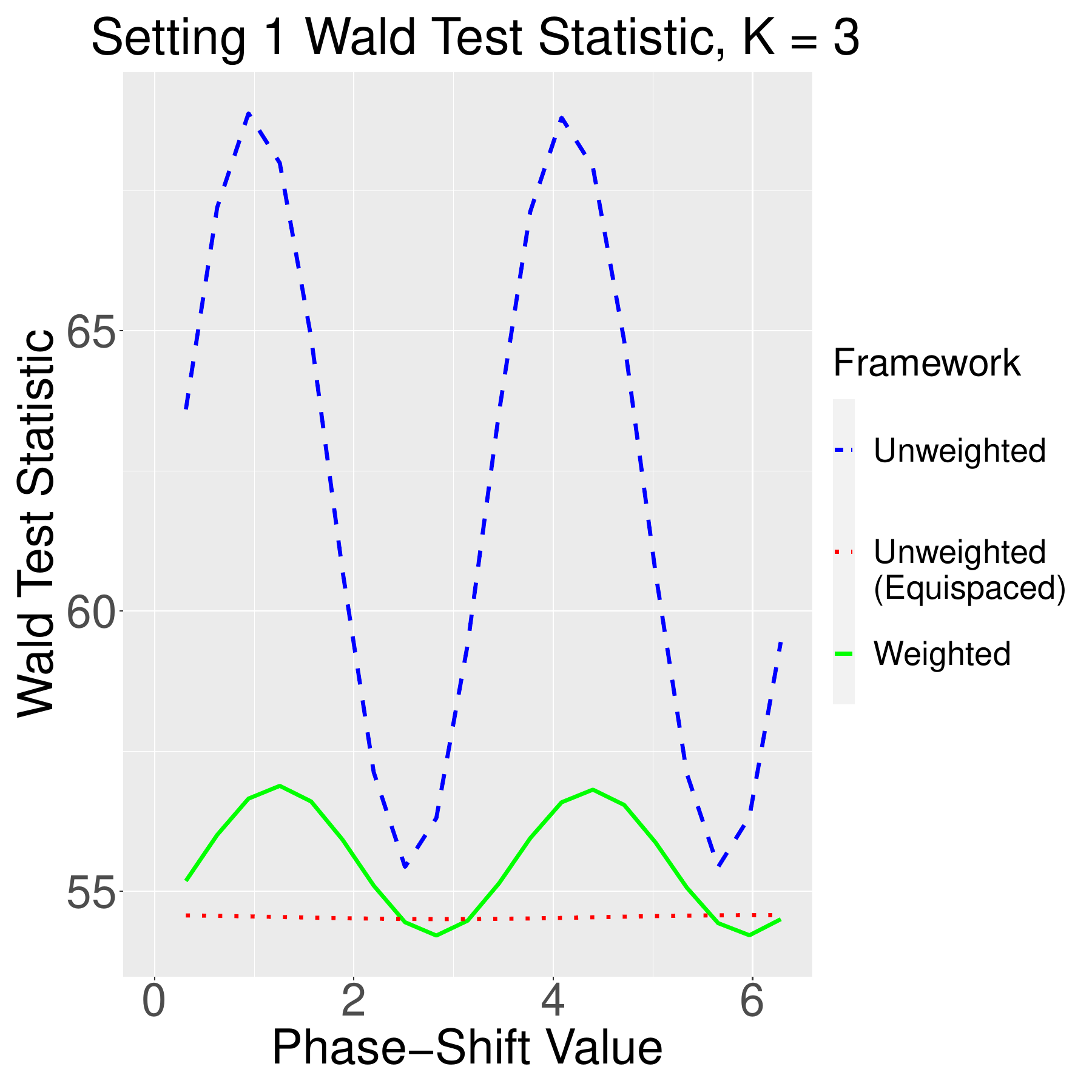}
\includegraphics[scale=0.175]{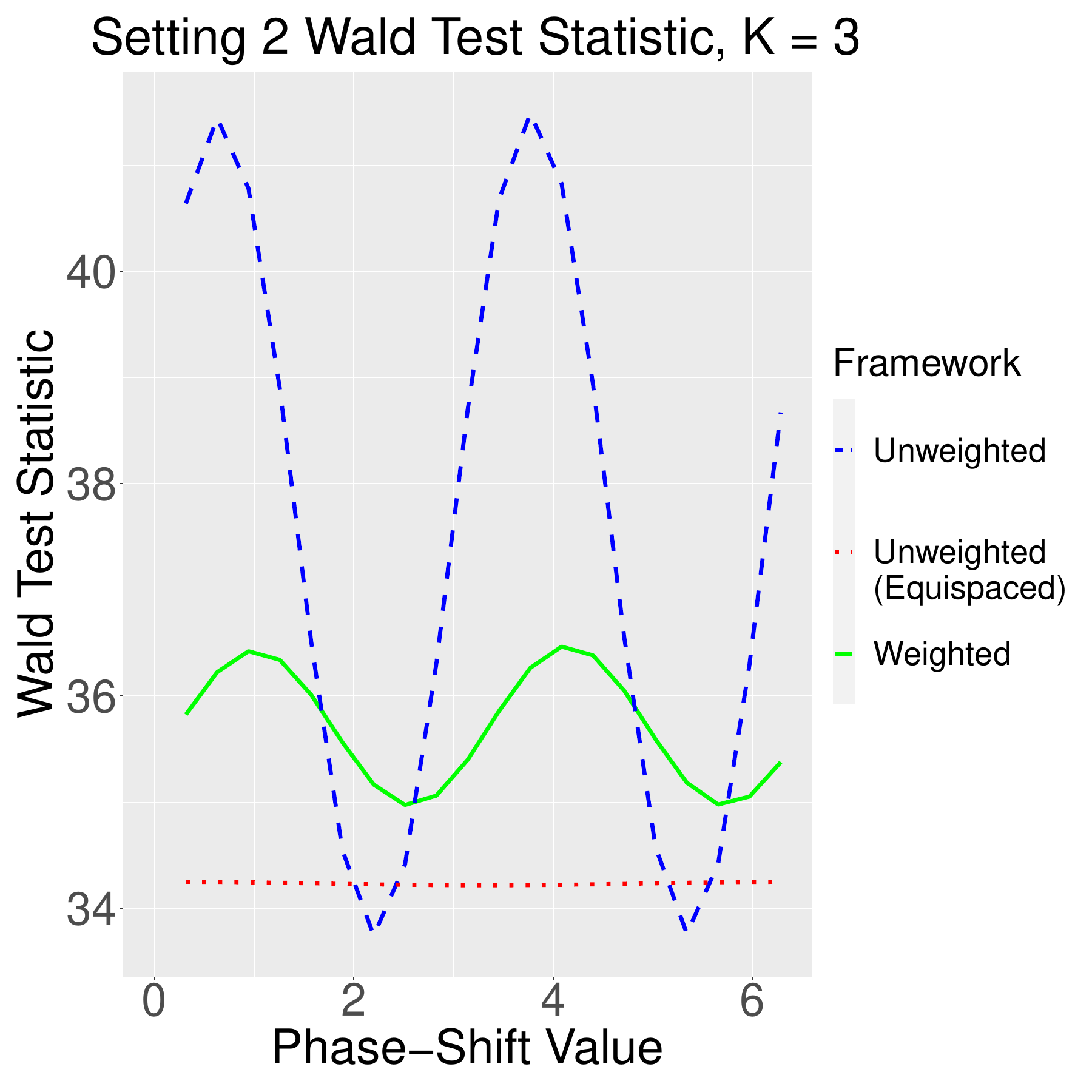} 
\includegraphics[scale=0.175]{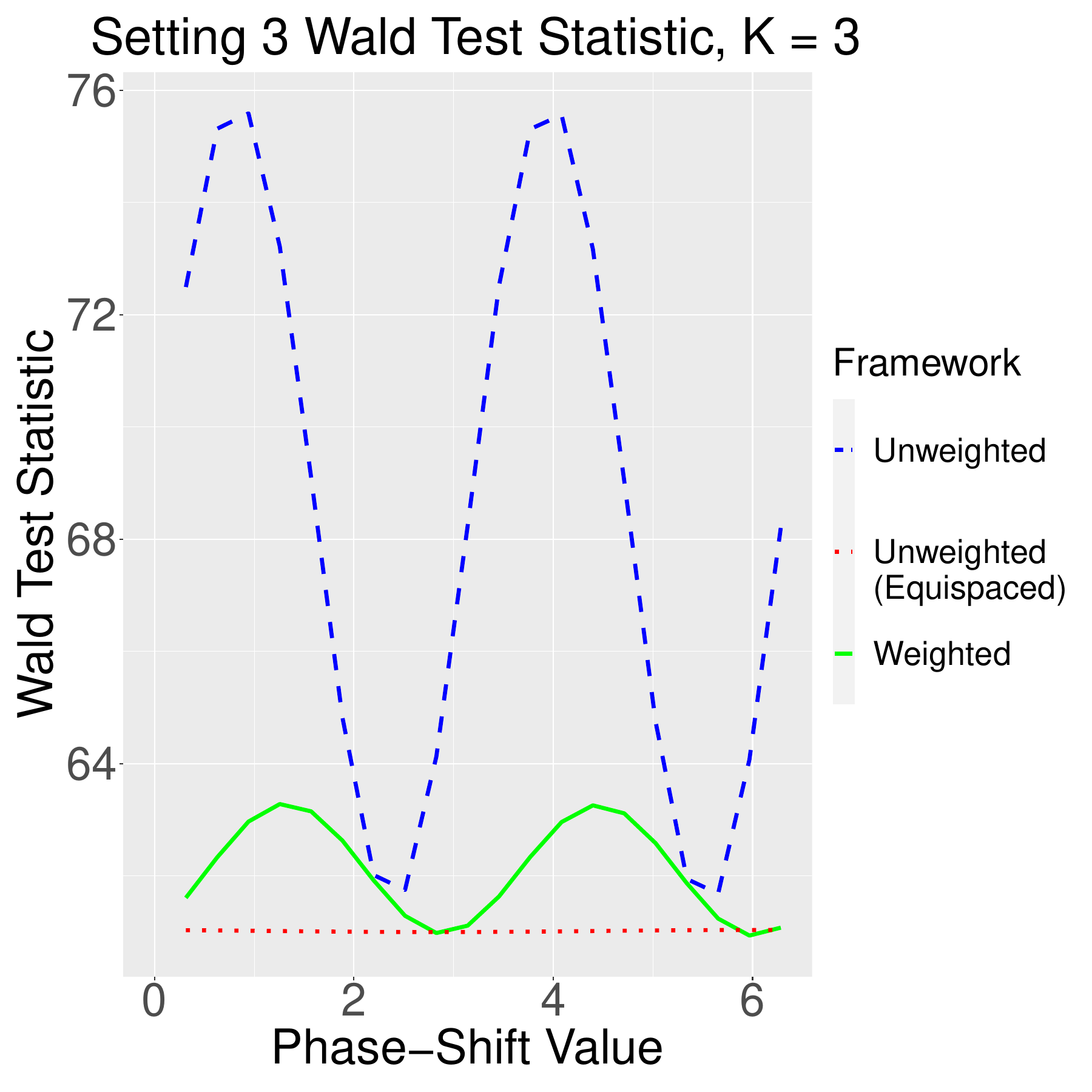} \\
\includegraphics[scale=0.175]{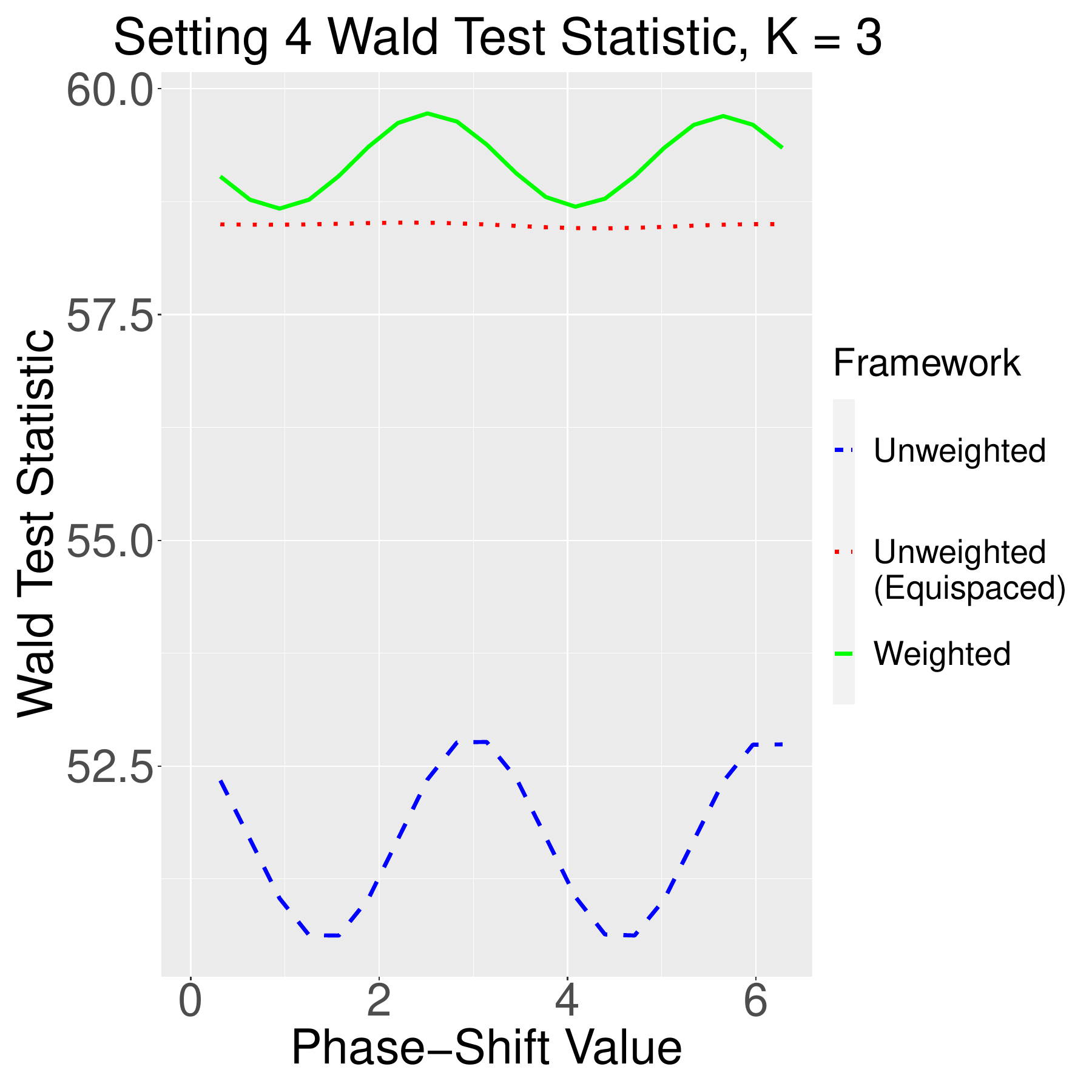} 
\includegraphics[scale=0.175]{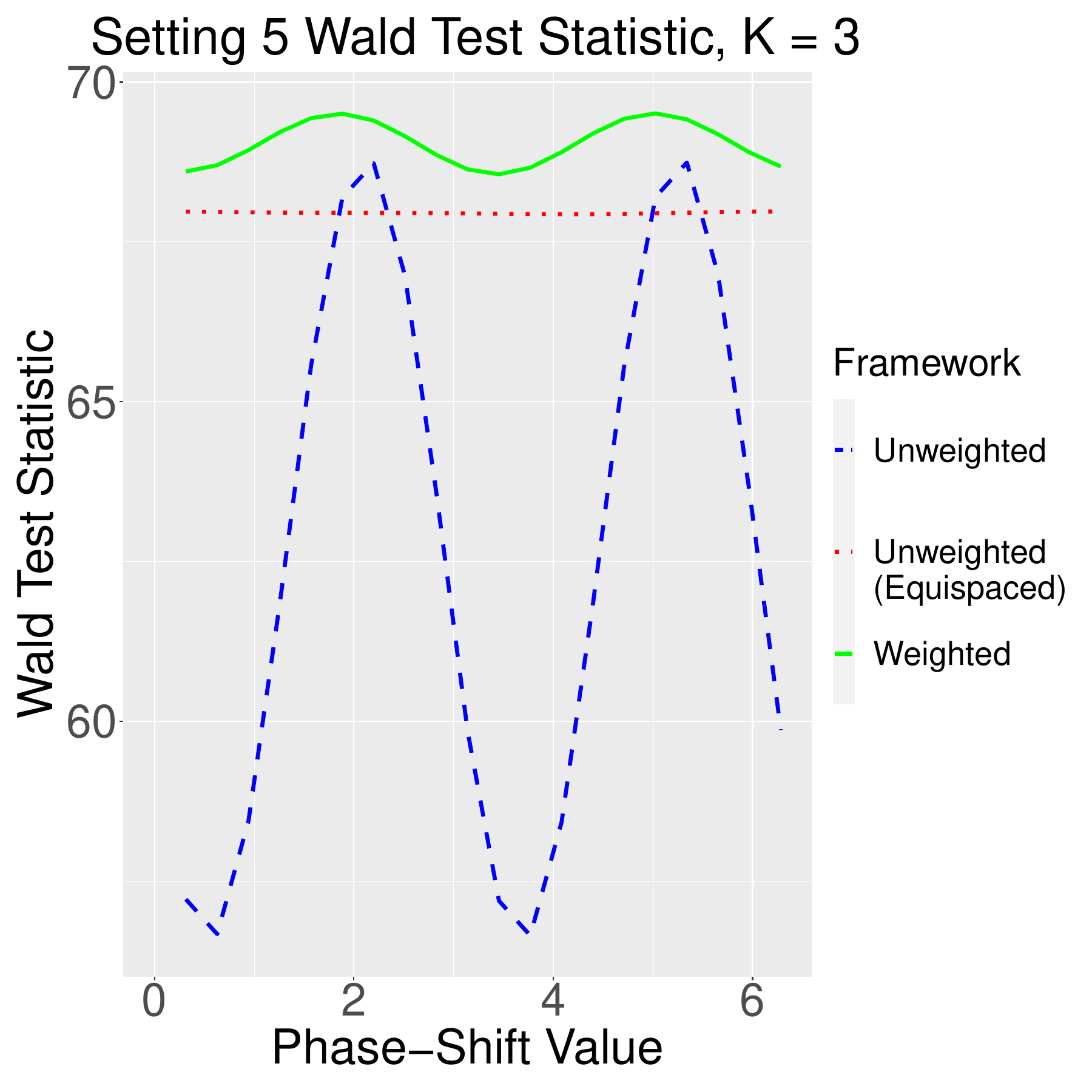} \\
\includegraphics[scale=0.175]{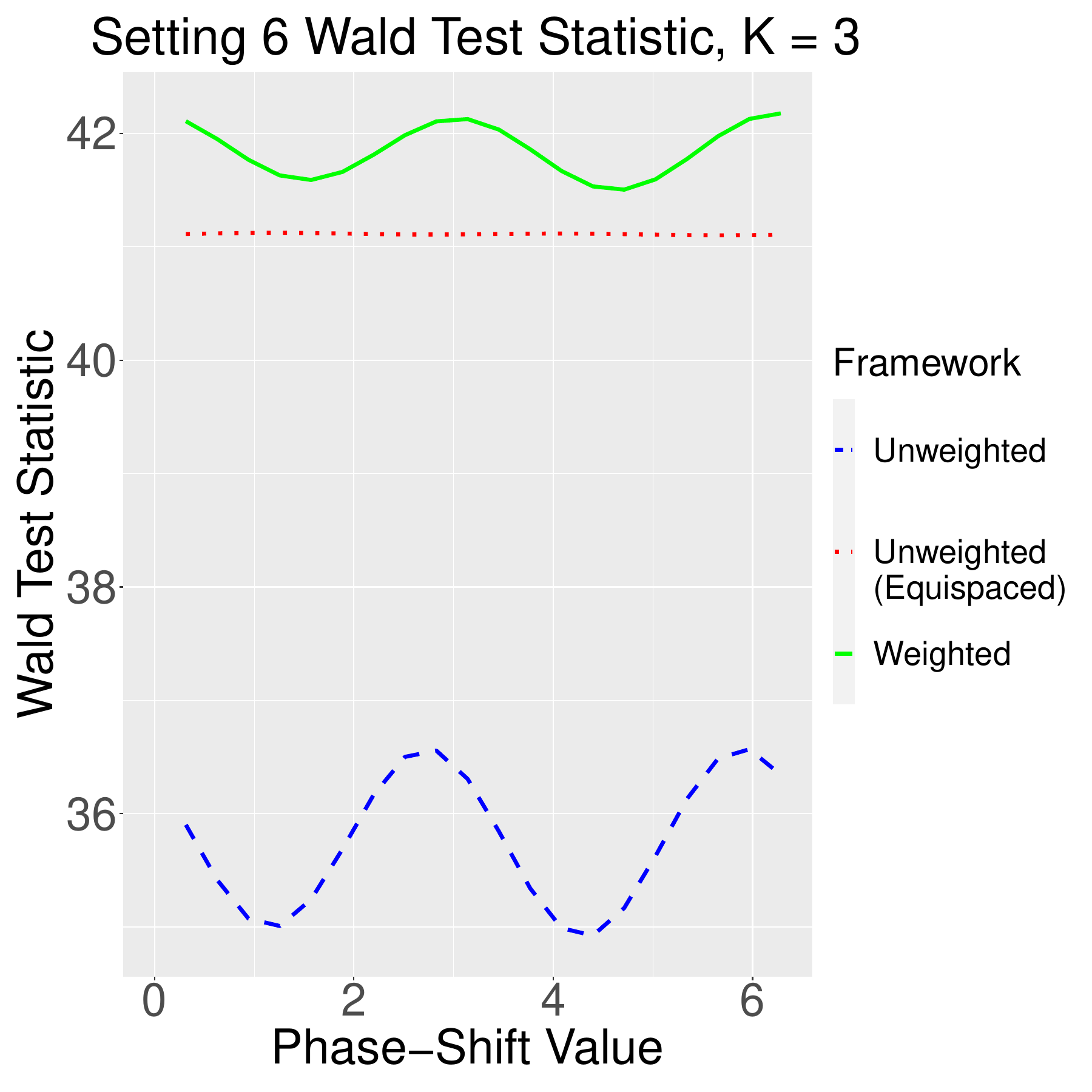}
\includegraphics[scale=0.175]{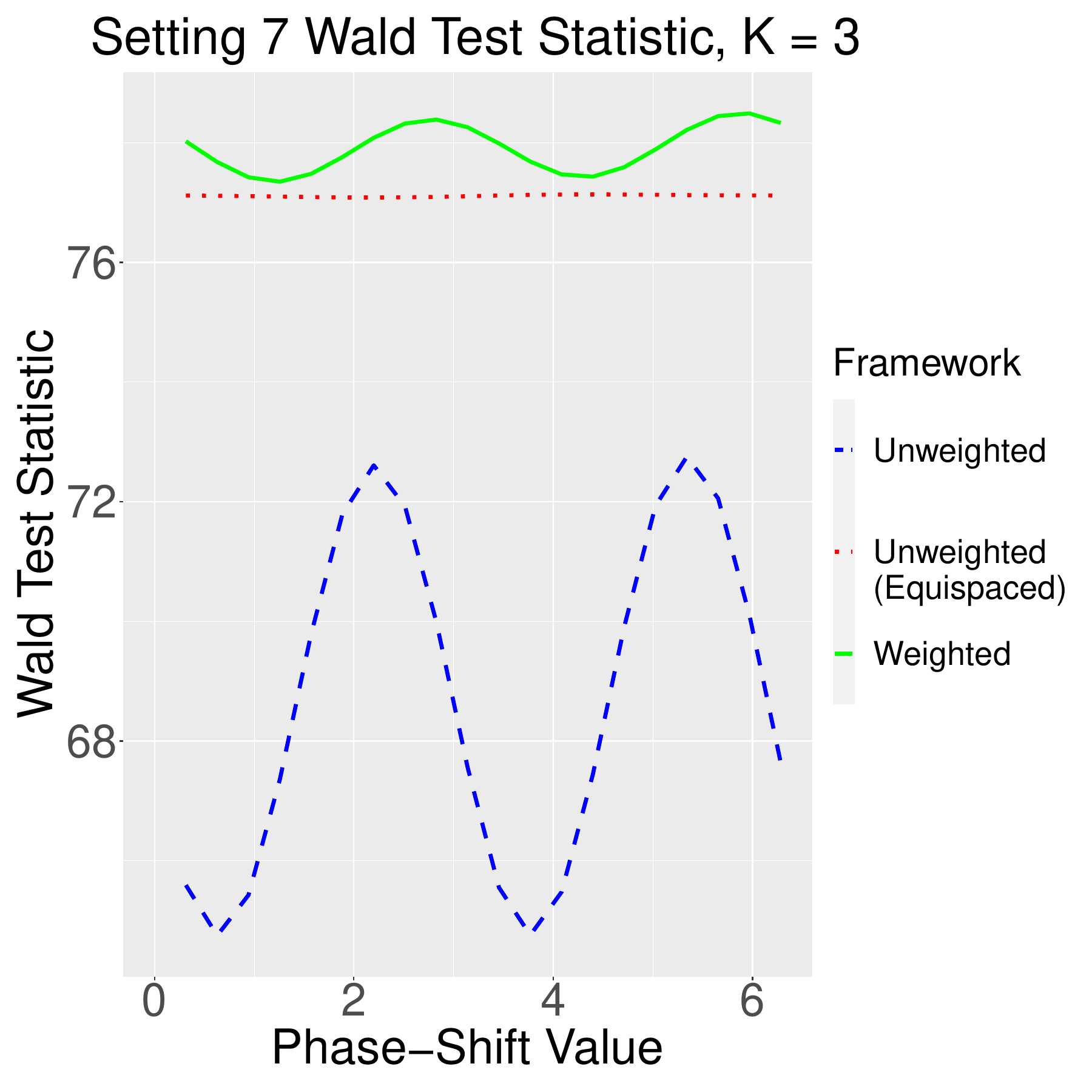}
\caption{Wald test statistics divided by sample size across 250,000 trials for third-order trigonometric regression ($K=3$). Empirically, the unweighted regression with sampled data has greater variability in numeric value for the computed statistic than the corresponding statistic computed from the weighted regression as the phase-shift estimand varies from $0$ to $2\pi$. }
    \label{fig:phases3}
\end{figure*}

Figures \ref{fig:phases1}, \ref{fig:phases2}, and \ref{fig:phases3} presents the mean Wald test statistic computed across different $\phi_{\text{base}}$ values with $K=1, 2, 3$ over 250,000 simulation trials, respectively. Each figure illustrates that the variability in computed test statistics obtained from the unweighted regression is notably larger than the variability of the weighted regression and unweighted regression where data was sampled from an equispaced design. Table \ref{tab:CoV} further examines this phenomenon by presenting the coefficients of variation derived from each mean test statistic quantity as $\phi_{\text{base}}$ is varied. Across each simulation setting, we observe that the unweighted regression yields coefficients of variation larger than the weighted regression when data are not sampled from an equispaced design. It is also observed in these experiments that the unweighted regression can produce test statistics that are larger than those produced from the weighted regression. In particular, for simulation settings 1-3, the unweighted regression produces larger Wald test statistics than the weighted regression in Figures \ref{fig:phases2} ($K=2$) and \ref{fig:phases3} ($K=3$). Corresponding results for $F$-test statistics are presented in Section 1 of the supplementary material, which yield similar conclusions.

\begin{figure*}[!h]
\centering
\includegraphics[scale=0.175]{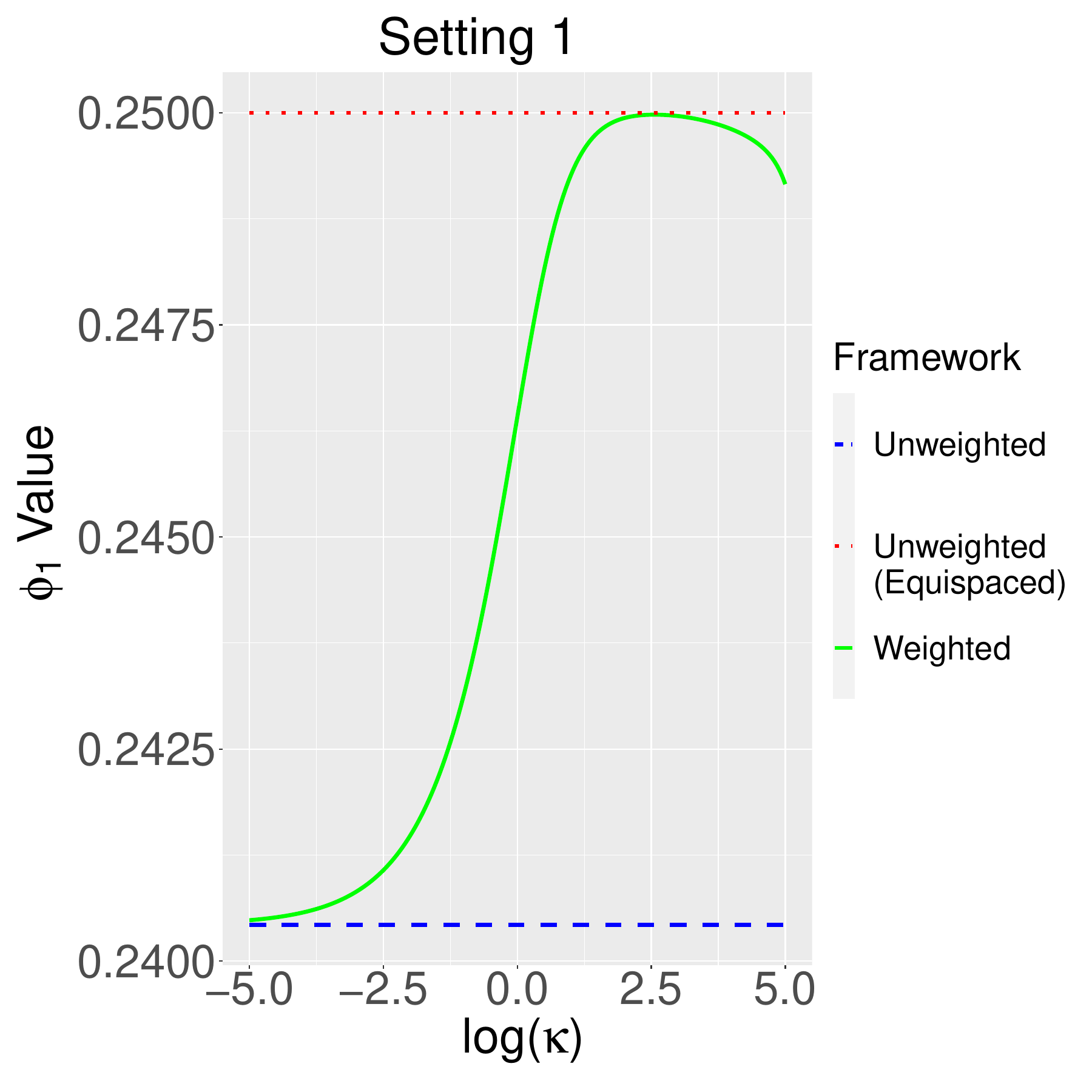}
\includegraphics[scale=0.175]{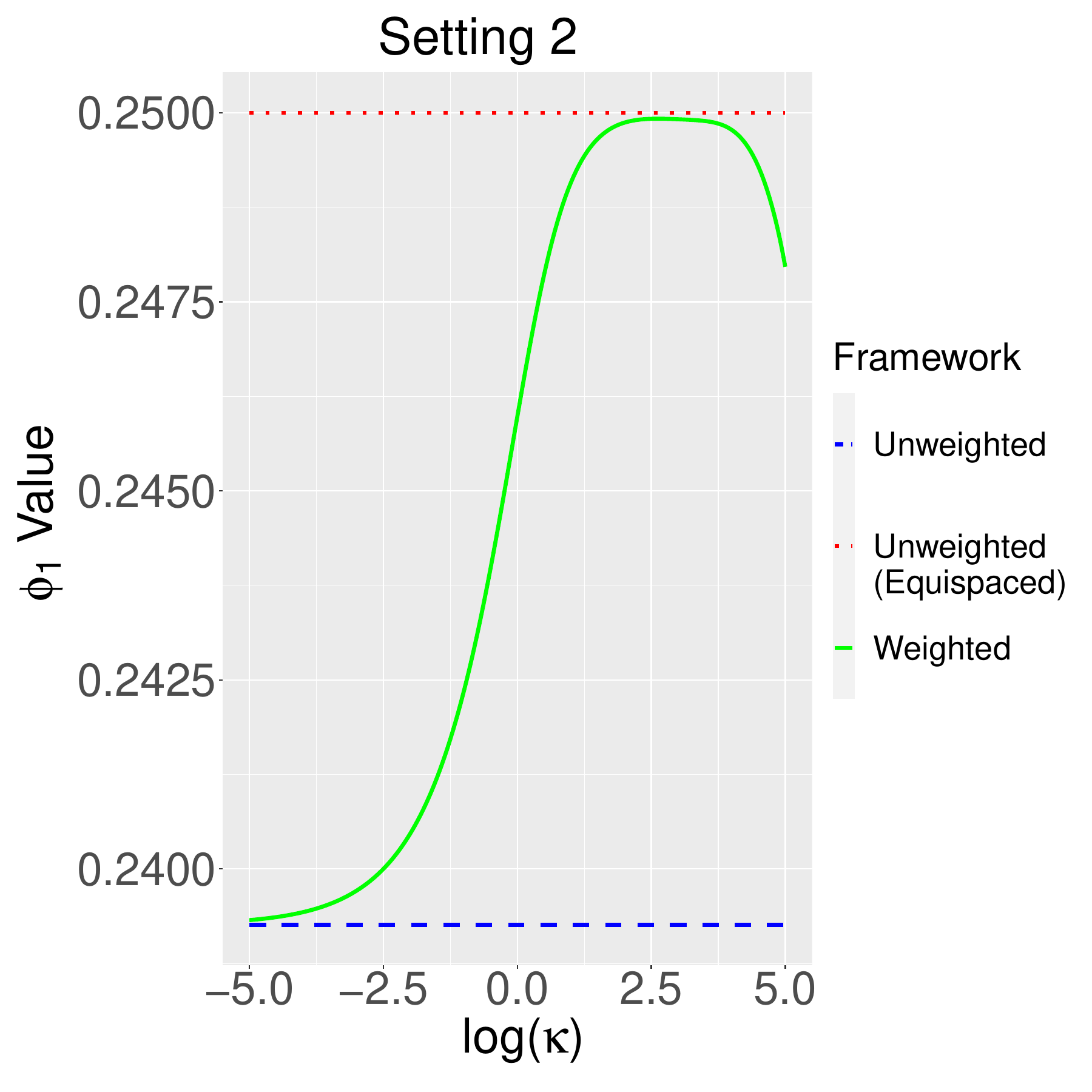}
\includegraphics[scale=0.175]{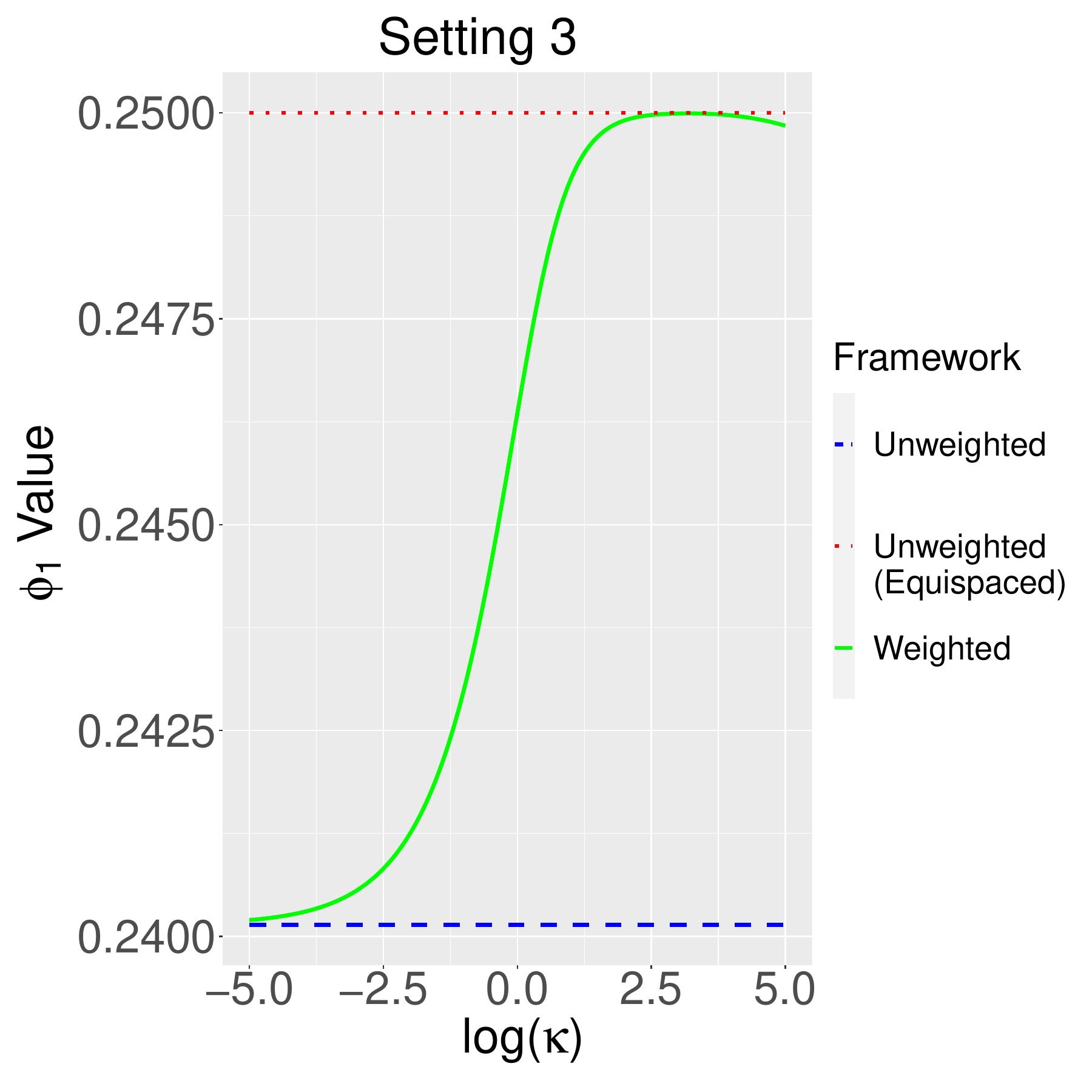} \\
\includegraphics[scale=0.175]{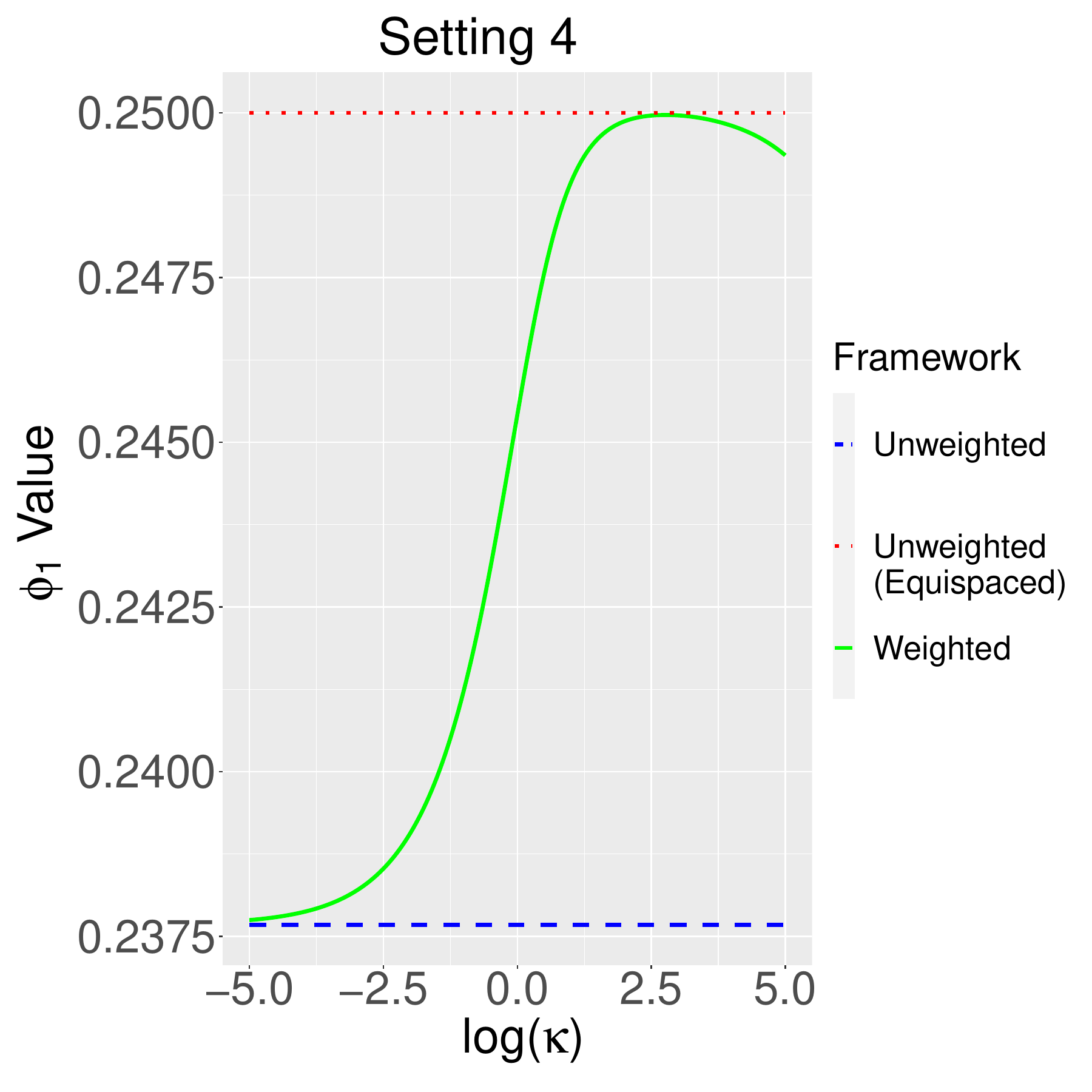}
\includegraphics[scale=0.175]{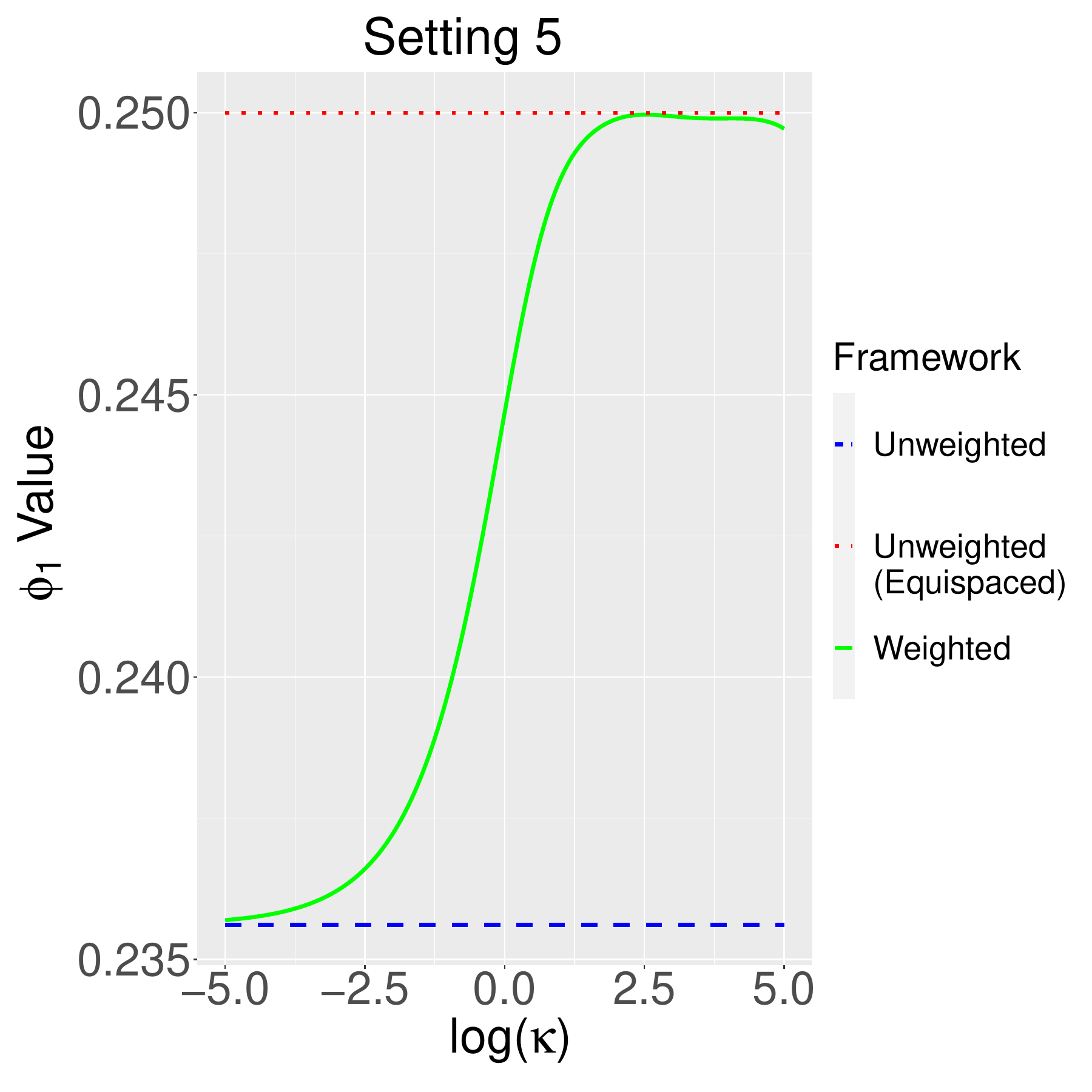} \\
\includegraphics[scale=0.175]{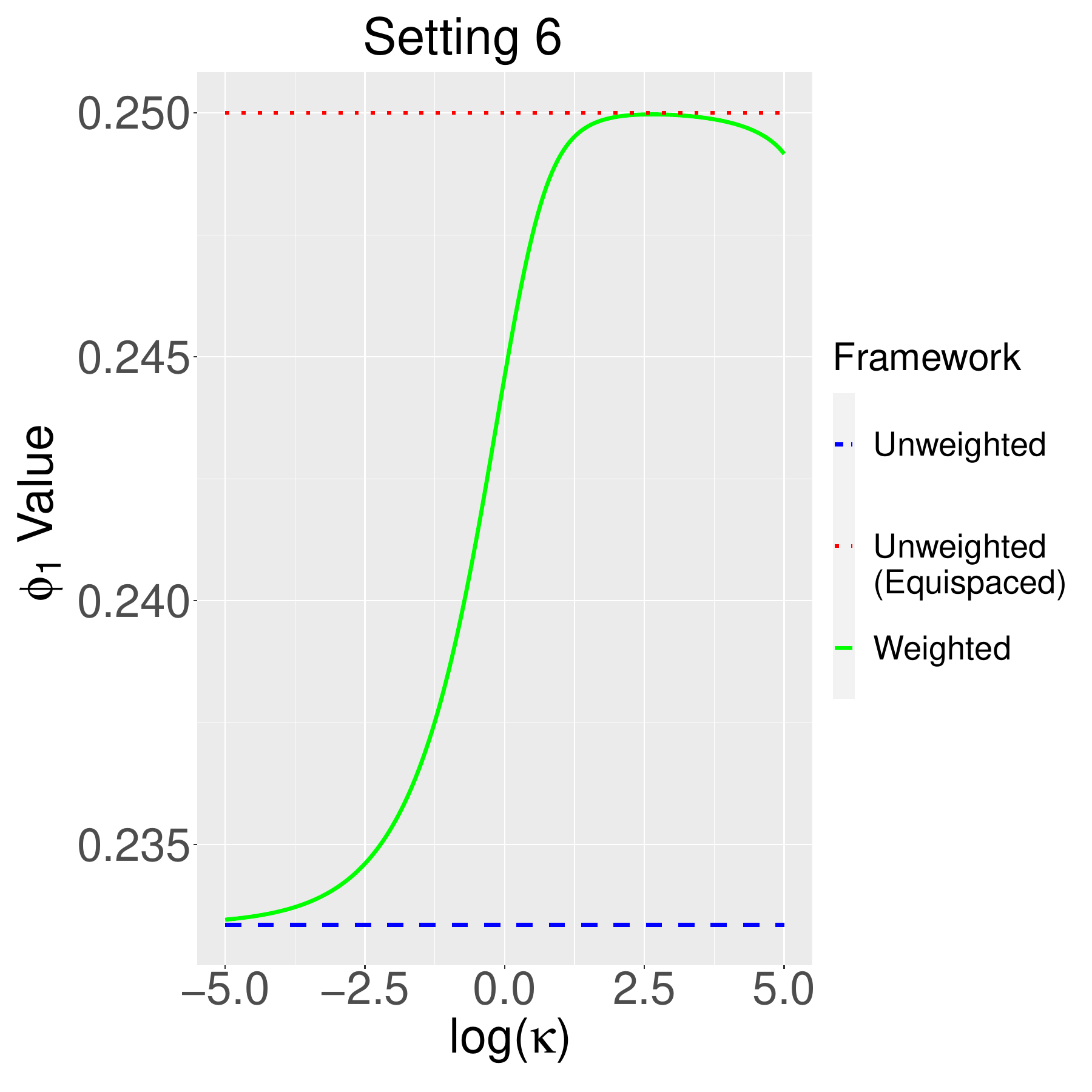} 
\includegraphics[scale=0.175]{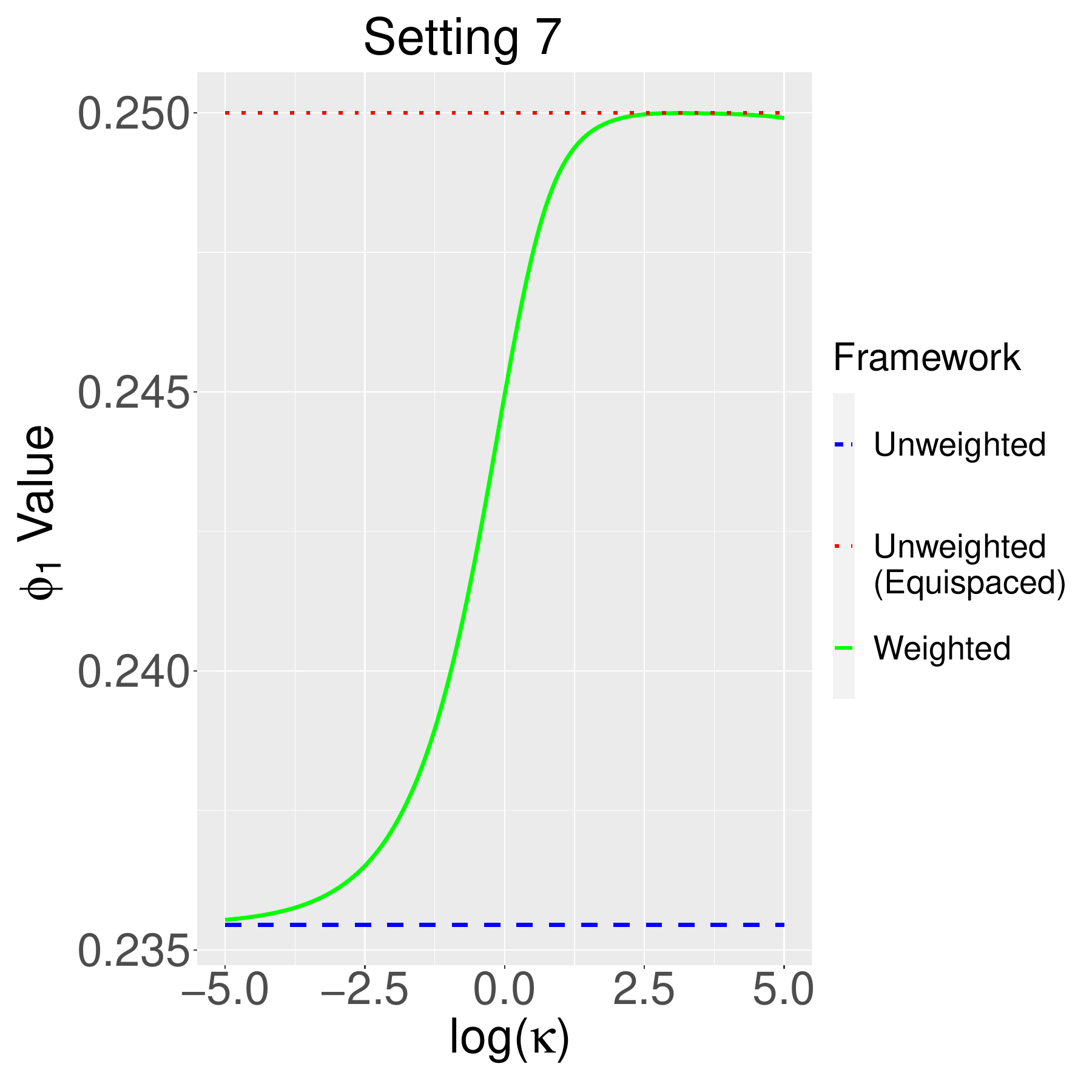}

\caption{Values of $\phi_1$ from (\ref{eq:opt_CV}) as the concentration hyperparameter $\kappa$ is varied for $K=1$. For every simulation setting, the value $\kappa_{\text{opt}}$ obtains a $\phi_1$ value that approximately equals $1/4$, which is $D$-optimal. The red dotted line denotes the value obtained from a $D$-optimal design, and the blue dashed line denotes the value obtained from the unweighted regression given observed time data. }
    \label{fig:opt}
\end{figure*}

Figure \ref{fig:opt} presents the results obtained from the optimization problem in (\ref{eq:opt_CV}) for different values of the hyperparameter $\kappa$. This optimization problem relates to maximizing the $D$-optimality criterion, which obtains a maximum value of $1/4^K$. For each simulation setting, the selected value $\kappa_{\text{opt}}$ consistently obtains the optimal value of $1/4$ when $K=1$. Additionally, Figure \ref{fig:opt} includes the corresponding $D$-optimality criterion value computed from the observed data, which is used by the unweighted regression. The $D$-optimality criterion value is consistently smaller than $1/4$ when computed from observed data.

\section{Illustrations with Circadian Biology Data Sets} \label{sec:4}

We consider the same seven sample populations from the three publicly available circadian biology data sets utilized in the simulation study in Section \ref{sec:3} for illustrations. The gene expression data from these data sets have been processed by \cite{Huang2024} and consist of 7,615 genes after processing. It is noted that some genes in the Archer data set have missing expression data, and these genes are removed for illustration: the control samples from the Archer data set consist of 4,475 genes, the experimental samples from the Archer data set consist of 4,599 genes, and the control samples and the experimental samples from the Archer data set combined consist of 3,689 genes. Given the prevalence of first-order trigonometric regression in practice \citep{Archer2014, Cornelissen2014, delolmo2022, Kitsos1988, MllerLevet2013}, we only consider fitting this regression for each gene in these illustrations.

The illustrations follow a similar setup to the simulation study in Section \ref{sec:3}, where the Wald test statistics and $F$-test statistics computed from an unweighted regression model are compared against those from a weighted regression model. However, it is emphasized that each gene exhibits distinct oscillatory behavior over time. To account for the distinct oscillatory behavior displayed by each gene, we adopt a linear model-based approach for illustration. In these illustrations, the covariate of the linear model is specified as the test statistic derived from the unweighted regression, while the corresponding test statistic from the weighted regression is specified as the response variable. No intercept term is specified for this linear model and the resulting regression yields a single regression parameter estimate, $\beta$. We utilize $\beta$ to assess the relationship between test statistics obtained from the weighted and unweighted regressions. For example, if $\beta=1$, a linear relationship exists between the test statistics from both models. If $\beta > 1$, the test statistics from the weighted regression are consistently larger than those from the unweighted regression. If $\beta < 1$, the test statistics from the weighted regression are consistently smaller than those from the unweighted regression.

Figure \ref{fig:app1} displays scatter plots and linear model fits to data from each of the seven sample populations. For each sample population, the linear model parameter estimate $\beta > 1$, which indicates that test statistics computed from the unweighted regression are larger than those computed from the unweighted regression. Specifically, across each sample population the parameter estimate $\beta$ varied from $1.064$ to $1.198$. This corresponds to a mean increase in test statistic magnitude ranging from $6.4\%$ to $19.8\%$ when employing weighted regression compared to unweighted regression. Table \ref{tab:diff} summarizes these results further, and shows that the majority of genes for each sample population are larger for the weighted regression than they are for the unweighted regression.

\section{Discussion} \label{sec:5}

In this paper, we propose a weighted regression for modelling gene expression given time when data are not obtained from an equispaced design. Specifically, the weights in this regression are the normalized reciprocals of estimates derived from a kernel density estimator for the time of sample collection, which increases the contribution of samples obtained at underrepresented time points. The simulation study in Section \ref{sec:3} demonstrates that the weighted regression consistently yields a $D$-optimality criterion value comparable to a value that would be obtained in an equispaced design. The simulation study also indicates that the weighted regression does not necessarily produce larger test statistics than an unweighted regression depending on the phase-shift estimands of the true model in finite-sample settings, it does consistently mitigate variability in test statistics computed at different phase-shift estimands. Given that the phase-shift estimands of genes and the marginal density for the time of sample collection can vary across study cohorts, mitigating this variability could improve consistency in the identification of genes for further analysis. Further, across all seven sample populations in the real data set, the weighted regression consistently produced larger test statistics than the unweighted regression. 

It is essential to acknowledge the limitations of this weighted regression. One limitation is revealed in the simulation studies, which indicate that the weighted regression mitigates variability across different phase-shift estimands, but it does not eliminate it entirely. This relates to properties of kernel density estimation, which indicate that as sample size increases, the kernel density estimator converges towards its estimand and would asymptotically remove variability. We particularly recommend applying this method to larger cohorts where more samples are collected. A second limitation is that the methodology assumes that data are sampled over an entire 24 hour period. While this assumption is reasonable for many circadian biology studies, and guidelines for circadian biology studies advocate for consistent sample collection for 48 hours \citep{Hughes2017}, we recognize that this could be infeasible for some cohorts. However, this weighted regression would still increase the contribution of samples at underrepresented time points, and a concentration hyperparameter can be identified from (\ref{eq:opt_CV}) to facilitate an appropriate weighting. 

This paper presents opportunities for future work. One direction concerns investigating the behavior of other frameworks proposed for identifying circadian genes when data are not obtained from an equispaced design \citep{Mei2020}. If these frameworks are also sensitive to the behavior of the marginal distribution for the time of sample collection, one could identify similar weighting schemes to mitigate the sensitivity of these frameworks to suboptimal designs. Another direction is that the experimental designs of circadian biology studies are some times curated to consider multiple cohorts for investigation. One could extend the weighted regression for hypothesis testing of how gene expression changes across cohorts in these studies \citep{Parsons2019}.

\vspace*{1pc}

\noindent {\bf{Acknowledgements.}}
The authors would like to thank Tavish McDonald at Lawrence Livermore National Laboratory for productive discussions that led to the derivation in Appendix \ref{app:B}.

\

\noindent {\bf{Conflict of Interest.}}
The authors have declared no conflict of interest.

\

\noindent {\bf{Supporting Information.}}
Additional simulation and real data results related to $F$-statistic outputs are provided in the supplementary material of this article.

\

\begin{figure*}[!h]
\centering
\includegraphics[scale=0.175]{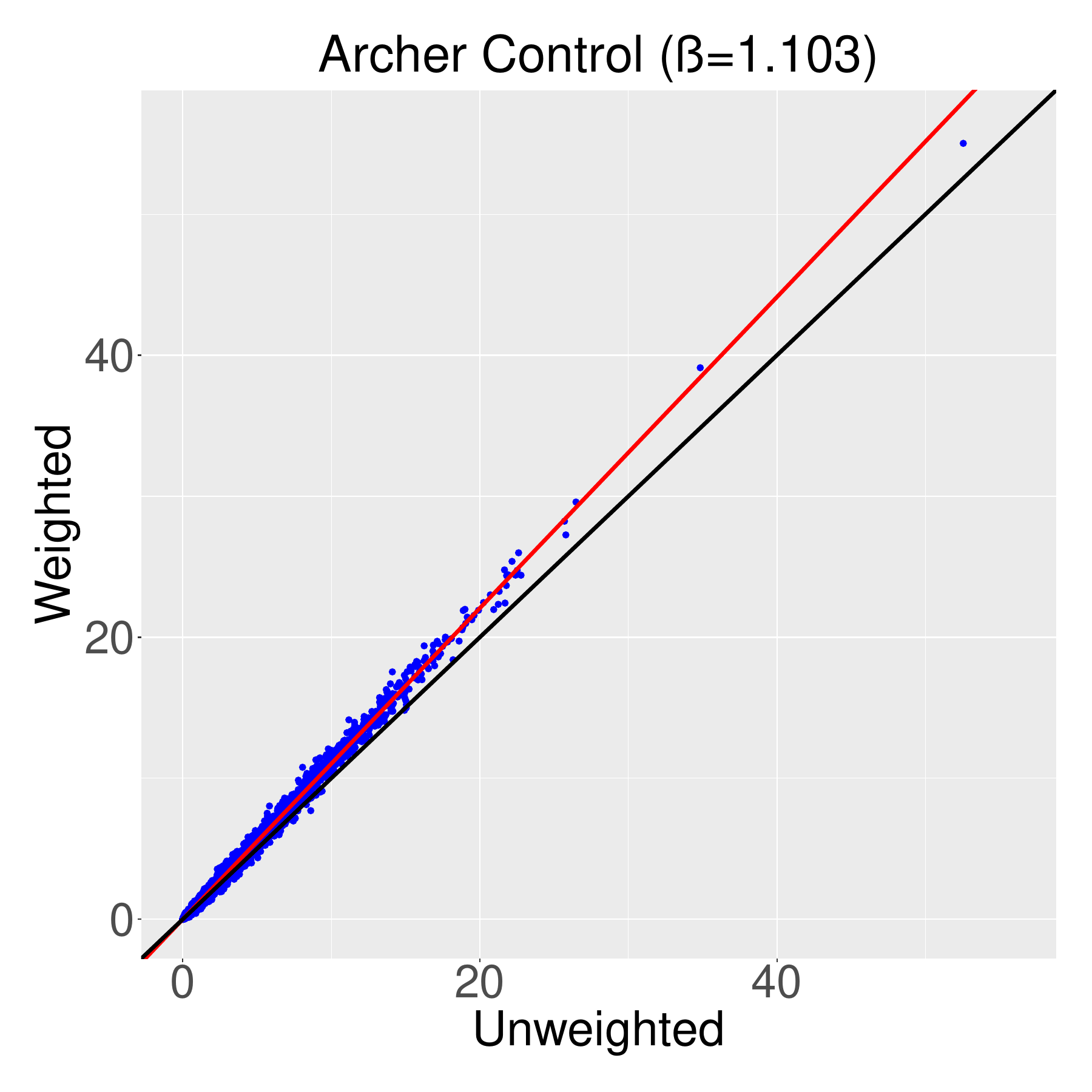}
\includegraphics[scale=0.175]{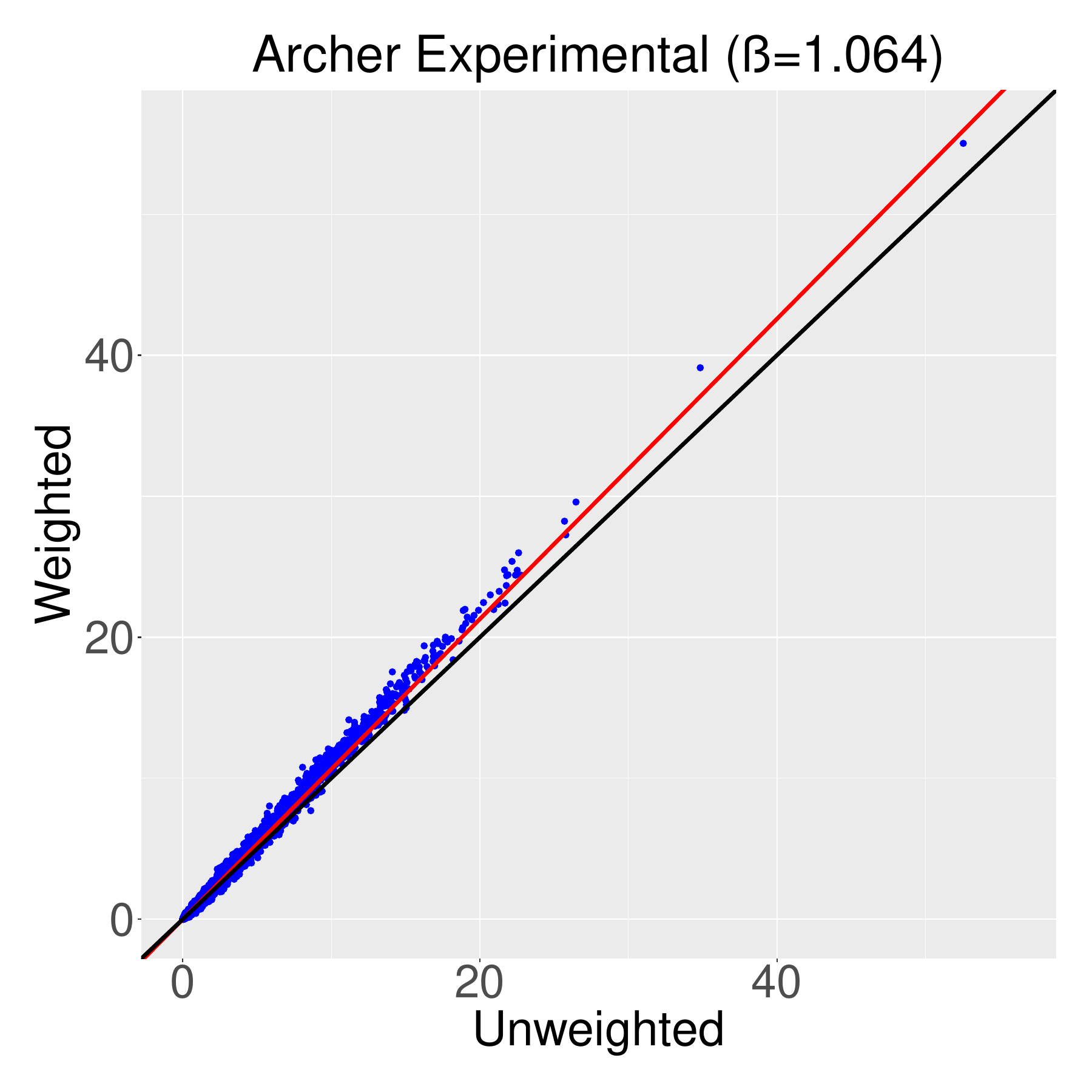}
\includegraphics[scale=0.175]{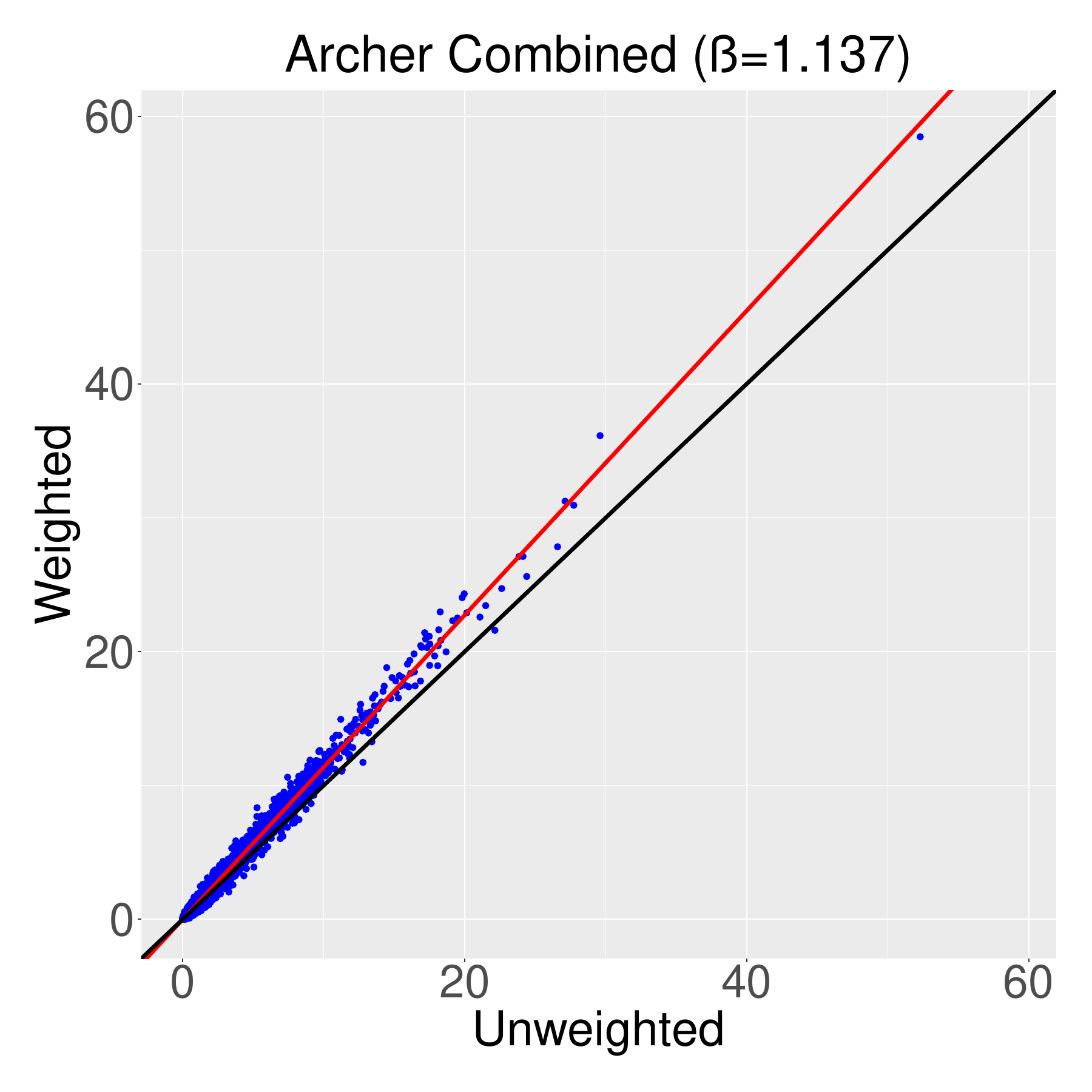} \\
\includegraphics[scale=0.175]{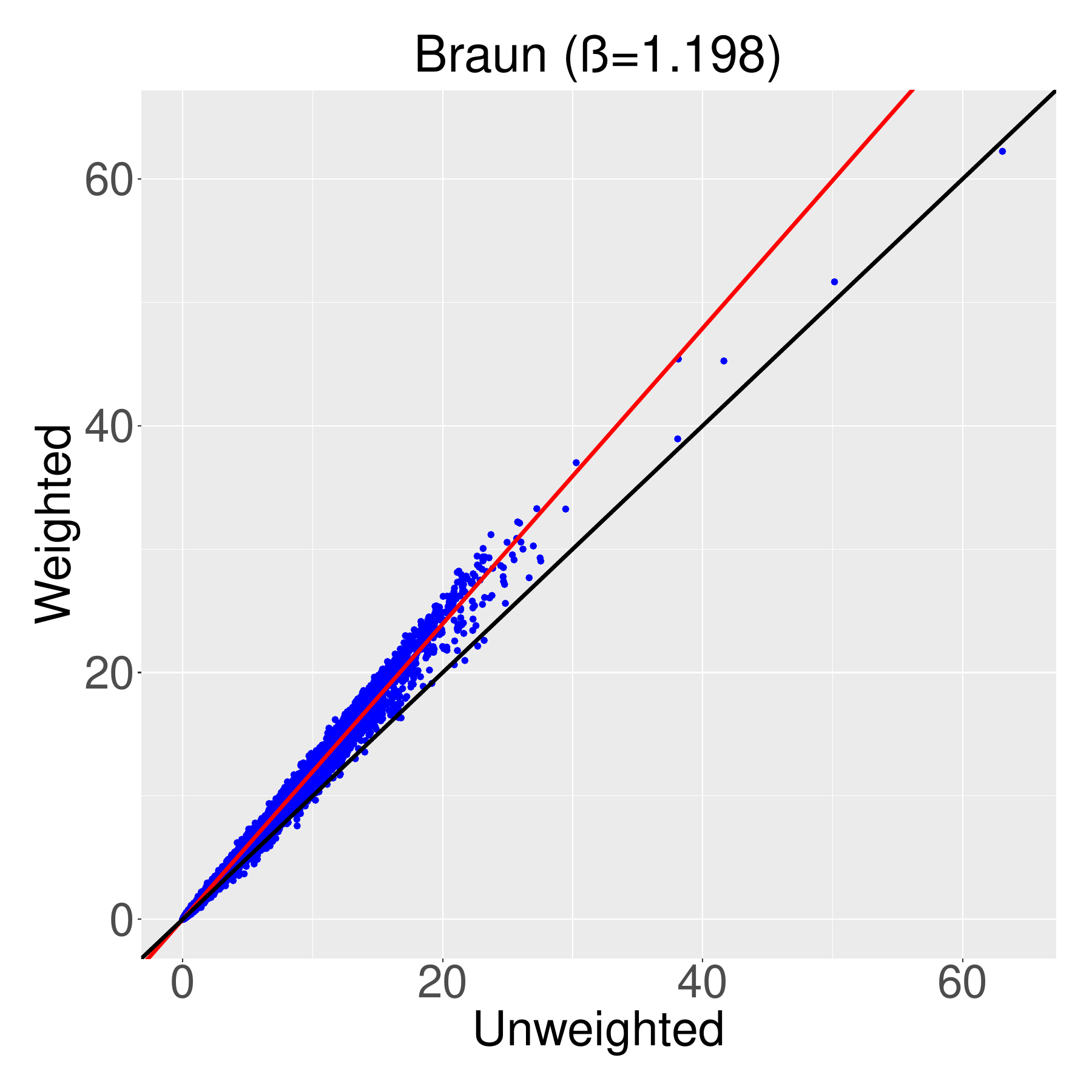}
\includegraphics[scale=0.175]{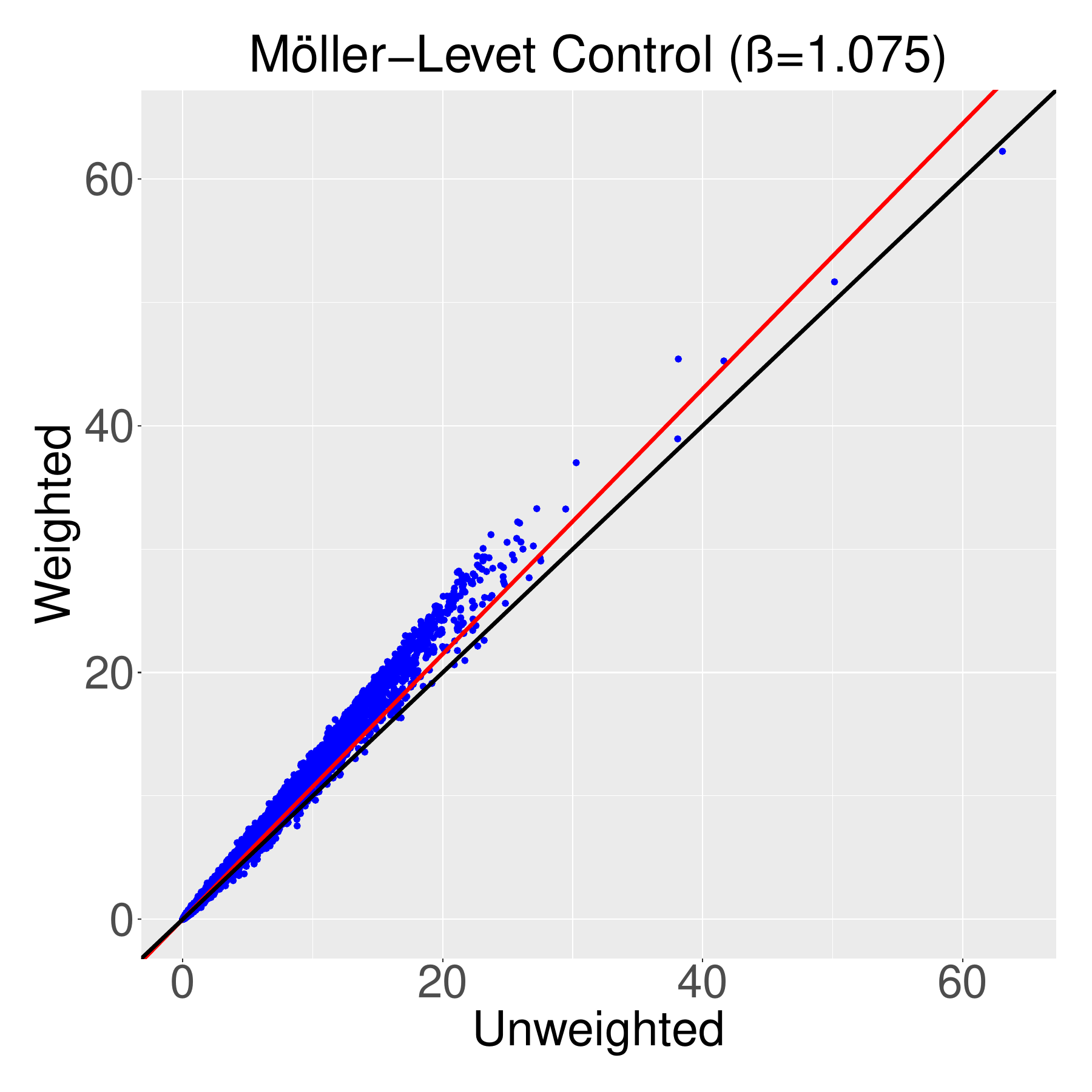} \\
\includegraphics[scale=0.175]{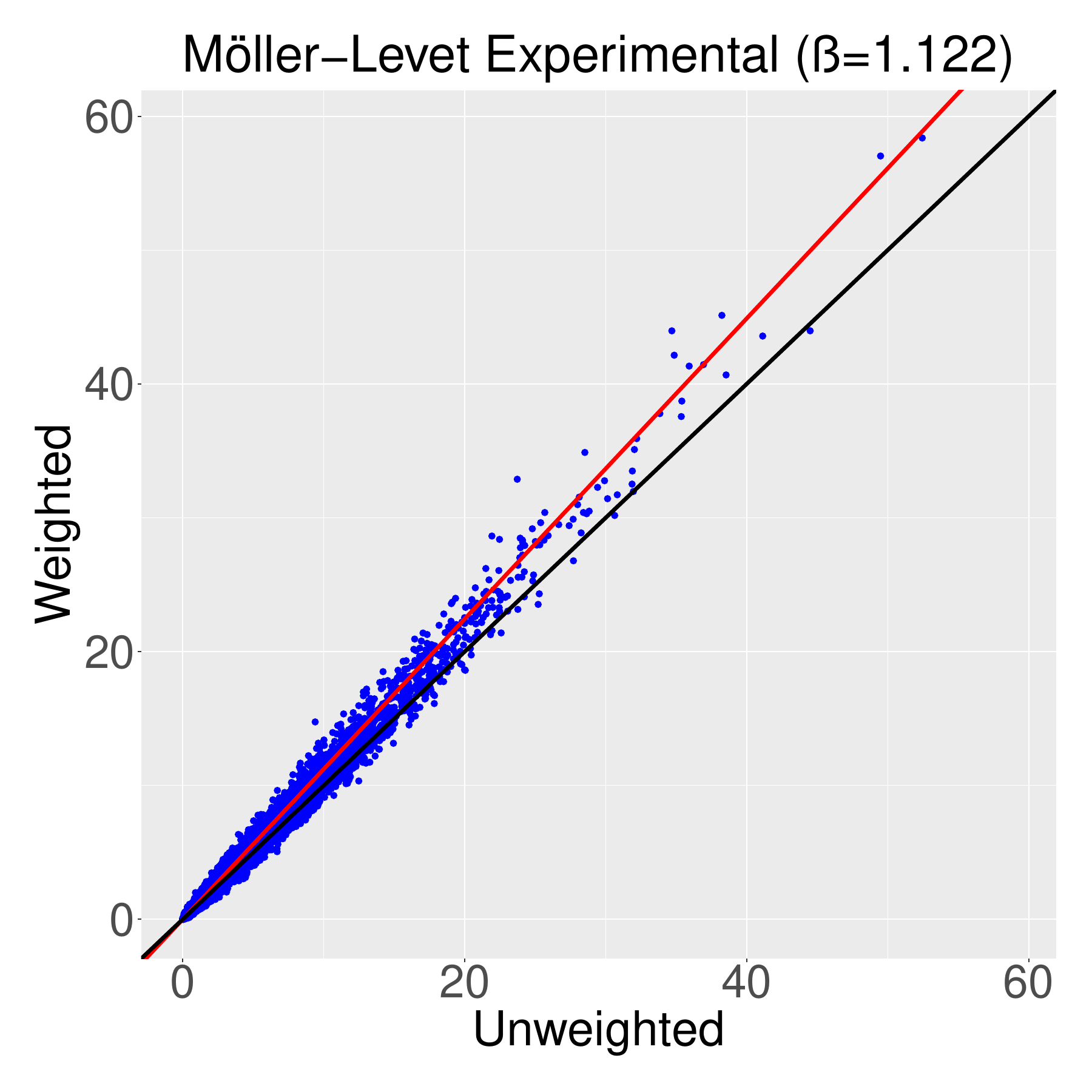}
\includegraphics[scale=0.175]{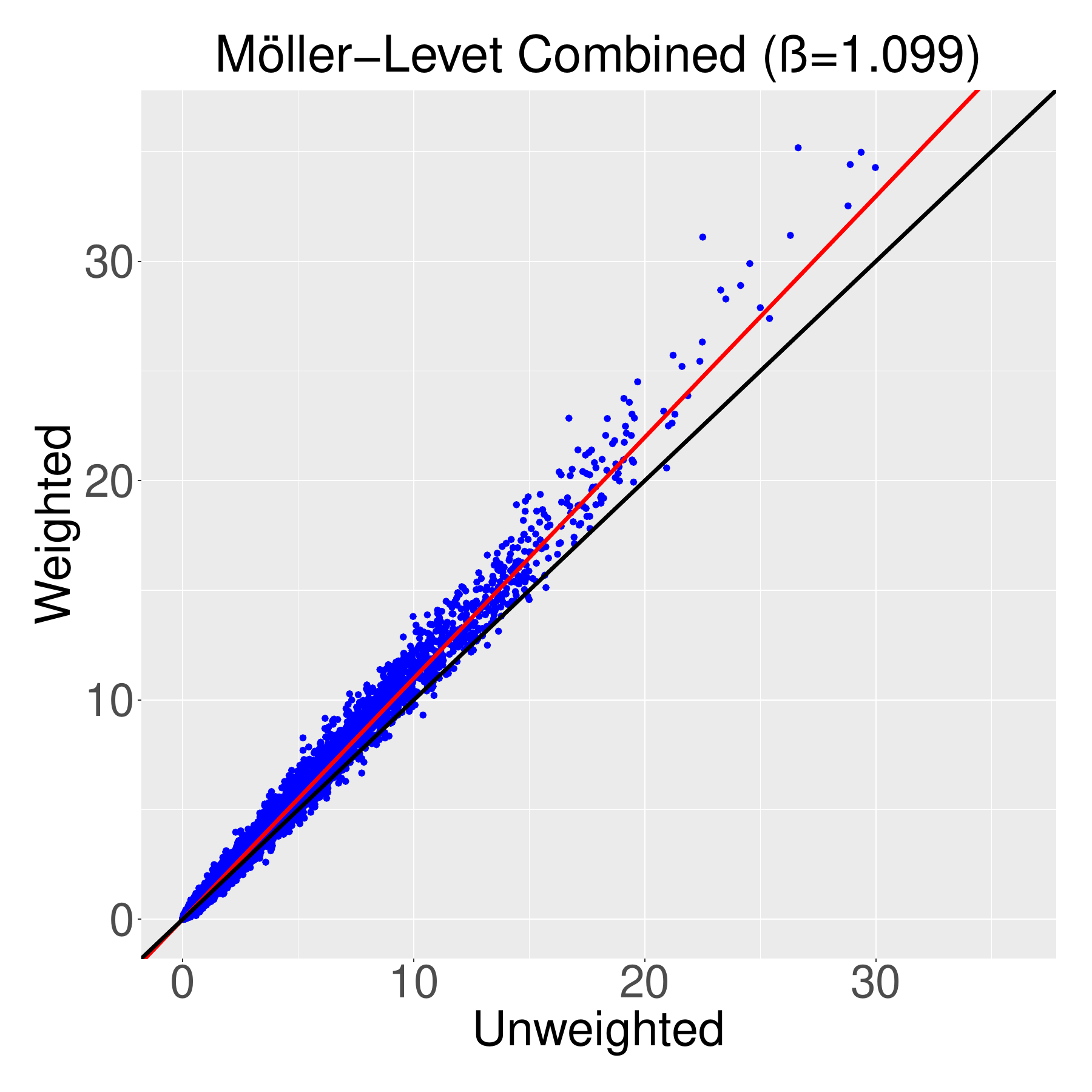}
\caption{Wald test statistics computed from each sample population. The black line represents a reference where $\beta=1$, and the red line indicates the estimated regression model fit. A blue point represents a Wald test statistic obtained for a specific gene from the weighted and unweighted regressions. The weighted regression consistently produces larger test statistics than the unweighted regression.  }
    \label{fig:app1}
\end{figure*}

\noindent {\bf{Data Availability Statement}}
Code scripts for reproducing the linear models in Figure \ref{fig:app1} and gene counts in Table \ref{tab:diff} are available at
\textcolor{blue}{\href{https://bitbucket.org/michaelgorczyca/weighted_trigonometric_regression/}{https://bitbucket.org/michaelgorczyca/weighted\_trigonometric\_regression/}}.

\newpage
\clearpage

\clearpage
\newpage

\clearpage
\newpage

\clearpage
\newpage

\clearpage
\newpage

\begin{sidewaystable*}[!h]
	\caption{Coefficients of variation for test statistics computed in each simulation setting. The unweighted regression consistently produces the largest coefficient of variation across all seven simulation settings and types of trigonometric regression models considered ($K=1,2,3$).} \label{tab:CoV}
 \centering
		\begin{tabular}{|c|c|c|c|c|c|c|}
			\hline
    \multirow{3}{*}{Setting} & \multicolumn{6}{c|}{Coefficient of Variation ($K=1$)} \\
   \cline{2-7}
    & \multicolumn{3}{c|}{Wald Test} & \multicolumn{3}{c|}{$F$-Test} \\
   \cline{2-7}
    & Unweighted & Unweighted (Equispaced) & Weighted & Unweighted & Unweighted (Equispaced) & Weighted \\ 
   \hline
    1 & $5.777 \times 10^{-2}$ & $1.112 \times 10^{-3}$ & $5.490 \times 10^{-3}$ & $5.777 \times 10^{-2}$ & $1.112 \times 10^{-3}$ & $5.490 \times 10^{-3}$ \\
2 & $4.992 \times 10^{-2}$ & $1.284 \times 10^{-3}$ & $1.209 \times 10^{-2}$ & $4.992 \times 10^{-2}$ & $1.284 \times 10^{-3}$ & $1.209 \times 10^{-2}$ \\
3 & $5.309 \times 10^{-2}$ & $6.135 \times 10^{-4}$ & $4.154  \times 10^{-3}$ & $5.309 \times 10^{-2}$ & $6.135 \times 10^{-4}$ & $4.154 \times 10^{-3}$ \\
4 & $5.553 \times 10^{-2}$ & $1.158 \times 10^{-3}$ & $7.338 \times 10^{-3}$ & $5.553 \times 10^{-2}$ & $1.158 \times 10^{-3}$ & $7.338 \times 10^{-3}$ \\
5 & $6.255 \times 10^{-2}$ & $9.599 \times 10^{-4}$ & $5.803 \times 10^{-3}$ & $6.255 \times 10^{-2}$ & $9.599 \times 10^{-4}$ & $5.803 \times 10^{-3}$ \\
6 & $3.871 \times 10^{-2}$ & $9.476 \times 10^{-4}$ & $6.103 \times 10^{-3}$ & $3.871 \times 10^{-2}$ & $9.476 \times 10^{-4}$ & $6.103 \times 10^{-3}$ \\
7 & $5.105 \times 10^{-2}$ & $5.924 \times 10^{-4}$ & $3.425 \times 10^{-3}$ & $5.105 \times 10^{-2}$ & $5.924 \times 10^{-4}$ & $3.425 \times 10^{-3}$ \\
\hline
\multirow{3}{*}{Setting} & \multicolumn{6}{c|}{Coefficient of Variation ($K=2$)} \\
   \cline{2-7}
    & \multicolumn{3}{c|}{Wald Test} & \multicolumn{3}{c|}{$F$-Test} \\
   \cline{2-7}
    & Unweighted & Unweighted (Equispaced) & Weighted & Unweighted & Unweighted (Equispaced) & Weighted \\ 
   \hline
    1 & $7.845 \times 10^{-2}$ & $6.670 \times 10^{-4}$ & $1.214 \times 10^{-2}$ & $7.845 \times 10^{-2}$ & $6.670 \times 10^{-4}$ & $1.214 \times 10^{-2}$ \\
2 & $2.716\times 10^{-2}$ & $6.837\times 10^{-4}$ & $1.131\times 10^{-2}$ & $2.716\times 10^{-2}$ & $6.837\times 10^{-4}$ & $1.131\times 10^{-2}$ \\
3 & $5.337\times 10^{-2}$ & $2.400\times 10^{-4}$ & $1.089\times 10^{-2}$ & $5.337\times 10^{-2}$ & $2.400\times 10^{-4}$ & $1.089\times 10^{-2}$ \\
4 & $2.080\times 10^{-2}$ & $5.811\times 10^{-4}$ & $1.969\times 10^{-3}$ & $2.080\times 10^{-2}$ & $5.811\times 10^{-4}$ & $1.969\times 10^{-3}$ \\
5 & $5.562\times 10^{-2}$ & $3.068\times 10^{-4}$ & $9.857\times 10^{-3}$ & $5.562\times 10^{-2}$ & $3.068\times 10^{-4}$ & $9.857\times 10^{-3}$ \\
6 & $2.029\times 10^{-2}$ & $3.716\times 10^{-4}$ & $5.295\times 10^{-3}$ & $2.029\times 10^{-2}$ & $3.716\times 10^{-4}$ & $5.295\times 10^{-3}$ \\
7 & $3.420\times 10^{-2}$ & $2.184\times 10^{-4}$ & $4.600\times 10^{-3}$ & $3.420\times 10^{-2}$ & $2.184\times 10^{-4}$ & $4.600\times 10^{-3}$ \\
\hline
\multirow{3}{*}{Setting} & \multicolumn{6}{c|}{Coefficient of Variation ($K=3$)} \\
   \cline{2-7}
    & \multicolumn{3}{c|}{Wald Test} & \multicolumn{3}{c|}{$F$-Test} \\
   \cline{2-7}
    & Unweighted & Unweighted (Equispaced) & Weighted & Unweighted & Unweighted (Equispaced) & Weighted \\ 
   \hline
1 & $7.864\times 10^{-2}$ & $4.738\times 10^{-4}$ & $1.722\times 10^{-2}$ & $7.864\times 10^{-2}$ & $4.738\times 10^{-4}$ & $1.722\times 10^{-2}$ \\
2 & $7.446\times 10^{-2}$ & $3.506\times 10^{-4}$ & $1.505\times 10^{-2}$ & $7.446\times 10^{-2}$ & $3.506\times 10^{-4}$ & $1.505\times 10^{-2}$ \\
3 & $7.567\times 10^{-2}$ & $2.186\times 10^{-4}$ & $1.359\times 10^{-2}$ & $7.567\times 10^{-2}$ & $2.186\times 10^{-4}$ & $1.359\times 10^{-2}$ \\
4 & $1.567\times 10^{-2}$ & $3.357\times 10^{-4}$ & $6.310\times 10^{-3}$ & $1.567\times 10^{-2}$ & $3.357\times 10^{-4}$ & $6.310\times 10^{-3}$ \\
5 & $7.092\times 10^{-2}$ & $1.957\times 10^{-4}$ & $4.892\times 10^{-3}$ & $7.092\times 10^{-2}$ & $1.957\times 10^{-4}$ & $4.892\times 10^{-3}$ \\
6 & $1.647\times 10^{-2}$ & $1.596\times 10^{-4}$ & $5.358\times 10^{-3}$ & $1.647\times 10^{-2}$ & $1.596\times 10^{-4}$ & $5.358\times 10^{-3}$ \\
7 & $4.187\times 10^{-2}$ & $2.205\times 10^{-4}$ & $4.967\times 10^{-3}$ & $4.187\times 10^{-2}$ & $2.205\times 10^{-4}$ & $4.967\times 10^{-3}$ \\
\hline
\end{tabular}
\end{sidewaystable*}

\clearpage
\newpage

\clearpage
\newpage

\clearpage
\newpage

\begin{table}[h]
	\centering
	\caption{Number of genes where the weighted regression produced a larger Wald test statistic than the unweighted regression, and vice versa. Across all seven sample populations, the weighted regression consistently produced larger test statistics for more genes.}
	\label{tab:diff}
		\begin{tabular}{|c|c|c|c|}
      \hline
      Sample Population & Weighted $>$ Unweighted & Weighted $\leq$ Unweighted    & Total  \\
      \hline
        Archer Control & 3,526 & 949 & 4,475 \\	
		Archer Experimental & 2,772 & 1,827 & 4,599 \\
        Archer Combined & 2,728 & 961 & 3,689 \\	
        Braun & 6,766 & 849 & 7,615 \\	
        M\"{o}ller-Levet Control & 5,249 & 2,366 & 7,615 \\	
		M\"{o}ller-Levet Experimental & 5,887 & 1,728 & 7,615 \\
        M\"{o}ller-Levet Combined & 5,796 & 1,819 & 7,615 \\	
\hline
\end{tabular}%
\end{table}

\clearpage
\newpage

\appendix

\section{Derivation for Inverse of Variance Submatrix in Section \ref{sec:2.2}} \label{app:A}
To simplify presentation, we drop the superscript $(g)$ and define $V = \text{Var}(Y, X; \hat{\theta})$. Recall that $\sigma^2 = 1$ for the motivating example in Section \ref{sec:2.2}. Finally, we borrow the notation of \citet{Kent1982} for partitioning the inverse of the variance matrix:
\begin{align*}
V^{-1} &= \begin{bmatrix}
V_{\theta_0, \theta_0} &  V_{\theta_0, \gamma} \\
 V_{\gamma, \theta_0} & V_{\gamma, \gamma}
\end{bmatrix}^{-1} \\
&= \mathbb{E}\left\{ f(X)f(X)^T\right\} \\
&= \left[
\begin{array}{c|c c}
    1 & \mathbb{E}\left\{\sin\left(\frac{\pi X}{12}\right)\right\} & \mathbb{E}\left\{\cos\left(\frac{\pi X}{12}\right)\right\} \\[0.5ex]
    \hline  \\[-2ex]
    \mathbb{E}\left\{\sin\left(\frac{\pi X}{12}\right)\right\} & \mathbb{E}\left\{\sin^2\left(\frac{\pi X}{12}\right)\right\} & \mathbb{E}\left\{\sin\left(\frac{\pi X}{12}\right)\cos\left(\frac{\pi X}{12}\right)\right\} \\[0.5ex] 
    \mathbb{E}\left\{\cos\left(\frac{\pi X}{12}\right)\right\} & \mathbb{E}\left\{\sin\left(\frac{\pi X}{12}\right)\cos\left(\frac{\pi X}{12}\right)\right\} & \mathbb{E}\left\{\cos^2\left(\frac{\pi X}{12}\right)\right\} 
    \end{array}
\right] \\
&=\begin{bmatrix}
V^{\theta_0, \theta_0} &  V^{\theta_0, \gamma} \\
 V^{\gamma, \theta_0} & V^{\gamma, \gamma}
\end{bmatrix}.
\end{align*}
By definition of the Schur complement, $V_{\gamma, \gamma}$ in (\ref{eq:2}) can be expressed as
\begin{align*}
V_{\gamma, \gamma} &= (V^{\gamma, \gamma} -V^{\gamma, \theta_0}V^{\theta_0, \theta_0} V^{\theta_0, \gamma})^{-1} \\
&=\Bigg(\begin{bmatrix}
    \mathbb{E}\left\{\sin^2\left(\frac{\pi X}{12}\right)\right\} & \mathbb{E}\left\{\sin\left(\frac{\pi X}{12}\right)\cos\left(\frac{\pi X}{12}\right)\right\} \\[0.5ex] 
    \mathbb{E}\left\{\sin\left(\frac{\pi X}{12}\right)\cos\left(\frac{\pi X}{12}\right)\right\} & \mathbb{E}\left\{\cos^2\left(\frac{\pi X}{12}\right)\right\} 
    \end{bmatrix} \\
    &\quad - \begin{bmatrix}
    \mathbb{E}\left\{\sin\left(\frac{\pi X}{12}\right)\right\} \\[0.5ex] \mathbb{E}\left\{\cos\left(\frac{\pi X}{12}\right)\right\}
    \end{bmatrix} 
    \begin{bmatrix}
    \mathbb{E}\left\{\sin\left(\frac{\pi X}{12}\right)\right\} & \mathbb{E}\left\{\cos\left(\frac{\pi X}{12}\right)\right\}
    \end{bmatrix} \Bigg)^{-1} \\
    &=\Bigg(\begin{bmatrix}
    \mathbb{E}\left\{\sin^2\left(\frac{\pi X}{12}\right)\right\} & \mathbb{E}\left\{\sin\left(\frac{\pi X}{12}\right)\cos\left(\frac{\pi X}{12}\right)\right\} \\[0.5ex]
    \mathbb{E}\left\{\sin\left(\frac{\pi X}{12}\right)\cos\left(\frac{\pi X}{12}\right)\right\} & \mathbb{E}\left\{\cos^2\left(\frac{\pi X}{12}\right)\right\}
    \end{bmatrix}  \\
    &\quad - \begin{bmatrix}
     \mathbb{E}\left\{\sin\left(\frac{\pi X}{12}\right)\right\}^2 & \mathbb{E}\left\{\sin\left(\frac{\pi X}{12}\right)\right\}\mathbb{E}\left\{\cos\left(\frac{\pi X}{12}\right)\right\} \\[0.5ex] 
    \mathbb{E}\left\{\sin\left(\frac{\pi X}{12}\right)\right\}\mathbb{E}\left\{\cos\left(\frac{\pi X}{12}\right)\right\} & \mathbb{E}\left\{\cos\left(\frac{\pi X}{12}\right)\right\}^2
    \end{bmatrix}\Bigg)^{-1},
\end{align*}
which implies
\begin{align*}
V_{\gamma, \gamma}^{-1} &=\begin{bmatrix}
    \mathbb{E}\left\{\sin^2\left(\frac{\pi X}{12}\right)\right\} & \mathbb{E}\left\{\sin\left(\frac{\pi X}{12}\right)\cos\left(\frac{\pi X}{12}\right)\right\} \\[0.5ex] 
    \mathbb{E}\left\{\sin\left(\frac{\pi X}{12}\right)\cos\left(\frac{\pi X}{12}\right)\right\} & \mathbb{E}\left\{\cos^2\left(\frac{\pi X}{12}\right)\right\}
    \end{bmatrix}  \\
    &\quad - \begin{bmatrix}
     \mathbb{E}\left\{\sin\left(\frac{\pi X}{12}\right)\right\}^2 & \mathbb{E}\left\{\sin\left(\frac{\pi X}{12}\right)\right\}\mathbb{E}\left\{\cos\left(\frac{\pi X}{12}\right)\right\} \\[0.5ex] 
    \mathbb{E}\left\{\sin\left(\frac{\pi X}{12}\right)\right\}\mathbb{E}\left\{\cos\left(\frac{\pi X}{12}\right)\right\} & \mathbb{E}\left\{\cos\left(\frac{\pi X}{12}\right)\right\}^2
    \end{bmatrix}.
\end{align*}

\noindent Now, from \citet[page 376]{Abramowitz1965}, the modified Bessel function of the first kind, or $I_{\psi}(\nu)$, can be expressed as
\begin{align*}
I_{\psi}(\nu) = \frac{1}{\pi}\int_0^{\pi}\text{exp}\{\nu \text{cos}(Z)\}\text{cos}(\psi Z) dZ
\end{align*}
when $\psi$ is an integer. This definition for $I_{\psi}(\nu)$ indicates that $\mathbb{E}\{\text{cos}(\pi X/12)\}$, $\mathbb{E}\{\text{cos}^2(\pi X/12)\}$, and $\mathbb{E}\{\text{sin}^2(\pi X/12)\}$ can each be expressed as
\begin{align*}
\mathbb{E}\left\{\text{cos}\left(\frac{\pi X}{12}\right)\right\}&=\frac{I_1(1)}{I_0(1)}, \\
\mathbb{E}\left\{\text{cos}^2\left(\frac{\pi X}{12}\right)\right\}&=\frac{1}{2}+\frac{\mathbb{E}\{\text{cos}(\pi X/6)\}}{2}=\frac{1}{2}+\frac{I_2(1)}{2I_0(1)}, \\ \mathbb{E}\left\{\text{sin}^2\left(\frac{\pi X}{12}\right)\right\}&= \frac{1}{2}-\frac{I_2(1)}{2I_0(1)},
\end{align*}
as the support of $X$ is on the interval $[0, 24]$. Given that $\text{sin}(X)$ and $\text{sin}(X)\text{cos}(X)$ are anti-symmetric functions whose expectations evaluate to zero when a probability density function is symmetric around zero, we conclude
\begin{align*}
V_{\gamma, \gamma}^{-1} &=\begin{bmatrix}
    \frac{1}{2}-\frac{I_2(1)}{2I_0(1)} & 0 \\ 
    0 & \frac{1}{2}+\frac{I_2(1)}{2I_0(1)} -\frac{I_1(1)^2}{I_0(1)^2}
    \end{bmatrix}.
\end{align*}

\section{Derivation for Upper Bound on Equation \ref{eq:opt}} \label{app:B}

Let $w$ be an $N$-dimensional vector of non-negative weights such that $\sum_{j=1}^N w_j = 1$. Define the matrix
\begin{align*}
    W_N(X, Y; \theta) = \sum_{j=1}^N w_j f(X_j)f(X_j)^T.
\end{align*}
Finally, recall Hadamard's inequality for a positive semi-definite matrix $Z$,
\begin{align*}
\text{det}(Z) \leq \prod_{j=1}^pZ_{j,j},
\end{align*}
where $Z_{j,j}$ denotes the $j$-th diagonal element of $Z$ \citep[Theorem 11]{Raski2017}. To upper bound $\text{det}\{W_N(X, Y; \theta)\}$, consider the pair of diagonal elements $\{\sum_{j=1}^N w_j \text{sin}^2(kX_j), \sum_{j=1}^N w_j \text{cos}^2(kX_j)\}$ for some arbitrary $k$. We find
\begin{align*}
\left\{\sum_{j=1}^N w_j \text{sin}^2(kX_j)\right\}\left\{\sum_{j=1}^N w_j \text{cos}^2(kX_j)\right\} &  = \left[\left\{ \sum_{j=1}^N w_j \right\} - \left\{\sum_{j=1}^N w_j \text{cos}^2(kX_j)\right\}\right]\left\{\sum_{j=1}^N w_j \text{cos}^2(kX_j)\right\} \\
&  = \left[ 1 - \left\{\sum_{j=1}^N w_j \text{cos}^2(kX_j)\right\}\right]\left\{\sum_{j=1}^N w_j \text{cos}^2(kX_j)\right\} \\
&  = \left\{\sum_{j=1}^N w_j \text{cos}^2(kX_j)\right\} - \left\{\sum_{j=1}^N w_j \text{cos}^2(kX_j)\right\}^2.
\end{align*}
Now, let $U = \sum_{j=1}^N w_j \text{cos}^2(kX_j)$, and define the function $g(U) = U-U^2$. It is apparent that $g(U)$ is a concave function that is maximized at $U=1/2$, with $g(1/2) = 1/4$, which implies
\begin{align*}
\left\{\sum_{j=1}^N w_j \text{sin}^2(kX_j)\right\}\left\{\sum_{j=1}^N w_j \text{cos}^2(kX_j)\right\} &\leq \frac{1}{4}.
\end{align*}
Given that $W_N(X, Y; \theta)$ is a $(2K+1)\times (2K+1)$ matrix where its first diagonal element equals one, it follows that $\text{det}\{W_N(X, Y; \theta)\}\leq 1/4^K$. Note from (\ref{eq:equi_mat}) that the matrix $A_N(Y, X; \theta) = \text{diag}(1, 1/2,\ldots, 1/2)$ when data are sampled from an equispaced design, with $\text{det}\{A_N(Y, X; \theta)\} = 1/4^K$. We can conclude that $\text{det}\{W_N(X, Y; \theta)\}$ is upper bounded by $\text{det}\{A_N(Y, X; \theta)\}$ when $A_N(Y, X; \theta)$ is computed with data sampled from an equispaced design.

\bibliographystyle{apalike} 
\bibliography{bibliography}

\begin{thebibliography}{}

\bibitem[Abramowitz and Stegun, 1965]{Abramowitz1965}
Abramowitz, M. and Stegun, I.~A., editors (1965).
\newblock {\em Handbook of mathematical functions}.
\newblock Dover Books on Mathematics. Dover Publications, Mineola, NY.

\bibitem[Aguet et~al., 2020]{gtex2020}
Aguet, F., Anand, S., Ardlie, K.~G., Gabriel, S., Getz, G.~A., Graubert, A., Hadley, K., Handsaker, R.~E., Huang, K.~H., Kashin, S., Li, X., MacArthur, D.~G., Meier, S.~R., Nedzel, J.~L., Nguyen, D.~T., Segr{\`{e}}, A.~V., Todres, E., Balliu, B., Barbeira, A.~N., Battle, A., Bonazzola, R., Brown, A., Brown, C.~D., Castel, S.~E., Conrad, D.~F., Cotter, D.~J., Cox, N., Das, S., de~Goede, O.~M., Dermitzakis, E.~T., Einson, J., Engelhardt, B.~E., Eskin, E., Eulalio, T.~Y., Ferraro, N.~M., Flynn, E.~D., Fresard, L., Gamazon, E.~R., Garrido-Mart{\'{\i}}n, D., Gay, N.~R., Gloudemans, M.~J., Guig{\'{o}}, R., Hame, A.~R., He, Y., Hoffman, P.~J., Hormozdiari, F., Hou, L., Im, H.~K., Jo, B., Kasela, S., Kellis, M., Kim-Hellmuth, S., Kwong, A., Lappalainen, T., Li, X., Liang, Y., Mangul, S., Mohammadi, P., Montgomery, S.~B., Mu{\~{n}}oz-Aguirre, M., Nachun, D.~C., Nobel, A.~B., Oliva, M., Park, Y., Park, Y., Parsana, P., Rao, A.~S., Reverter, F., Rouhana, J.~M., Sabatti, C., Saha, A., Stephens, M., Stranger, B.~E.,
  Strober, B.~J., Teran, N.~A., Vi{\~{n}}uela, A., Wang, G., Wen, X., Wright, F., Wucher, V., Zou, Y., Ferreira, P.~G., Li, G., Mel{\'{e}}, M., Yeger-Lotem, E., Barcus, M.~E., Bradbury, D., Krubit, T., McLean, J.~A., Qi, L., Robinson, K., Roche, N.~V., Smith, A.~M., Sobin, L., Tabor, D.~E., Undale, A., Bridge, J., Brigham, L.~E., Foster, B.~A., Gillard, B.~M., Hasz, R., Hunter, M., Johns, C., Johnson, M., Karasik, E., Kopen, G., Leinweber, W.~F., McDonald, A., Moser, M.~T., Myer, K., Ramsey, K.~D., Roe, B., Shad, S., Thomas, J.~A., Walters, G., Washington, M., Wheeler, J., Jewell, S.~D., Rohrer, D.~C., Valley, D.~R., Davis, D.~A., Mash, D.~C., Branton, P.~A., Barker, L.~K., Gardiner, H.~M., Mosavel, M., Siminoff, L.~A., Flicek, P., Haeussler, M., Juettemann, T., Kent, W.~J., Lee, C.~M., Powell, C.~C., Rosenbloom, K.~R., Ruffier, M., Sheppard, D., Taylor, K., Trevanion, S.~J., Zerbino, D.~R., Abell, N.~S., Akey, J., Chen, L., Demanelis, K., Doherty, J.~A., Feinberg, A.~P., Hansen, K.~D., Hickey, P.~F.,
  Jasmine, F., Jiang, L., Kaul, R., Kibriya, M.~G., Li, J.~B., Li, Q., Lin, S., Linder, S.~E., Pierce, B.~L., Rizzardi, L.~F., Skol, A.~D., Smith, K.~S., Snyder, M., Stamatoyannopoulos, J., Tang, H., Wang, M., Carithers, L.~J., Guan, P., Koester, S.~E., Little, A.~R., Moore, H.~M., Nierras, C.~R., Rao, A.~K., Vaught, J.~B., and Volpi, S. (2020).
\newblock The {GTEx} consortium atlas of genetic regulatory effects across human tissues.
\newblock {\em Science}, 369(6509):1318--1330.

\bibitem[Almon et~al., 2008]{Almon2008}
Almon, R.~R., Yang, E., Lai, W., Androulakis, I.~P., DuBois, D.~C., and Jusko, W.~J. (2008).
\newblock Circadian variations in rat liver gene expression: Relationships to drug actions.
\newblock {\em Journal of Pharmacology and Experimental Therapeutics}, 326(3):700–716.

\bibitem[Archer et~al., 2014]{Archer2014}
Archer, S.~N., Laing, E.~E., M\"{o}ller-Levet, C.~S., van~der Veen, D.~R., Bucca, G., Lazar, A.~S., Santhi, N., Slak, A., Kabiljo, R., von Schantz, M., Smith, C.~P., and Dijk, D.-J. (2014).
\newblock Mistimed sleep disrupts circadian regulation of the human transcriptome.
\newblock {\em Proceedings of the National Academy of Sciences}, 111(6).

\bibitem[Boos, 1992]{Boos1992}
Boos, D.~D. (1992).
\newblock On generalized score tests.
\newblock {\em The American Statistician}, 46(4):327.

\bibitem[Boos and Stefanski, 2013]{Boos2013}
Boos, D.~D. and Stefanski, L.~A. (2013).
\newblock {\em Essential statistical inference}.
\newblock Springer Texts in Statistics. Springer, New York, NY, 2013 edition.

\bibitem[Braun et~al., 2018]{Braun2018}
Braun, R., Kath, W.~L., Iwanaszko, M., Kula-Eversole, E., Abbott, S.~M., Reid, K.~J., Zee, P.~C., and Allada, R. (2018).
\newblock Universal method for robust detection of circadian state from gene expression.
\newblock {\em Proceedings of the National Academy of Sciences}, 115(39).

\bibitem[Braun et~al., 2019]{Braun2019}
Braun, R., Kath, W.~L., Iwanaszko, M., Kula-Eversole, E., Abbott, S.~M., Reid, K.~J., Zee, P.~C., and Allada, R. (2019).
\newblock Reply to laing et al.: Accurate prediction of circadian time across platforms.
\newblock {\em Proceedings of the National Academy of Sciences}, 116(12):5206–5208.

\bibitem[Carlucci et~al., 2019]{Carlucci2019}
Carlucci, M., Kriščiūnas, A., Li, H., Gibas, P., Koncevičius, K., Petronis, A., and Oh, G. (2019).
\newblock Discorhythm: an easy-to-use web application and r package for discovering rhythmicity.
\newblock {\em Bioinformatics}, 36(6):1952–1954.

\bibitem[Ceglia et~al., 2018]{Ceglia2018}
Ceglia, N., Liu, Y., Chen, S., Agostinelli, F., Eckel-Mahan, K., Sassone-Corsi, P., and Baldi, P. (2018).
\newblock {CircadiOmics}: circadian omic web portal.
\newblock {\em Nucleic Acids Research}, 46(W1):W157--W162.

\bibitem[Chan et~al., 2017]{Chan2017}
Chan, S., Zhang, L., Rowbottom, L., McDonald, R., Bjarnason, G.~A., Tsao, M., Barnes, E., Danjoux, C., Popovic, M., Lam, H., DeAngelis, C., and Chow, E. (2017).
\newblock Effects of circadian rhythms and treatment times on the response of radiotherapy for painful bone metastases.
\newblock {\em Annals of Palliative Medicine}, 6(1):14–25.

\bibitem[Chauhan et~al., 2017]{Chauhan2017}
Chauhan, R., Chen, K.-F., Kent, B.~A., and Crowther, D.~C. (2017).
\newblock Central and peripheral circadian clocks and their role in alzheimer’s disease.
\newblock {\em Disease Models \& Mechanisms}, 10(10):1187–1199.

\bibitem[Cheung et~al., 2003]{Cheung2003}
Cheung, V.~G., Conlin, L.~K., Weber, T.~M., Arcaro, M., Jen, K.-Y., Morley, M., and Spielman, R.~S. (2003).
\newblock Natural variation in human gene expression assessed in lymphoblastoid cells.
\newblock {\em Nature Genetics}, 33(3):422--425.

\bibitem[Cornelissen, 2014]{Cornelissen2014}
Cornelissen, G. (2014).
\newblock Cosinor-based rhythmometry.
\newblock {\em Theoretical Biology and Medical Modelling}, 11(1).

\bibitem[Cox and Takahashi, 2019]{Cox2019}
Cox, K.~H. and Takahashi, J.~S. (2019).
\newblock Circadian clock genes and the transcriptional architecture of the clock mechanism.
\newblock {\em Journal of Molecular Endocrinology}, 63(4):R93–R102.

\bibitem[Crnko et~al., 2019]{Crnko2019}
Crnko, S., Du~Pré, B.~C., Sluijter, J. P.~G., and Van~Laake, L.~W. (2019).
\newblock Circadian rhythms and the molecular clock in cardiovascular biology and disease.
\newblock {\em Nature Reviews Cardiology}, 16(7):437–447.

\bibitem[del Olmo et~al., 2022]{delolmo2022}
del Olmo, M., Sp\"{o}rl, F., Korge, S., J\"{u}rchott, K., Felten, M., Grudziecki, A., de~Zeeuw, J., Nowozin, C., Reuter, H., Blatt, T., Herzel, H., Kunz, D., Kramer, A., and Ananthasubramaniam, B. (2022).
\newblock Inter-layer and inter-subject variability of diurnal gene expression in human skin.
\newblock {\em {NAR} Genomics and Bioinformatics}, 4(4).

\bibitem[Dette and Studden, 1993]{Dette1993}
Dette, H. and Studden, W.~J. (1993).
\newblock Geometry of e-optimality.
\newblock {\em The Annals of Statistics}, 21(1).

\bibitem[{Di Marzio} et~al., 2021]{DiMarzio2021}
{Di Marzio}, M., Fensore, S., Panzera, A., and Taylor, C.~C. (2021).
\newblock Density estimation for circular data observed with errors.
\newblock {\em Biometrics}, 78(1):248--260.

\bibitem[Di~Marzio et~al., 2009]{DiMarzio2009}
Di~Marzio, M., Panzera, A., and Taylor, C.~C. (2009).
\newblock Local polynomial regression for circular predictors.
\newblock {\em Statistics \& Probability Letters}, 79(19):2066–2075.

\bibitem[Dibner and Schibler, 2015]{Dibner2015}
Dibner, C. and Schibler, U. (2015).
\newblock Circadian timing of metabolism in animal models and humans.
\newblock {\em Journal of Internal Medicine}, 277(5):513–527.

\bibitem[Federov, 1972]{Federov1972}
Federov, V.~V. (1972).
\newblock {\em Theory of optimal experiments}.
\newblock Springer, New York.

\bibitem[Gentry et~al., 2021]{Gentry2021}
Gentry, N.~W., Ashbrook, L.~H., Fu, Y.-H., and Ptáček, L.~J. (2021).
\newblock Human circadian variations.
\newblock {\em Journal of Clinical Investigation}, 131(16).

\bibitem[Gorczyca et~al., 2024]{Gorczyca2024}
Gorczyca, M., McDonald, T., and Sefas, J. (2024).
\newblock A corrected score function framework for modelling circadian gene expression.
\newblock {\em arXiv:2401.01998}.

\bibitem[Halberg et~al., 2013]{Halberg2013}
Halberg, F., Powell, D., Otsuka, K., Watanabe, Y., Beaty, L.~A., Rosch, P., Czaplicki, J., Hillman, D., Schwartzkopff, O., and Cornelissen, G. (2013).
\newblock Diagnosing vascular variability anomalies, not only mesor-hypertension.
\newblock {\em American Journal of Physiology-Heart and Circulatory Physiology}, 305(3):H279–H294.

\bibitem[Haus, 2009]{Haus2009}
Haus, E. (2009).
\newblock Chronobiology in oncology.
\newblock {\em International Journal of Radiation Oncology*Biology*Physics}, 73(1):3–5.

\bibitem[Hoffman et~al., 2019]{Hoffman2019}
Hoffman, G.~E., Bendl, J., Voloudakis, G., Montgomery, K.~S., Sloofman, L., Wang, Y.-C., Shah, H.~R., Hauberg, M.~E., Johnson, J.~S., Girdhar, K., Song, L., Fullard, J.~F., Kramer, R., Hahn, C.-G., Gur, R., Marenco, S., Lipska, B.~K., Lewis, D.~A., Haroutunian, V., Hemby, S., Sullivan, P., Akbarian, S., Chess, A., Buxbaum, J.~D., Crawford, G.~E., Domenici, E., Devlin, B., Sieberts, S.~K., Peters, M.~A., and Roussos, P. (2019).
\newblock Commonmind consortium provides transcriptomic and epigenomic data for schizophrenia and bipolar disorder.
\newblock {\em Scientific Data}, 6(1).

\bibitem[Huang and Braun, 2024]{Huang2024}
Huang, Y. and Braun, R. (2024).
\newblock Platform-independent estimation of human physiological time from single blood samples.
\newblock {\em Proceedings of the National Academy of Sciences}, 121(3).

\bibitem[Hughes et~al., 2017]{Hughes2017}
Hughes, M.~E., Abruzzi, K.~C., Allada, R., Anafi, R., Arpat, A.~B., Asher, G., Baldi, P., de~Bekker, C., Bell-Pedersen, D., Blau, J., Brown, S., Ceriani, M.~F., Chen, Z., Chiu, J.~C., Cox, J., Crowell, A.~M., DeBruyne, J.~P., Dijk, D.-J., DiTacchio, L., Doyle, F.~J., Duffield, G.~E., Dunlap, J.~C., Eckel-Mahan, K., Esser, K.~A., FitzGerald, G.~A., Forger, D.~B., Francey, L.~J., Fu, Y.-H., Gachon, F., Gatfield, D., de~Goede, P., Golden, S.~S., Green, C., Harer, J., Harmer, S., Haspel, J., Hastings, M.~H., Herzel, H., Herzog, E.~D., Hoffmann, C., Hong, C., Hughey, J.~J., Hurley, J.~M., de~la Iglesia, H.~O., Johnson, C., Kay, S.~A., Koike, N., Kornacker, K., Kramer, A., Lamia, K., Leise, T., Lewis, S.~A., Li, J., Li, X., Liu, A.~C., Loros, J.~J., Martino, T.~A., Menet, J.~S., Merrow, M., Millar, A.~J., Mockler, T., Naef, F., Nagoshi, E., Nitabach, M.~N., Olmedo, M., Nusinow, D.~A., Pt{\'{a}}{\v{c}}ek, L.~J., Rand, D., Reddy, A.~B., Robles, M.~S., Roenneberg, T., Rosbash, M., Ruben, M.~D., Rund, S.~S., Sancar,
  A., Sassone-Corsi, P., Sehgal, A., Sherrill-Mix, S., Skene, D.~J., Storch, K.-F., Takahashi, J.~S., Ueda, H.~R., Wang, H., Weitz, C., Westermark, P.~O., Wijnen, H., Xu, Y., Wu, G., Yoo, S.-H., Young, M., Zhang, E.~E., Zielinski, T., and Hogenesch, J.~B. (2017).
\newblock Guidelines for genome-scale analysis of biological rhythms.
\newblock {\em Journal of Biological Rhythms}, 32(5):380--393.

\bibitem[Hughes et~al., 2009]{Hughes2009}
Hughes, M.~E., DiTacchio, L., Hayes, K.~R., Vollmers, C., Pulivarthy, S., Baggs, J.~E., Panda, S., and Hogenesch, J.~B. (2009).
\newblock Harmonics of circadian gene transcription in mammals.
\newblock {\em PLoS Genetics}, 5(4):e1000442.

\bibitem[Kent, 1982]{Kent1982}
Kent, J.~T. (1982).
\newblock Robust properties of likelihood ratio test.
\newblock {\em Biometrika}, 69(1):19.

\bibitem[Kitsos et~al., 1988]{Kitsos1988}
Kitsos, C.~P., Titterington, D.~M., and Torsney, B. (1988).
\newblock An optimal design problem in rhythmometry.
\newblock {\em Biometrics}, 44(3):657--671.

\bibitem[Koike et~al., 2012]{Koike2012}
Koike, N., Yoo, S.-H., Huang, H.-C., Kumar, V., Lee, C., Kim, T.-K., and Takahashi, J.~S. (2012).
\newblock Transcriptional architecture and chromatin landscape of the core circadian clock in mammals.
\newblock {\em Science}, 338(6105):349–354.

\bibitem[Laing et~al., 2019]{Laing2019}
Laing, E.~E., M\"{o}ller-Levet, C.~S., Archer, S.~N., and Dijk, D.-J. (2019).
\newblock Universal and robust assessment of circadian time?
\newblock {\em Proceedings of the National Academy of Sciences}, 116(12):5205–5205.

\bibitem[Lee, 2010]{Lee2010}
Lee, A. (2010).
\newblock Circular data.
\newblock {\em WIREs Computational Statistics}, 2(4):477–486.

\bibitem[Lemos et~al., 2006]{Lemos2006}
Lemos, D.~R., Downs, J.~L., and Urbanski, H.~F. (2006).
\newblock Twenty-four-hour rhythmic gene expression in the rhesus macaque adrenal gland.
\newblock {\em Molecular Endocrinology}, 20(5):1164–1176.

\bibitem[Li et~al., 2013]{Li2013}
Li, J.~Z., Bunney, B.~G., Meng, F., Hagenauer, M.~H., Walsh, D.~M., Vawter, M.~P., Evans, S.~J., Choudary, P.~V., Cartagena, P., Barchas, J.~D., Schatzberg, A.~F., Jones, E.~G., Myers, R.~M., Watson, S.~J., Akil, H., and Bunney, W.~E. (2013).
\newblock Circadian patterns of gene expression in the human brain and disruption in major depressive disorder.
\newblock {\em Proceedings of the National Academy of Sciences}, 110(24):9950–9955.

\bibitem[Lowrey and Takahashi, 2004]{Lowrey2004}
Lowrey, P.~L. and Takahashi, J.~S. (2004).
\newblock Mammalian circadian biology: Elucidating genome-wide levels of temporal organization.
\newblock {\em Annual Review of Genomics and Human Genetics}, 5(1):407–441.

\bibitem[Lowrey and Takahashi, 2011]{Lowrey2011}
Lowrey, P.~L. and Takahashi, J.~S. (2011).
\newblock {\em Genetics of Circadian Rhythms in Mammalian Model Organisms}, page 175–230.
\newblock Elsevier.

\bibitem[Mei et~al., 2020]{Mei2020}
Mei, W., Jiang, Z., Chen, Y., Chen, L., Sancar, A., and Jiang, Y. (2020).
\newblock Genome-wide circadian rhythm detection methods: systematic evaluations and practical guidelines.
\newblock {\em Briefings in Bioinformatics}, 22(3).

\bibitem[M\"{o}ller-Levet et~al., 2013]{MllerLevet2013}
M\"{o}ller-Levet, C.~S., Archer, S.~N., Bucca, G., Laing, E.~E., Slak, A., Kabiljo, R., Lo, J. C.~Y., Santhi, N., von Schantz, M., Smith, C.~P., and Dijk, D.-J. (2013).
\newblock Effects of insufficient sleep on circadian rhythmicity and expression amplitude of the human blood transcriptome.
\newblock {\em Proceedings of the National Academy of Sciences}, 110(12).

\bibitem[Moser and Lin, 1992]{Moser1992}
Moser, B.~K. and Lin, Y.-R. (1992).
\newblock Equivalence of the corrected {F}-test and the weighted least squares procedure.
\newblock {\em The American Statistician}, 46(2):122–124.

\bibitem[Mure et~al., 2018]{Mure2018}
Mure, L.~S., Le, H.~D., Benegiamo, G., Chang, M.~W., Rios, L., Jillani, N., Ngotho, M., Kariuki, T., Dkhissi-Benyahya, O., Cooper, H.~M., and Panda, S. (2018).
\newblock Diurnal transcriptome atlas of a primate across major neural and peripheral tissues.
\newblock {\em Science}, 359(6381).

\bibitem[Parsons et~al., 2019]{Parsons2019}
Parsons, R., Parsons, R., Garner, N., Oster, H., and Rawashdeh, O. (2019).
\newblock Circacompare: a method to estimate and statistically support differences in mesor, amplitude and phase, between circadian rhythms.
\newblock {\em Bioinformatics}, 36(4):1208–1212.

\bibitem[Pukelsheim, 2006]{Pukelsheim2006}
Pukelsheim, F. (2006).
\newblock {\em Optimal Design of Experiments}.
\newblock Springer Texts in Statistics. Society for Industrial and Applied Mathematics, Philadelphia, PA.

\bibitem[Rijo-Ferreira and Takahashi, 2019]{RijoFerreira2019}
Rijo-Ferreira, F. and Takahashi, J.~S. (2019).
\newblock Genomics of circadian rhythms in health and disease.
\newblock {\em Genome Medicine}, 11(1).

\bibitem[Rockman and Kruglyak, 2006]{Rockman2006}
Rockman, M.~V. and Kruglyak, L. (2006).
\newblock Genetics of global gene expression.
\newblock {\em Nature Reviews Genetics}, 7(11):862--872.

\bibitem[Ruben et~al., 2018]{Ruben2018}
Ruben, M.~D., Wu, G., Smith, D.~F., Schmidt, R.~E., Francey, L.~J., Lee, Y.~Y., Anafi, R.~C., and Hogenesch, J.~B. (2018).
\newblock A database of tissue-specific rhythmically expressed human genes has potential applications in circadian medicine.
\newblock {\em Science Translational Medicine}, 10(458).

\bibitem[Różański et~al., 2017]{Raski2017}
Różański, M., Wituła, R., and Hetmaniok, E. (2017).
\newblock More subtle versions of the hadamard inequality.
\newblock {\em Linear Algebra and its Applications}, 532:500–511.

\bibitem[Schernhammer et~al., 2003]{Schernhammer2003}
Schernhammer, E.~S., Laden, F., Speizer, F.~E., Willett, W.~C., Hunter, D.~J., Kawachi, I., Fuchs, C.~S., and Colditz, G.~A. (2003).
\newblock Night-shift work and risk of colorectal cancer in the nurses’ health study.
\newblock {\em JNCI Journal of the National Cancer Institute}, 95(11):825–828.

\bibitem[Sigurdardottir et~al., 2012]{Sigurdardottir2012}
Sigurdardottir, L.~G., Valdimarsdottir, U.~A., Fall, K., Rider, J.~R., Lockley, S.~W., Schernhammer, E., and Mucci, L.~A. (2012).
\newblock Circadian disruption, sleep loss, and prostate cancer risk: A systematic review of epidemiologic studies.
\newblock {\em Cancer Epidemiology, Biomarkers \& Prevention}, 21(7):1002–1011.

\bibitem[Stranger et~al., 2007]{Stranger2007}
Stranger, B.~E., Nica, A.~C., Forrest, M.~S., Dimas, A., Bird, C.~P., Beazley, C., Ingle, C.~E., Dunning, M., Flicek, P., Koller, D., Montgomery, S., Tavar{\'{e}}, S., Deloukas, P., and Dermitzakis, E.~T. (2007).
\newblock Population genomics of human gene expression.
\newblock {\em Nature Genetics}, 39(10):1217--1224.

\bibitem[Tong, 1976]{Tong1976}
Tong, Y.~L. (1976).
\newblock Parameter estimation in studying circadian rhythms.
\newblock {\em Biometrics}, 32(1):85.

\bibitem[Tovin et~al., 2012]{Tovin2012}
Tovin, A., Alon, S., Ben-Moshe, Z., Mracek, P., Vatine, G., Foulkes, N.~S., Jacob-Hirsch, J., Rechavi, G., Toyama, R., Coon, S.~L., Klein, D.~C., Eisenberg, E., and Gothilf, Y. (2012).
\newblock Systematic identification of rhythmic genes reveals camk1gb as a new element in the circadian clockwork.
\newblock {\em PLoS Genetics}, 8(12):e1003116.

\bibitem[Walker et~al., 2020]{Walker2020}
Walker, W.~H., Walton, J.~C., DeVries, A.~C., and Nelson, R.~J. (2020).
\newblock Circadian rhythm disruption and mental health.
\newblock {\em Translational Psychiatry}, 10(1).

\bibitem[Wittenbrink et~al., 2018]{Wittenbrink2018}
Wittenbrink, N., Ananthasubramaniam, B., M\"{u}nch, M., Koller, B., Maier, B., Weschke, C., Bes, F., de~Zeeuw, J., Nowozin, C., Wahnschaffe, A., Wisniewski, S., Zaleska, M., Bartok, O., Ashwal-Fluss, R., Lammert, H., Herzel, H., Hummel, M., Kadener, S., Kunz, D., and Kramer, A. (2018).
\newblock High-accuracy determination of internal circadian time from a single blood sample.
\newblock {\em Journal of Clinical Investigation}, 128(9):3826--3839.

\bibitem[Yang et~al., 2020]{Yang2020}
Yang, Y., Li, Y., Sancar, A., and Oztas, O. (2020).
\newblock The circadian clock shapes the arabidopsis transcriptome by regulating alternative splicing and alternative polyadenylation.
\newblock {\em Journal of Biological Chemistry}, 295(22):7608–7619.

\bibitem[Zhang et~al., 2014]{Zhang2014}
Zhang, R., Lahens, N.~F., Ballance, H.~I., Hughes, M.~E., and Hogenesch, J.~B. (2014).
\newblock A circadian gene expression atlas in mammals: Implications for biology and medicine.
\newblock {\em Proceedings of the National Academy of Sciences}, 111(45):16219–16224.

\bibitem[Zong et~al., 2023]{Zong2023}
Zong, W., Seney, M.~L., Ketchesin, K.~D., Gorczyca, M.~T., Liu, A.~C., Esser, K.~A., Tseng, G.~C., McClung, C.~A., and Huo, Z. (2023).
\newblock Experimental design and power calculation in omics circadian rhythmicity detection using the cosinor model.
\newblock {\em Statistics in Medicine}, 42(18):3236--3258.

\end{thebibliography}

\end{document}